\documentclass[fleqn,usenatbib]{mnras}

\usepackage{newtxtext,newtxmath}
\usepackage[T1]{fontenc}

\DeclareRobustCommand{\VAN}[3]{#2}
\let\VANthebibliography\thebibliography
\def\thebibliography{\DeclareRobustCommand{\VAN}[3]{##3}\VANthebibliography}

\usepackage{graphicx}	
\usepackage{amsmath}	
\usepackage{caption}
\usepackage{fancyhdr}
\usepackage{hyperref}
\usepackage{makecell}
\usepackage[table]{xcolor}
\usepackage{colortbl}
\usepackage{enumitem} 
\usepackage{comment}
\usepackage{needspace}

\graphicspath{ {./figures/} }

\definecolor{darkgreen}{rgb}{0.0,0.65,0.0}

\definecolor{mypurple}{rgb}{0.6,0.0,0.8}

\title[The ThunderKAT Radio:X-ray Plane]{The Homogeneous MeerKAT and \emph{Swift}/XRT X-ray Binary Radio:X-ray Plane}

\author[J. Crook-Mansour et al.]{
Justine Crook-Mansour$^{1}$\thanks{E-mail: justine.crook-mansour@physics.ox.ac.uk},
Rob Fender$^{1, 2}$,
Andrew Hughes$^{1}$, 
Sara Motta$^{1,3}$, 
Patrick A. Woudt$^{2}$,\newauthor 
Arash Bahramian$^{4}$,
Melania Del Santo$^{5}$, 
Zuobin Zhang$^{1}$, 
Thomas D. Russell$^{5}$,
Jakob van den Eijnden$^{6}$, \newauthor 
Joe Bright$^{1}$, 
David Williams-Baldwin$^{7}$,
Francesco Carotenuto$^{8}$, 
St\'{e}phane Corbel$^{9}$,
Fraser J. Cowie$^{1}$,\newauthor
Alex Andersson$^{1}$, 
Noa Grollimund$^{9}$,
James Matthews$^{1}$,
Kelebogile Gasealahwe$^{2,10}$,\newauthor 
Itumeleng Monaleng$^{2,10}$,
Lauren Rhodes$^{11,12}$,
Payaswini Saikia$^{13}$,
Katie Savard$^{1}$, 
Evangelia Tremou$^{14}$,\newauthor 
Xian Zhang$^{15}$
\\
$^{1}$Astrophysics, Department of Physics, University of Oxford, Keble Road, Oxford, OX1 3RH, UK\\
$^{2}$ Department of Astronomy, University of Cape Town, Private Bag X3, Rondebosch 7701, South Africa \\
$^{3}$ Istituto Nazionale di Astrofisica (INAF), Osservatorio Astronomico di Brera, via E. Bianchi 46, 23807 Merate (LC), Italy\\
$^{4}$ International Centre for Radio Astronomy Research, Curtin University, GPO Box U1987, Perth, WA 6845, Australia\\
$^{5}$ INAF, Istituto di Astrofisica Spaziale e Fisica Cosmica, Via U. La Malfa 153, I-90146 Palermo, Italy\\
$^{6}$ Anton Pannekoek Institute for Astronomy, University of Amsterdam, Science Park 904, 1098 XH Amsterdam, The Netherlands \\
$^{7}$ Jodrell Bank Centre for Astrophysics, Dept. of Physics \& Astronomy, University of Manchester, Manchester M13 9PL, UK \\
$^{8}$ INAF, Osservatorio Astronomico di Roma, Via Frascati 33, I-00078 Monte Porzio Catone, Italy \\
$^{9}$Universit\'e Paris Cit\'e, Universit\'e Paris-Saclay, CEA, CNRS, AIM, F-91190 Gif-sur-Yvette, France\\
$^{10}$South African Astronomical Observatory, PO Box 9, Observatory, Cape Town 7935, South Africa \\
$^{11}$Department of Physics, McGill University, 3600 rue University, Montr\'eal, QC H3A 2T8, Canada\\
$^{12}$Trottier Space Institute, McGill University, 3550 rue University, Montr\'eal, QC H3A 2A7, Canada\\
$^{13}$Department of Astronomy, Yale University, PO Box 208101, New Haven, CT 06520-8101, USA \\
$^{14}$National Radio Astronomy Observatory, P.O. Box O, Socorro, NM 87801, USA\\
$^{15}$College of Physics, Guizhou University, Guiyang 550025, China
}

\date{Accepted XXX. Received YYY; in original form ZZZ}

\pubyear{\the\year{}}

\begin{document}
\label{firstpage}
\pagerange{\pageref{firstpage}--\pageref{lastpage}}
\maketitle


\begin{abstract}During the hard and quiescent spectral states in X-ray binaries, a non-linear correlation is observed between radio and X-ray luminosities, providing a valuable tool to probe the connection between accretion and jet production. This relation was originally thought to define a single `standard' correlation spanning several orders of magnitude in X-ray luminosity, and was extended to active galactic nuclei by including a mass term. However, subsequent studies revealed a more complex picture, with some sources deviating from the standard correlation and instead populating distinct tracks. To date, all large studies of the radio:X-ray plane have combined data from multiple telescopes, introducing uncertainties due to differing instrument systematics and flux conversions between observing frequencies, thereby complicating comparisons and limiting constraints. ThunderKAT was a five-year programme on the MeerKAT radio telescope that monitored X-ray binaries in outburst, and ran alongside \emph{Swift}KAT which provided quasi-simultaneous \emph{Swift}/XRT X-ray coverage. We present the full set of light curves from these programmes, comprising 948 radio and 1029 X-ray data points. An important finding is the frequent detection of unresolved radio emission during the soft state, likely dominated by previously launched jet ejecta. Using these data, we construct the largest, observationally homogeneous X-ray binary radio:X-ray plane to date. We relate these results to the physical mechanisms proposed to drive inter-source diversity, and outline directions for future observational and theoretical work. This paper is accompanied by a public data release of the ThunderKAT and \emph{Swift}KAT measurements and a compiled radio:X-ray plane, available through an interactive website. 
\end{abstract}



\section{Introduction}

X-ray binaries (XRBs) are X-ray sources in which a compact object -- either a neutron star (NS) or black hole (BH) -- accretes matter from a stellar companion (donor star). When the XRB has a low-mass donor star ($\lesssim$1$M_\odot$), called a \emph{low-mass XRB} (LMXB), the mass transfer typically occurs through Roche lobe overflow. Accretion may give rise to highly collimated, relativistic outflows, referred to as \emph{`jets'}, which are broadly classified as either steady \emph{`compact jets’} or transient \emph{`ejecta’} (see, e.g., \citealt{fender_2006} for a review). These outflows are analogous to those observed from accreting supermassive BHs in the centres of galaxies (i.e., Active Galactic Nuclei, AGN). In both BH and NS systems, jets may be powered by disc-anchored magnetic fields \citep{blandford_1982}. Alternatively, they may arise from the extraction of rotational energy from the compact object -- via the Blandford-Znajek process in BHs \citep{blandford_1977}, or through analogous mechanisms in NSs (e.g., \citealt{parfrey_2016, das_2022}). In NS systems, disc-magnetosphere coupling can also launch outflows in a `propeller' mode (e.g., \citealt{illarionov_1975, romanova_2005}). 

In \emph{high-mass XRBs} (HMXBs; see reviews by \citealt{kretschmar_2019}; \citealt{fornasini_2023}), the donor is a massive ($\gtrsim$10$M_\odot$), typically O/B star. In these systems, accretion is primarily wind-fed, though transient or persistent accretion discs may form. Jets are also occasionally observed, especially in BH systems such as Cygnus X-1 (e.g., \citealt{stirling_2001}). 

Many LMXBs are transient systems that spend large portions of their lifetimes in a \emph{quiescent state} (QS), characterised by X-ray luminosities of $\sim$$10^{30}{-}10^{33}$ erg s$^{-1}$ and very low mass accretion rates onto the compact object \citep{remillard_2006}. These periods are occasionally interrupted by outbursts that are often detected by all-sky X-ray monitors, and usually last from weeks to months, but in some cases years (e.g., \citealt{tetarenko_2016}). During outbursts of BH LMXBs, the accretion flow undergoes transitions between distinct states, defined by their X-ray spectral and timing properties (e.g., \citealt{remillard_2006, belloni_motta_2016, de_marco_2022}), which are accompanied by correlated changes in the radio properties. This evolution is commonly illustrated using hardness-intensity diagrams (HIDs); see, e.g., Figure 7 of \cite{fender_2004c}.

At the start of a canonical outburst, as the X-ray flux increases, the source resides in the so-called \emph{hard state} (HS), during which the X-ray spectrum is dominated by a power law with an exponential cut-off at high energies. This emission is generally interpreted as arising from inverse Compton scattering of soft seed photons in a hot, optically thin plasma, commonly referred to as a corona (e.g., \citealt{narayan_2005, mcClintock_2006,  poutanen_2014}). The seed photons may originate either from a cool, geometrically thin accretion disc truncated at large radii, or from synchrotron radiation produced by hot electrons in the inner flow near the BH. During the HS, the source's power density spectra (PDS) exhibit strong variability, with a root mean square (RMS) amplitude typically of $\gtrsim$30 per cent (e.g., \citealt{belloni_2010}). Quasi-periodic oscillations (QPOs; see \citealt{ingram_motta_2019} for a review) are also observed during intervals of relatively high flux. In addition, the source produces a continuously-launched compact jet, observed as milli-arcsecond-scale structure in a few systems, such as the BH XRBs Cygnus X-1 \citep{stirling_2001}, GRS 1915+105 \citep{dhawan_2000}, and Swift J1727.8--1613 \citep{wood_2024}. The emission from these jets is thought to arise from the superposition of multiple partially self-absorbed synchrotron components, which results in a flat or slightly inverted radio spectrum ($\alpha \gtrsim 0$, where the flux density $F_\nu$ scales with frequency $\nu$ as $F_\nu \propto \nu^\alpha$) up to a break frequency where the spectrum becomes optically thin ($\alpha \sim -0.6$), typically in the sub-mm or near-infrared regimes (e.g., \citealt{corbel_fender_2002, russell_2014}). Some models additionally suggest that synchrotron and synchrotron self-Compton emission from the base of the compact jet may contribute significantly or even dominate the X-ray emission at low luminosities (e.g., \citealt{markoff_2001, markoff_2003, markoff_2005, falcke_2004, russell_2010, laurent_2011}).

As the outburst progresses, the X-ray flux increases (sometimes close to the Eddington luminosity, $L_\text{Edd}$\footnote{The maximum radiative output at which outward radiation pressure balances gravitational attraction. For LMXBs, this is typically $\sim$$10^{38}{-}10^{39}$ erg s$^{-1}$.}), the spectrum progressively softens, and the source transitions through the \emph{intermediate state} (IMS), which is often subdivided into the hard and soft IMS (HIMS and SIMS, respectively). During this state transition, the system undergoes significant changes in its X-ray emission -- reflecting an evolution of the accretion flow -- including changes in the properties of the observed QPOs (e.g., \citealt{belloni_2010, motta_2011, miller_jones_2012b, wood_2025}). Around the same time, the compact jet is thought to quench (e.g., \citealt{russell_2019_quench, russell_2020_quench}), and the system frequently exhibits short-timescale ($\lesssim$ days) radio flares, which are commonly associated with the ejection of discrete knots or blobs of material (e.g., \citealt{fender_2004c, fender_2009, miller_jones_2012b, russell_2019_quench,  bright_2020, wood_2021}). Many of these flares show a spectral evolution from optically thick to thin emission, typically interpreted as synchrotron self-absorption in expanding ejecta (e.g., \citealt{fender_2019}). However, discrete ejecta are not always detected following radio flares, which could reflect observational limitations -- such as Doppler deboosting or insufficient angular resolution -- or may indicate that not all XRBs produce discrete ejecta during this state transition (e.g., \citealt{rushton_2012, paragi_2013}). When detected, ejecta typically display steep, optically thin radio emission that is decoupled from the accretion flow, and they may become spatially resolved as they propagate away from the compact radio core (e.g., \citealt{mirabel_1994, hjellming_1995, corbel_2002, rushton_2017, russell_2019_quench, bright_2020, carotenuto_2021a, bahramian_2023}). Despite extensive observational study, the physical mechanism(s) responsible for launching these ejecta -- and for producing the associated radio flaring -- are not well understood. Some studies have suggested that the breakup of the compact jet and launch of ejecta may be related to the propagation of an internal shock caused when faster-moving plasma catches up with slower-moving plasma (e.g., \citealt{kaiser_2000, jamil_2010}), or due to the ejection of the corona (e.g., \citealt{rodriguez_2003, rodriguez_2008}). It has also been suggested that compact jet quenching may be related to the jet acceleration zone -- traced observationally by the jet spectral break -- becoming disconnected from the system, with its outward propagation giving rise to the observed ejecta \citep{russell_2020_quench, echibur_2024}.

Following the IMS, the source typically enters the \emph{soft state} (SS), during which the X-ray emission is dominated by black-body-like radiation from a geometrically thin, optically thick accretion disc extending down close to the innermost stable circular orbit around the BH \citep{shakura_1973}. The X-ray variability is correspondingly weak, with an RMS amplitude of $\lesssim$1 per cent (e.g., \citealt{belloni_2010}). During this state, the compact jet is generally observed to be strongly suppressed (e.g., \citealt{fender_1999b, corbel_2000, coriat_2011, rushton_2016, russell_2019_quench}). 

Thereafter, as the accretion rate drops and the X-ray luminosity begins to fade, the system transitions back to the HS, during which the compact jet is re-established (e.g., \citealt{fender_2009, miller_jones_2012b}), and then it finally returns to quiescence. 

This chain of events is considered typical. However, some sources display \textit{`failed'} outbursts that do not complete a full HS$\rightarrow$SS transition, either remaining in the HS or only transitioning as far as the IMS (e.g., \citealt{brocksopp_2001, curran_2014, tetarenko_2016, delsanto16, bassi19}). Others exhibit more complex outburst behaviour, including oscillations between the SS and HS that are sometimes accompanied by multiple radio flares and discrete ejecta (e.g., \citealt{homan_2001, russell_2022}), and/or rebrightenings following the main outburst (e.g., \citealt{parikh_2019, zhang_2019}).

For NS LMXBs, the spectral-timing states are historically named differently from those of BHs, although the HS/IMS/SS classification also works well (e.g., \citealt{van_der_klis_2006, munoz_darias_2014}). Weakly-magnetised NSs accreting above $\sim$0.01 $L_\text{Edd}$ can be divided into two classes based on their tracks in the X-ray colour-colour diagram: the \emph{Z} and \emph{atoll} sources \citep{hasinger_1989}. The difference between these classes and their various sub-classes is thought to be driven by the instantaneous mass accretion rate \citep{lin_2009a, homan_2010}. Atolls are broadly analogous to BHs accreting at $\lesssim$0.5 $L_\text{Edd}$, while Z-sources are associated with near-Eddington accretion \citep{munoz_darias_2014}. Another class of NS LMXBs is the \emph{accreting millisecond X-ray pulsars (AMXPs)}, in which coherent millisecond X-ray pulsations can be observed during outburst states (see \citealt{patruno_2021} for a review). A related subclass is the transitional millisecond pulsars (tMSPs), which switch between accretion-powered and rotation-powered states (see \citealt{papitto_2022} for a review).

\subsection{The Radio:X-ray Plane}

Despite the diversity of XRB phenomenology, clear empirical patterns emerge across these systems. One of the most prominent is the correlation between radio and X-ray luminosities for XRBs in the HS and QS, which provides strong evidence for a coupling between the compact jet (traced by radio emission) and the inner accretion flow or jet base (traced by X-rays), although the underlying physics is not yet fully understood.

The first evidence for this correlation was reported for the BH LMXB GX~339$-$4 \citep{hanni_1998, corbel_2000, corbel_2003a}. Shortly thereafter, \citet{gallo_2003} compiled data for ten BH candidates and proposed a `universal' radio:X-ray correlation in the HS and QS of the form $L_R \propto L_X^{\beta}$, where $L_R$ and $L_X$ denote the radio and X-ray luminosities, respectively, and $\beta$ is the power-law index or correlation coefficient. Studies measured coefficients in the range $\beta \sim 0.5{-}0.7$ for various sources \citep{corbel_2003a, gallo_2003, gallo_2006}, and this correlation became known as the \emph{`standard' track}. These and later observations appeared to confirm that the correlation extends down to very low quiescent luminosities (e.g., \citealt{gallo_2003, gallo_2006, corbel_2008,  corbel_2013, plotkin_2017a, tremou_2020}), although this remains debated (e.g., \citealt{rodriguez_2020}).

Contemporaneously, \citet{merloni_2003} and \citet{falcke_2004} showed that once BH mass is included, the correlation can be extended to AGN, implying that similar mechanisms couple accretion and ejection processes in BHs over $\sim$9 orders of magnitude in mass. Furthermore, it was shown that this correlation is not simply an artifact of plotting two distance-dependent values (see, e.g., \citealt{merloni_2006}). Shortly after, \citet{migliari_2006} suggested that a similar correlation exists for NS XRBs, but with a steeper correlation coefficient ($\beta \sim 1.4$) and fainter $L_R$ for a given $L_X$ than seen in BH systems. In addition, a similar correlation was shown to hold between optical-infrared and X-ray fluxes in both BH and NS systems \citep{homan_2005b, russell_2006, russell_2007, coriat_2009}. 

Over the years, virtually all newly identified jetted accreting sources were placed on the $L_R$--$L_X$ plane when quasi-simultaneous radio and X-ray observations were available. As these new observations accumulated, it became clear that some BH XRBs did not follow the standard $L_R$--$L_X$ correlation, casting doubt over its universality \citep{xue_2007}. Examples of these so-called \emph{`outliers'} include GRS 1716--249 \citep{bassi19}, H1743--322 \citep{jonker_2010, coriat_2011}, IGR J17497$-$2821 \citep{rodriguez_2007}, Swift J1753.5--0127 \citep{cadolle_2007, soleri_2010}, and XTE J1650--500 \citep{corbel_2004}. Some of these sources follow a steep high-luminosity branch (typically $\beta \gtrsim 1$) as well as a lower-luminosity \emph{`transitional'} branch that rejoins the standard track near quiescence, and are usually labelled as \emph{`hybrid’} \citep{xie_2016}. Numerous such examples (often only partially sampled) have been reported: GRS 1739$-$278 \citep{xie_2020}, H1743$-$322 \citep{coriat_2011, williams_2020}, MAXI J1659$-$152 \citep{jonker_2012}, MAXI J1535$-$571 \citep{russell_2019_quench}, MAXI J1348$-$630 \citep{carotenuto_2021b}, Swift J1727.8$-$1613 \citep{hughes_2025_lrlx}, Swift J1753.5$-$0127 \citep{plotkin_2017}, and XTE J1752$-$223 \citep{ratti_2012}. In fact, it has been suggested that all outliers -- or perhaps even all LMXBs -- belong to the hybrid class (e.g., \citealt{koljonen_2019, carotenuto_2021b}). These sources clearly demonstrate that the jet–accretion coupling in XRBs is more complex than originally thought. Although the origin of the distinct tracks remains uncertain, they likely reflect variations in underlying system parameters and a physical evolution of the accretion flow and/or jet (e.g., \citealt{coriat_2011}). 

To date, all large-scale (global) studies of the $L_R$--$L_X$ plane have relied on \emph{heterogeneous} data sets compiled from multiple telescopes. The most comprehensive sample assembled so far is that of \citet{bahramian_2022}. However, such inhomogeneity introduces several sources of systematic uncertainty. In particular, when considering X-ray measurements, cross-calibration differences between instruments can lead to flux discrepancies that are sometimes as high as $\gtrsim$10--20 per cent, which may introduce systematic offsets in inferred luminosities (e.g., \citealt{tsujimoto_2011, plucinsky_2012, madsen_2017}). These effects are further compounded by variations in data reduction and analysis procedures, which are difficult to quantify. For the radio data, when constructing the global plane, flux densities are frequently scaled to a common frequency by assuming a flat spectral index in cases where it is not directly measured. However, observational studies have demonstrated that even in the HS, flat radio spectra are not ubiquitous (e.g., \citealt{fender_2001_a, corbel_2013, russell_2014,  espinasses_2018}), resulting in significant errors when such an assumption is made. Taken together, these factors highlight that the $L_R$--$L_X$ plane constructed using inhomogeneous data is subject to significant and poorly constrained systematics, which inherently limit the interpretive power of the plane and restrict its utility for advancing a physical understanding of the accretion--jet coupling. As such, it is important that a \emph{homogeneous} sample is compiled, so that stringent constraints can be placed on $L_R$--$L_X$ correlations.

In this paper, we present the largest homogeneous $L_R$--$L_X$ plane to date, constructed using data from a single radio telescope and single X-ray telescope, and using a consistent approach to pair contemporaneous observations. Section \ref{sec: data} introduces our data sample. In Section \ref{sec: light_curves}, we present the corresponding light curves, highlight the numerous detections of core radio emission in the SS, and estimate the kinetic energy feedback from jets over the course of our monitoring campaign. Sections \ref{sec: coupling} and \ref{sec: clustering} respectively introduce the compiled $L_R$–$L_X$ plane and the results of a clustering analysis on this plane. In Section \ref{sec: linreg}, we describe our linear regression scheme, and apply this to both the full sample and individual sources. Section \ref{sec: discussion} places our $L_R$–$L_X$ results in the broader context of the literature. Finally, in Section \ref{sec: conclu}, we summarise our findings and outline directions for future work.


\begin{table*}
    \centering
    \caption{\label{tab: system_properties} Relevant properties for the X-ray binaries in our sample. In column 2, we list the nature of the compact object in the system: a black hole (BH; confirmed through dynamical studies), BH candidate (BHC), or neutron star (NS). For the NS class, we also indicate whether the object is an accreting millisecond pulsar (AMXP). All the sources in our sample are low-mass X-ray binaries, except for Vela X-1, which is a high-mass X-ray binary (HMXB). In columns 3--5, we respectively list the adopted distance to the system, the distance distribution used in our Monte Carlo analyses, and the method(s) used to derive the distance, along with the relevant reference(s) (see Appendix \ref{app: sources} for details regarding the assumed distance ranges).}
    \begin{tabular}{lcccc}
        \hline 
        Source name & Type & \makecell{Nominal \\distance [kpc]} & Distance distribution [kpc] & Distance method and references\\
        \hline \hline
        1A 1744$-$361   & NS  & 8.00 & Uniform: [1.00, 9.00] & \makecell{Unknown; no photospheric radius expansion during burst.$^{[1]}$}  \\ \arrayrulecolor{black!5} \hline
        4U 1543$-$47   &  BH  & 5.00 &  Normal: $5.00\pm 2.00$ & Gaia parallax.$^{[2]}$     \\ \hline
        4U 1630$-$47   &  BHC  &  11.50 & Normal: $11.50\pm1.00$ & Dust-scattering halo.$^{[3]}$    \\ \hline
        Aquila X-1 & NS & 6.00 & Normal: $6.00 \pm 2.00$  & Near-infrared photometry.$^{[4]}$ \\ \hline
        Centaurus X-4   & NS  & 1.87 & Normal: $1.87\pm0.75$ & Gaia parallax.$^{[5]}$  \\ \hline
        EXO 1846$-$031   &  BHC & 4.50 & Uniform: [2.40, 7.50] & State transition luminosity.$^{[6]}$   \\ \hline
        GRS 1739$-$278   & BHC  & 8.00 & Uniform: [6.00, 8.50] & \makecell{Using absorption column estimate; possibly in Galactic bulge.$^{[7]}$}    \\ \hline
        GX 339$-$4   &  BH  & 10.00 & Uniform: [8.00, 12.00] & Evolutionary model.$^{[8]}$  \\ \hline
        H1743$-$322  & BHC  & 8.50 & Normal: $8.50\pm0.80$ & Kinematic modelling of jets.$^{[9]}$  \\ \hline
        IGR J17091$-$3624   &  BHC & 14.00 & Uniform: [11.00, 17.00] & State transition luminosity.$^{[10]}$  \\ \hline
        MAXI J1348$-$630   & BHC  &  2.20 & Normal: $2.20\pm0.60$ &  HI absorption.$^{[11]}$    \\ \hline
        MAXI J1631$-$472  & BHC & 5.10 & Uniform: [4.50, 6.50] & \makecell{Optical observed vs theoretical fluxes; X-ray disc-continuum fitting.$^{[12]}$}     \\ \hline
        MAXI J1803$-$298  & BHC  & 8.00 & Uniform: [6.00, 12.00]& Proximity to Galactic centre.$^{[13]}$  \\ \hline
        MAXI J1807+132  & NS  & 6.30 & Uniform: [4.20, 8.40] & \makecell{Absolute magnitude vs orbital period correlation.$^{[14]}$}    \\ \hline
        MAXI J1810$-$222  &  BHC & 8.00 & Uniform: [6.00, 17.00] & \makecell{Unknown; Galactic mass density model.$^{[15]}$}  \\ \hline
        MAXI J1816$-$195  & NS (AMXP)  & 6.00 & Uniform: [1.00, 8.70] & \makecell{Unknown; no photospheric radius expansion during burst.$^{[16]}$}   \\ \hline
        MAXI J1820+070   & BH  & 2.96 & Normal: $2.96\pm0.33$ & Radio parallax.$^{[17]}$   \\ \hline
        SAX J1808.4$-$3658  & NS (AMXP)  & 2.70 &Normal: $2.70\pm 0.30$ & Modelling X-ray burst.$^{[18]}$ \\ \hline
        SAX J1810.8$-$2609  &  NS & 4.90 & Normal: $4.90\pm0.30$ & Modelling X-ray burst.$^{[19]}$   \\ \hline
        Swift J1727.8$-$1613  &  BH & 5.50 & Normal: $5.50\pm1.40$  & Near-ultraviolet reddening and optical photometry.$^{[20]}$  \\ \hline
        Swift J1842.5$-$1124  & BHC  & 8.12 & Normal: $8.12\pm2.20$ & State transition luminosity.$^{[21]}$  \\ \hline
        XTE J1701$-$462  & NS  & 8.80 &Normal: $8.80\pm1.32$ & Modelling X-ray burst.$^{[22]}$    \\
        \arrayrulecolor{black}
        \hline 
        \multicolumn{5}{l}{\textbf{Not used for the $L_R$--$L_X$ analysis:}} \\ \hline  
        Circinus X-1   & NS  & 9.40 & Normal: $9.40\pm1.00$ &  X-ray dust scattering light echoes.$^{[23]}$  \\ \arrayrulecolor{black!10}\hline
        GRS 1915+105  & BH  & 9.40 & Normal: $9.40 \pm 1.00$ & 3D kinematics.$^{[24]}$  \\  \hline
        Swift J1728.9$-$3613 & BHC  & 8.40 & Uniform: [7.60, 9.20] & Absorption column integration.$^{[25]}$ \\ \hline
        Swift J1858.6$-$0814  &  NS & 12.80 & Uniform: [9.00, 18.00]  &  Modelling X-ray burst.$^{[26]}$   \\  \hline
        Vela X-1  & NS (HMXB)  & 1.99 & Normal: $1.99\pm0.13$ & Gaia parallax.$^{[27]}$   \\ 
        \arrayrulecolor{black} 
    \end{tabular}
    \hypersetup{hidelinks}
    \caption*{References: \small\textit{[1]: \citet{bhattacharyya_2006}; [2]: \citet{zhang_2025b}; [3]: \citet{kalemci_2018, kalemci_2025}; [4]: \citet{mata_sanchez_2017}; [5]: \citet{van_den_eijnden_2022b}; [6]: \citet{williams_2022}; [7]: \cite{greiner_1996}, \cite{mereminskiy_2019}; [8]: \citet{zdziarski_x-ray_2019}; [9]: \citet{steiner_2012}; [10]: \citet{rodriguez_2011}; [11]: \citet{chauhan_2021}; [12]:  \citet{monageng_2021}; \citet{rout_2023}; \citet{zdziarski_2025}; [13]: \citet{shidatsu_2022}; [14]: \citet{saavedra_2025}; [15]: \citet{russell_2022}; [16]: \citet{chen_2022}, \citet{bult_2022b}, \citet{mandal_2023}; [17]: \citet{atri_2020}; [18]: \citet{galloway_2024}; [19]: \citet{natalucci_2000}; [20]: \citet{burridge_2025}; [21]: \citet{abdulghani_2024}; [22]: \citet{lin_2009b}; [23]: \citet{heinz_2015}; [24]: \citet{reid_2023}; [25]: \citet{balakrishnan_2023}; [26]: \citet{buisson_2020}; [27]: \citet{kretschmar_2021}}.}
\end{table*}

\section{Observations and Data Sample}\label{sec: data}

\subsection{The ThunderKAT Programme}

\textbf{T}he \textbf{HUN}t for \textbf{D}ynamic and \textbf{E}xplosive \textbf{R}adio transients with Meer\textbf{KAT} (ThunderKAT; \citealt{thunderkat_2018}; PIs: R. Fender and P. A. Woudt) was a five-year programme on the Meer Karoo Array Telescope (MeerKAT; \citealt{jonas_2016}), starting in September 2018. MeerKAT is located in the Karoo desert in South Africa, and is a precursor for the Square Kilometre Array Observatory (SKAO) SKA-Mid project. It consists of 64 antennas, each 13.5 m in diameter, and is currently equipped with L-band, S-band, and UHF-band receivers. The array features a dense core and a maximum baseline of $\sim$8 km, providing excellent snapshot \emph{uv} coverage. In the L-band ($0.856{-}1.712$ GHz; centred at 1.284 GHz, with a 856 MHz bandwidth), it covers a large field of view of 1.69 deg$^2$, and achieves a resolution of $\sim$5 arcsec ("). 

The ThunderKAT programme was designed to observe a range of transient systems in the radio band, in coordination with X-ray and optical monitoring programmes. Among its science goals was weekly L-band radio monitoring of bright, active XRBs, with the capacity for higher cadence observations. Over the course of the programme, approximately 60 BH and NS XRBs in outburst were observed. Notably, ThunderKAT observations of the field of GX 339$-$4 led to the discovery of the first transient with MeerKAT (\citealt{driessen_2020}). Subsequent searches using ThunderKAT data have yielded numerous additional transient discoveries (e.g., \citealt{driessen_2022a, driessen_2022b, andersson_2022, rowlinson_2022}). 

ThunderKAT ran in conjunction with the \emph{Swift}KAT programme (PI: S. Motta), which provided quasi-simultaneous $0.2{-}10$ keV X-ray data for the monitored XRBs using the Neil Gehrels Swift Observatory X-ray Telescope (hereafter referred to as \emph{Swift}/XRT; \citealt{gehrels_2004, burrows_swift_2005}). 

The successor of ThunderKAT is X-KAT, which has been in operation since September 2023. 

\subsection{Data Collection \& Preprocessing}

We compiled data for all the XRBs monitored during ThunderKAT, either via private communication with the collaboration members who reduced the data sets, or by reducing the data ourselves when this has not already been done. In addition, we included the 2023--2024 (X-KAT) observations of the BH XRB Swift J1727.8$-$1613. We also used data from the 2024 outbursts of the NS systems 1A 1744$-$361 and Aquila X-1, to augment the relatively small ThunderKAT NS sample. Sources with only radio non-detections, or with very few radio detections and no quasi-simultaneous X-ray data, were excluded from the sample and are listed in Appendix \ref{app: not_included}. Notably, although numerous ThunderKAT observations of MAXI J1848$-$015 were obtained, the source was excluded from our sample due to its faint, only marginally detected core radio emission and the lack of quasi-simultaneous \emph{Swift}/XRT coverage during any potential HS episode (owing to Sun constraints; \citealt{bahramian_2023}).

In Table \ref{tab: system_properties}, for each source in our sample, we list its name and accretor type, the assumed distance to the source, as well as the distance distribution employed in our analysis. Notably, the designation \emph{`BH candidate'} (BHC) indicates that the accretor is inferred to be a BH based on its X-ray spectral and timing properties, whereas \emph{`BH'} denotes a source whose nature has been confirmed through dynamical mass measurements. In Appendices \ref{app: source_properties} and \ref{app: sources}, for each source, we provide further information regarding its classification and the adopted distance range, as well as details regarding its data reduction.

As the data were compiled, it became clear that a variety of different approaches had been used in the data reduction processes. However, since our results are ultimately dominated by intrinsic scatter -- rather than the uncertainties of individual measurements -- we do not expect these methodological differences to significantly affect the conclusions presented in this paper. Notably, although the methods used to classify the spectral states of individual sources vary, data points with ambiguous state assignments are likely to be few, so do not affect our overall conclusions. Looking ahead, we intend to standardise the data-reduction approach within our collaboration. To this end, Appendix \ref{app: data_reduction} outlines key aspects of the data reduction procedures employed for our newly analysed sources, with the goal of incorporating the lessons learnt into future data reductions.  

All the radio data presented and used in this paper are the \emph{`core'} flux densities, meaning the flux density obtained by fitting an elliptical Gaussian at the XRB position. In other words, it is the unresolved central radio emission (i.e., at the location of the accretor), excluding any resolved ejecta components. We note, however, that in some observations, the central region may contain emission from ejecta that have not yet propagated far enough outward to be resolved at the angular resolution of MeerKAT. In contrast, for some observations, the core was not detected -- i.e., the flux density was below a noise threshold, which is usually set to 3$\sigma$, where $\sigma$ is the RMS in a nearby source-free region -- so we instead report 3$\sigma$ upper limits. 

The X-ray data used in this paper are \emph{`unabsorbed'} 1$-$10 keV fluxes, meaning that they have been corrected for the effects of interstellar absorption (i.e., the influence of hydrogen column density, $N_\mathrm{H}$, has been removed). Only a small number of X-ray upper limits are reported in this work (for IGR J17091–3624, MAXI J1348–630, and Swift J1727.8–1613), calculated following the procedures described in Appendix~\ref{app: sources}.

The data used in this paper are available for download in a machine-readable format on our website: \newline \url{https://thunderkat.physics.ox.ac.uk/}.

\subsection{Systematic Uncertainties}

For the radio data, following standard practice, for each flux density measurement, we added a 5 per cent systematic uncertainty in quadrature with the 68 per cent confidence interval (CI) statistical uncertainties, to account for the secular evolution of the calibrator(s). We performed this calculation in linear flux density space; while this assumes Gaussian errors in linear space (which is an approximation that may not always hold), it is reasonable in the absence of better constraints. For MeerKAT, some efforts to better quantify systematics exist (e.g., \citealt{driessen_2022}), but a detailed investigation is left to future work.

The X-ray flux statistical uncertainties in our data set are reported at either the 68 or 90 per cent confidence level, reflecting the differing conventions adopted in the original analyses. For the uncertainties quoted at the 90 per cent CI, we converted them to 68 per cent intervals by multiplying by 0.607 in linear flux space, thereby assuming normally distributed uncertainties in that space. In addition, we conservatively added a 10 per cent systematic uncertainty in quadrature in linear flux space to account for calibration issues (e.g., imperfect pile-up corrections) and the inhomogeneous data reduction methods employed. Although combining uncertainties in linear versus logarithmic flux space can lead to differences -- particularly at low signal-to-noise ratios (SNR) -- these are small, so do not affect the results presented here. We therefore adopted the linear-space approach as it is consistent with previous work (e.g., \citealt{hughes_2025_lrlx}).


\subsection{Luminosity Conversion}\label{sec: lum_conv}

\begin{figure*}
    \centering
    \includegraphics[width=\textwidth]{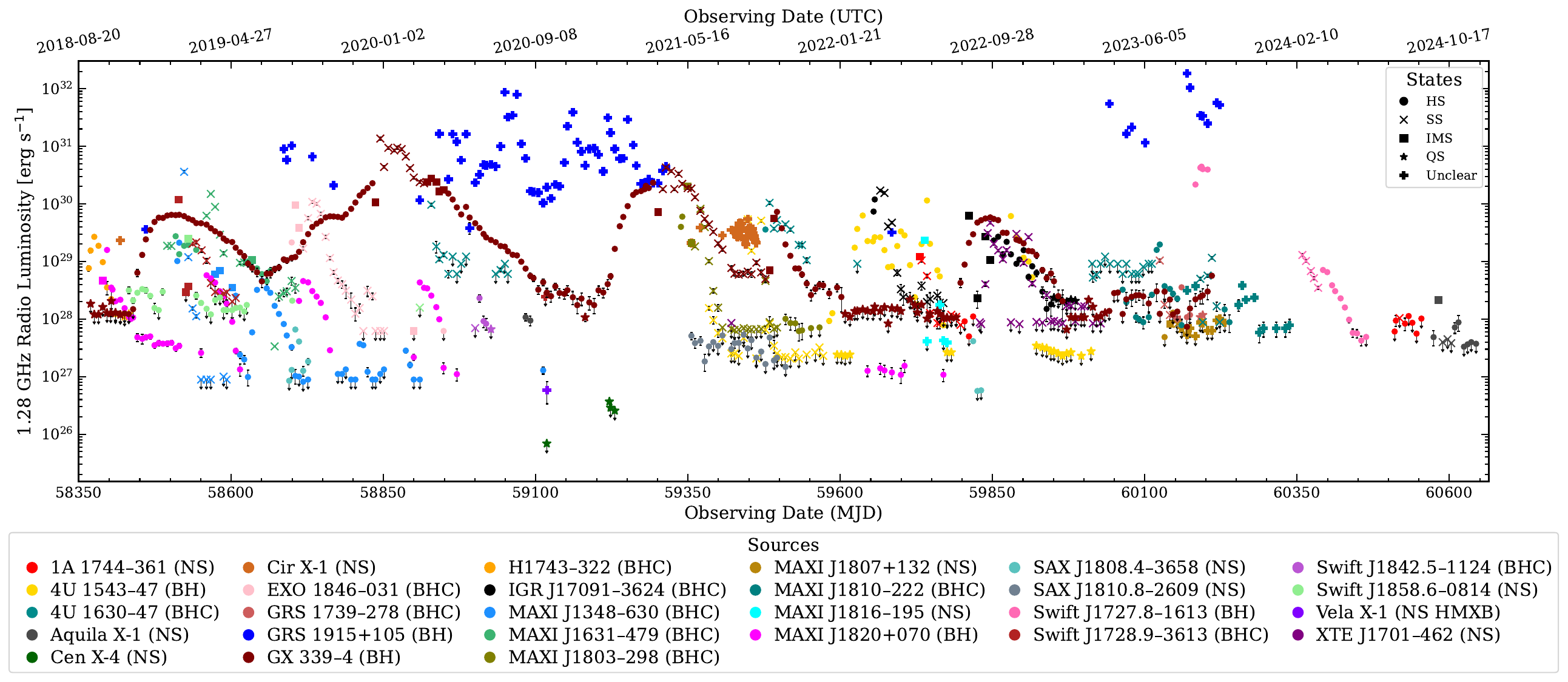}
    \caption{The 1.28 GHz MeerKAT luminosities ($L_R = \nu L_\nu$, with $\nu$ = 1.28 GHz) obtained during ThunderKAT (plus Swift J1727.8$-$1613, 1A 1744$-$361, and Aquila X-1 observations during X-KAT), converted from core flux densities using the nominal source distances listed in Table \ref{tab: system_properties}. The marker shapes indicate the spectral state at the time of the observation -- HS, SS, IMS, and QS refer respectively to the hard, soft, intermediate, and quiescent spectral states, while `Unclear' indicates that we did not have enough information to classify that observation. The data points are coloured by source name. For the low-mass X-ray binaries, the labels `BH', `BHC', and `NS' denote that the compact object in the system is a black hole, black hole candidate, or neutron star, respectively, while `HMXB' indicates that the source is a high-mass X-ray binary. Observations marked with downward arrows are 3$\sigma$ upper limits, where $\sigma$ is the image RMS in a source-free region near the XRB. The error bars include both statistical and systematic uncertainties, but exclude the uncertainty associated with the distance. This plot can be seen in an interactive format on our \href{https://thunderkat.physics.ox.ac.uk/lightcurves}{website}.}
    \label{fig: all_radio_luminosities}
\end{figure*}

The 1--10 keV X-ray luminosities were calculated as $L_X = 4\pi D^2 F$, where $D$ is the nominal distance to the source and $F$ is the unabsorbed 1--10 keV X-ray flux. To obtain equivalent units on both axes of the $L_R$--$L_X$ plane, we computed the \emph{scaled} monochromatic/specific radio luminosity using $L_R = 4\pi D^2 \nu F_\nu = \nu L_\nu$, where $\nu = 1.28$ GHz and $F_\nu$ is the MeerKAT L-band core flux density. This quantity corresponds to the integrated 1.28 GHz luminosity (i.e., the integral of the monochromatic luminosity up to 1.28 GHz) in cases where the spectral index is flat ($\alpha \sim 0$) from low frequencies up to 1.28 GHz, which is typically a good approximation for observations in the HS and QS. In contrast, for epochs in the IMS and SS -- when the radio emission is generally expected to be optically thin -- we emphasize that $L_R$ is not a reliable tracer of the integrated luminosity.

\section{Light Curves}\label{sec: light_curves}

In Figure \ref{fig: all_radio_luminosities}, we show all the L-band MeerKAT core flux density measurements obtained during ThunderKAT (plus observations of Swift J1727.8$-$1613, 1A 1744$-$361, and Aquila X-1 during X-KAT), converted to scaled monochromatic luminosities (i.e., $L_R = \nu L_\nu$ at 1.28 GHz) using the nominal distance estimates in Table \ref{tab: system_properties}. The data points are coloured by source name, and the markers indicate the spectral state of each observation: HS, SS, IMS, or QS. The spectral state designation `Unclear' indicates that we did not have enough information to classify that particular observation. Throughout this paper, arrows indicate 3$\sigma$ upper limits. The light curves can be seen in greater detail on our \href{https://thunderkat.physics.ox.ac.uk/lightcurves}{website}. Further information regarding the radio flux density evolution for each source in our sample can be found in Appendix \ref{app: sources}. 

\subsection{Soft State Emission}\label{subsec: SS_emission}

\subsubsection{Black Hole Systems}

It is generally accepted that as BH XRBs enter the SS, radio emission from the steady compact jet is strongly suppressed, as has been observed in numerous systems (e.g., \citealt{fender_1999b, corbel_2000, coriat_2011, rushton_2016, russell_2019_quench}). However, a number of sources exhibit faint radio detections in the SS, typically characterised by optically thin spectra and, in most cases, a monotonic decay of the emission (e.g., \citealt{corbel_2004, fender_2004c, brocksopp_2007,  fender_2009}). 

These SS detections are often core radio flares occurring shortly after the HS$\rightarrow$SS \emph{state transition}. Since this transition is generally associated with the launch of discrete ejecta, core radio detections during the IMS and early SS are commonly interpreted as emission from these ejecta, superimposed on a decaying core flux, where the two components are unresolved at the resolution of the telescope. In some cases, this interpretation is supported by quasi-simultaneous high-resolution observations that spatially resolve the ejecta. One of the clearest examples is the HS$\rightarrow$SS state transition flare by MAXI J1820+070, observed with the Arcminute Microkelvin Imager Large Array (AMI-LA) prior to our ThunderKAT monitoring campaign (Extended Data Figure 1 in \citealt{bright_2020}). In this case, spatially resolved ejecta were detected with the Very Long Baseline Array (VLBA) only $\sim$3 days into the IMS (Extended Data Figure 2), confirming that the unresolved AMI-LA core emission originated from transient ejecta. 

Alternatively, some studies propose that radio emission observed at the start of the SS may originate from the quenched compact jet. In this scenario, during the state transition, the jet break shifts to frequencies below the radio band as the jet’s acceleration zone recedes from the compact object, resulting in radio emission that is dominated by optically thin synchrotron radiation (e.g., \citealt{russell_2021}).

Additionally, radio \emph{reflares/rebrightenings} are sometimes observed during the SS, long after the HS$\rightarrow$SS transition. These events typically exhibit optically thin spectra and are therefore also generally associated with ejecta, although a number of proposed explanations exist. One possibility is that they arise from \emph{shocks} from previously-launched discrete ejecta, leading to in-situ particle acceleration and the release of internal energy. These include internal and/or reverse shocks within the ejecta, or forward shocks produced from ejecta-ISM interactions (e.g., \citealt{kaiser_2000, corbel_2002, corbel_2004, corbel_2005, hao_2009, jamil_2010, rushton_2017, bright_2020, espinasse_2020, matthews_2025}). Emission produced from shocks can persist for many months, owing to ongoing continuous particle acceleration, and consequently decline slowly. This gradually fading component may therefore account for the low-level radio emission that is sometimes observed during the SS.

Another possibility is that SS core radio reflares may arise from brief \emph{excursions} to the IMS during the SS (potentially missed due to poor X-ray coverage), possibly resulting in the temporary reactivation of the compact jet (e.g., \citealt{russell_2020, carotenuto_2025}) and/or the subsequent launch of new ejecta. For example, during the 1999 outburst of XTE J1859+226, the source displayed multiple quick excursions to the HIMS that were temporally correlated with radio flares, suggesting that multiple discrete ejecta may have been launched \citep{brockstopp_2002, fender_2009}. In some cases, short-timescale radio flaring in the IMS may represent the superposition of multiple closely-spaced ejection events, distinguishable only at millimetre/infrared frequencies (e.g, \citealt{tetarenko_2017}) or through Very Long Baseline Interferometry (e.g., \citealt{miller_jones_2019_v404}).

Finally, an alternative scenario is that some systems continue to produce weak compact jets even in the SS. For example, \cite{rushton_2012} suggested the presence of a compact jet in the SS of Cygnus X-1, although they also cautioned that the SS of this source may not be representative of BH XRBs more generally (see also \citealt{drappeau_2017}).

\begin{figure}
    \centering
    \includegraphics[width=\columnwidth]{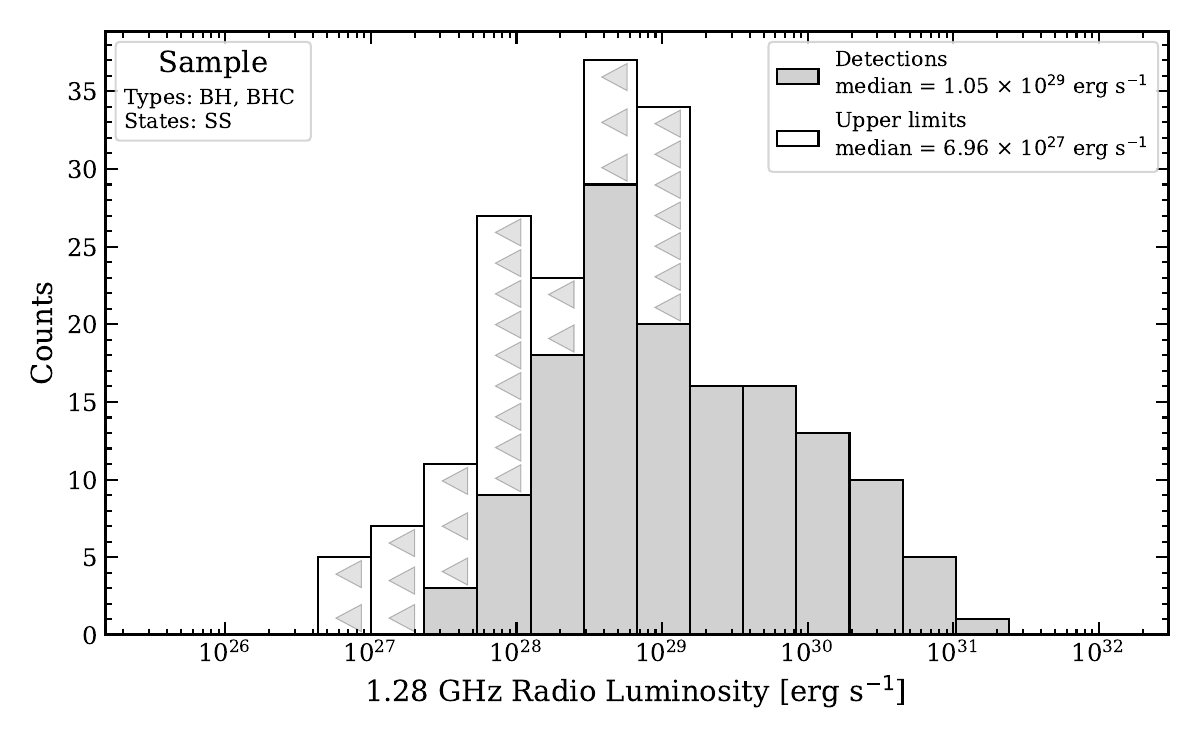}
    \caption{Histograms of the 1.28 GHz MeerKAT luminosities ($L_R = \nu L_\nu$, with $\nu$ = 1.28 GHz) for data points in the \emph{soft spectral state}, for the systems with black holes and black hole candidates in our sample. These luminosities were converted from core flux density measurements using the nominal source distance estimates in Table \ref{tab: system_properties}. The histograms are stacked so that detections (shown in grey) are plotted beneath the 3$\sigma$ upper limits (shown in white, with arrows pointing left to indicate the limit direction).}
    \label{fig: radio_SS_distribution}
\end{figure}

In our sample, we likewise detect significant core radio emission during the SS in some cases. In Figure \ref{fig: radio_SS_distribution}, we present the distribution of core radio luminosities ($L_R = \nu L_\nu$) for observations of BH and BHCs that are classified as being in the SS (see also the light curves in Figure \ref{fig: all_radio_luminosities}). We note that a small number of these observations may have been misclassified, although we expect such cases to be rare. 

A clear instance of a HS$\rightarrow$ SS flare is seen for MAXI J1348$-$630 at MJD $\sim$58523, shortly after the IMS (see Figure \ref{fig: all_radio_luminosities}, and Figure 2 in \citealt{carotenuto_2021a}), where the core radio emission ($L_R \sim 3.6 \times 10^{30}$ erg s$^{-1}$) is likely due to contamination from ejecta, as indicated by proper-motion fits to an ejection that was resolved at later times \citep{carotenuto_2022}. The source also displayed an optically thin SS reflare, $\sim$50 days after the state-transition flare (at MJDs $\sim$58573${-}$58582; \citealt{carotenuto_2021a, carotenuto_2025}). In this case, a discrete ejection was subsequently resolved and inferred to have been launched only a few days after the SS reflare. This close temporal association suggests that the SS reflare resulted from a brief excursion to the IMS, during which the compact jet may have been temporarily reactivated before discrete ejecta were launched \citep{carotenuto_2025}.

We also observed optically thin SS reflares for MAXI J1803$-$298, approximately 20 and 130 days after the state-transition flare (at MJDs $\sim$59370 and $\sim$59477, respectively; Espinasse et al. submitted). The former is attributed to ejecta-ISM shocks, while the latter is believed to arise from a short excursion to the IMS. 

Another example in our sample is the long-lived SS emission from Swift J1727.8$-$1613. This emission is optically thin, suggesting a jet ejecta origin, as confirmed through spatially resolved S-band observations of the ejecta (\citealt{hughes_2025_compre}; Carotenuto et al. in preparation). Furthermore, during the SS of GX 339$-$4 on MJDs 59335$-$59484, \cite{tremou_2026} observed an arcsecond-scale positional shift of the core emission in MeerKAT images, supporting the scenario that the persistently detected radio emission in the SS is due to propagating transient ejecta.

Therefore, it seems plausible that the SS emission observed for BH systems in our sample is dominated by discrete ejecta, and is therefore not associated with the compact jet emission seen in the HS and QS. Further evidence for this is provided in Section \ref{subsec: interp} (see Figure \ref{fig: LrLx_plane_interp_HS_vs_SS}), where we see that the SS data points occupy a distinct region in the $L_R$--$L_X$ plane compared to the HS/QS data, indicating that they represent distinct regimes.


\subsubsection{Neutron Star Systems}

For the NS LMXBs in our sample, there are also a number of SS epochs with detectable core radio flux densities. However, for these systems, it is less well established whether there is a universal SS quenching, partly due to the low SNR of many NS measurements. Clear cases of suppression have been observed (e.g., \citealt{tudose_2009a, miller_jones_2010, migliari_2011_conf, gusinskaia_2017, russell_2021}), yet other studies suggest weaker quenching or possibly inconsistent behaviour (e.g., \citealt{migliari_2004, migliari_2006, gusinskaia_2020}). These results suggest that additional parameters tied to the nature of the compact object play a role in governing jet formation, such as surface emission, boundary layers, or magnetic fields. For further discussion, see, e.g., Section 4.1 of \cite{gusinskaia_2017}, or Section 5.3.2 of \citealt{van_den_eijden_2021}.

\subsection{Energy Estimates}\label{sec: energy_estimates}

XRBs return energy to the ambient medium via (i) radiation, which is dominated by X-ray emission at high accretion rates, and (ii) kinetic feedback, predominantly from jets, but with additional contributions from winds \citep{fender_2016}. In this section, we calculate order-of-magnitude estimates for the kinetic energy feedback from the jets that were monitored during the ThunderKAT programme (using all the MeerKAT data shown in Figure~\ref{fig: all_radio_luminosities}).

\subsubsection{Compact Jet Energy Estimates}

For the continuously replenished, steady, compact jets that exist during the HS, we assume a simple conical jet model, a flat spectral index ($\alpha \approx 0$), and electrons with a fiducial energy index $p \approx 2$, which yields the following relation for the jet kinetic power for optically thick radio emission: $L_\text{jet} \propto L_R^{12/17}$ (e.g., \citealt{blandford_1979, falcke_1995, heinz_2003, coriat_2011}). 

In \citet{kording_2006_energy}, the normalisation of this relation is derived by assuming that the jet power is always a constant fraction of the accretion rate and by using systems with independently estimated accretion rates -- namely, NSs in the HS, and BHs in the HS near state transitions -- under the assumption that the accretion rate evolves smoothly across the transition. It is further assumed that the radiative efficiency in the radio band of BHs and NSs is similar, resulting in a normalisation that differs only by a small factor between the two populations. Using their results, we obtain that $L_\text{jet} \approx 10^{16} L_R^{12/17}$, where we have converted to the 1.28 GHz scaled luminosity ($\equiv L_R$) by assuming a flat spectral index. The calculated normalisation is similar to estimates by \cite{heinz_2005} obtained using both XRB and AGN data, hinting at the scale invariance between core radio luminosity and jet power (see also \citealt{fender_2016}). 

Independent constraints on the kinetic energy can also be estimated using jet-driven ISM structures, such as the large-scale ($\sim$5 parsec in diameter) bow-shock nebula surrounding the BH system Cygnus X-1 (e.g., \citealt{gallo_2005_cyg_x1, russell_2007_cyg_x1, sell_2015_cyg_x1, prabu_2025}), as well as similar lobe-like structures associated with other sources such as GRS 1758$-$258 (e.g., \citealt{mariani_2025}). These studies yield jet power normalisations that are broadly consistent, although the lower bound by \cite{mariani_2025} is significantly smaller than the others. We also note that a substantially lower jet power estimate is reported by \cite{atri_2025_cyg_x1}, which \cite{prabu_2025} attribute to the shock velocities considered in this work. Estimates from jet-driven ISM structures/lobes are inherently uncertain as they are somewhat model-dependent and rely on \emph{time-averaged} jet powers and estimates of the radio luminosity. By contrast, \cite{prabu_2025} derived an \emph{instantaneous} jet-power estimate of $\sim$$10^{37.3}$ erg s$^{-1}$ for Cygnus X-1 by modelling its jet-wind interactions, yielding a normalisation of $\sim$$8 \times 10^{16}$ erg s$^{-1}$. We note that, although Cygnus X-1 is a HMXB, it occupies a comparable (albeit slightly lower) region of the $L_R$--$L_X$ plane relative to standard-track LMXBs and exhibits qualitatively similar behaviour -- though its radio emission may be affected by free-free absorption and emission in the stellar wind (e.g., \citealt{gallo_2003, zdziarski_2011b, zdziarski_2012}). This suggests that a broadly similar normalisation may be appropriate. 

In the analysis that follows, we adopt the conservative normalisation from \cite{kording_2006_energy} for all sources in our sample. However, we emphasize that this estimate is approximate, especially due to the small number of sources considered in this study. Additionally, variations are expected across the sources due to differences such as the jet composition, radiative efficiency, spin, bulk speed, and viewing angle. In forthcoming work, we will leverage our expanded data set to place more robust constraints on the relationship between jet power and core radio luminosity.

To estimate the energy output of the HS compact jet for each source in our sample, we integrate the jet power relation over the HS MeerKAT core luminosity data, so that our estimate is given by: $E_{\rm jet, HS} \approx 10^{16}\int L_R(t)^{12/17}\, dt$, where $L_R(t)$ is the 1.28 GHz luminosity in erg~s$^{-1}$, calculated using the nominal source distances in Table \ref{tab: system_properties}. Before performing the numerical integration, we interpolate the radio light curves using logarithmic Akima interpolation, as described in Section \ref{subsec: interp} for the X-ray data. We also compare the jet energy estimates derived using this method with those obtained via trapezoidal integration of the non-interpolated data in both linear and logarithmic luminosity space. These approaches yield values of the same order of magnitude. In all cases, upper limits (3$\sigma$) are conservatively treated as effective zero values (zero in linear space, and a small fraction of $\sigma$ in logarithmic space). Additionally, to account for sparse temporal sampling and occasional gaps during HS intervals, for each data point $t$, if the previous or following data point is $\ge T$ away, we conservatively set the flux density at that time (i.e., either $t-T$ or $t+T$) to an effective zero, adopting $T=10$ days since we observe at $\sim$weekly cadence.

When calculating the HS jet energies, we assume (sometimes incorrectly) that the HS core flux density arises solely from the compact jet, although this assumption may not always hold. In particular, our data set includes several cases of rebrightenings/reflares during the HS, similar to the SS reflares described previously. These events may originate from shocks or discrete ejecta launched during brief transitions to the IMS -- potentially missed due to the low monitoring cadence -- rather than from the compact jet. In addition, the HS phases were not fully covered for many of our sources, particularly because the onset of the HS rise was frequently missed due to delays in triggering observations following the start of the outburst. Finally, we neglect Doppler (de-)boosting (see Section~\ref{sec: doppler}), assuming low jet Lorentz factors. Since most sources are likely Doppler deboosted (as compact jets are typically only mildly relativistic and more likely to be viewed at larger inclination angles relative to the line of sight), this renders our energy estimates conservative.

When conducting this analysis, we exclude the BH system GRS 1915+105 due to the poorly understood nature of its bright, frequent flaring during large portions of our monitoring period \citep{motta_2021}. Using the methods described, we estimate that the compact jets from the remaining sources in our sample return a total/combined kinetic energy of $\sim$$5 \times 10^{45}$ erg to the ISM over our monitoring period. The breakdown of contributions from individual sources is shown in Appendix \ref{app: energy_estimates}. 

\subsubsection{Discrete Ejecta Energy Estimates}

The total energy (internal plus kinetic) of discrete ejecta can be estimated using several approaches. The internal energy required to power observed synchrotron radiation can be inferred using the size of the emitting region and source distance, assuming equipartition between the electrons and magnetic fields in the jet plasma \citep{longair_1994}. Ejecta sizes can be measured directly when resolved (e.g., \citealt{bright_2020}). When unresolved, they can be inferred from assumed expansion speeds and ejection timescales (typically the rise time of the radio flare at the jet’s launch) or from the synchrotron self-absorption turnover in the radio spectrum (\citealt{fender_2019}; Cowie \& Fender submitted). Alternatively, ejecta energetics can be constrained via jet-ISM interactions (e.g., \citealt{teterenko_2018_grs1915}), kinematic modelling (e.g., \citealt{carotenuto_2022, carotenuto_2024}), joint radiative and kinematic modelling (e.g., \citealt{cooper_2025}), or simulations (e.g., \citealt{savard_2025}). 

In particular, \citet{savard_2025} performed observation-informed simulations of the ejecta of MAXI J1820+070, using initial conditions from \citet{bright_2020}, and inferred a kinetic energy of $\sim$$2\times10^{44}$ erg per ejection component at $\sim$90 days post-launch, noting that the energy was likely higher at launch (prior to deceleration). Their estimate is broadly consistent with results from blast wave modelling from \cite{carotenuto_2024}, although in both cases, the energy depends strongly on the adopted ISM density.

In the following, we discuss the transient ejecta resolved with MeerKAT in our sample (see Appendix \ref{app: sources} for details regarding each source). Two distinct ejecta -- assumed to have been launched at different times -- were resolved for 4U 1543$-$47 \citep{zhang_2025b}, and one for MAXI J1820+070 \citep{bright_2020}. For Swift J1727.8$-$1613 (\citealt{hughes_2025_compre}; Carotenuto et al. in preparation), MeerKAT L-band data revealed an extended radio component that was resolved into two distinct point sources in higher resolution S-band data, suggesting that some of the other L-band ejecta observed in our sample may similarly comprise multiple components. For MAXI J1348$-$630, two distinct ejecta were resolved using MeerKAT \citep{carotenuto_2021a}, with an additional partially-resolved ejection observed using data from the Australia Telescope Compact Array (ATCA; \citealt{carotenuto_2025}). Furthermore, one ejection was resolved for EXO 1846$-$031 using additional quasi-simultaneous data from the Karl G. Jansky Very Large Array (VLA; \citealt{williams_2022}), as well as one for GX 339$-$4 using ATCA data \citep{tremou_2026}, and another for MAXI J1803$-$298 using the Very Long Baseline Array (VLBA; \citealt{wood_2023}). In addition, although MAXI J1848$-$015 was excluded from our sample, as previously discussed, we note that its MeerKAT data exhibited two components of a (presumably) single ejection event that were monitored regularly over $\sim$500 days \citep{bahramian_2023}. 

Assuming that the ejecta kinetic energy estimated by \citet{savard_2025} for MAXI J1820+070 is typical for BH XRBs, and that each ejection event consists of two identical components launched in opposite directions from the core (i.e., $\sim$$4\times 10^{44}$ erg per ejection event), we estimate a minimum kinetic energy of $\sim$$4.4 \times 10^{45}$ erg for the $\sim$11 resolved ejecta during ThunderKAT. 

This is an extremely conservative lower limit, as it is likely that the sources launched many additional ejecta that were not detected -- either because they could not be resolved from the core at the resolution of MeerKAT, or were undetectable at the sensitivity of the telescope (sometimes due to beaming effects). In particular, we observed several HS$\rightarrow$SS core radio flares for which no ejecta were resolved, despite the fact that ejecta might have been launched during these transitions (e.g., the flare for MAXI J1631$-$479 on MJD $\sim$58565; \citealt{monageng_2021}). In addition, we emphasize the large uncertainty in the energy estimate for a single ejection event, which may also vary considerably between sources. Possible future work to improve upon this analysis includes exploring scaling relations between ejecta energies and their corresponding radio flares (e.g., \citealt{bacon_2026}), informed by simulations and modelling of sources exhibiting well-characterised flaring behaviour. 

In any case, the kinetic energy feedback in our sample is dominated by GRS 1915+105. As previously noted, it was excluded from our HS energy estimates due to the poorly understood nature of its flaring. In fact, naively employing our HS energy prescription yields an energy of $\sim$$10^{46}$ erg. This is equivalent to $\sim$25 ejecta with energies comparable to those of MAXI J1820+070, although the true number of ejections is likely much higher, as inferred from its very frequent radio flaring which becomes more apparent in higher-cadence observations (e.g., \citealt{motta_2021}). Consequently, the jet-driven kinetic energy from GRS 1915+105 alone is likely at least an order of magnitude greater than the combined contributions of all other sources in our sample. This is consistent with previous studies suggesting that the integrated kinetic feedback from XRBs in our Galaxy may be dominated by a small number of the most powerful systems, since rare, high-power jets may contribute significantly to the high-energy particle output (e.g., \citealt{heinz_2002, fender_2005, cooper_2020, wang_2025}).


\section{Radio:X-ray Coupling}\label{sec: coupling}

To construct the $L_R$--$L_X$ plane, we require quasi-simultaneous radio and X-ray measurements. For this analysis, GRS 1915+105 is not included, as its radio and X-ray flux have been decoupled since $\sim$2018 due to it being in an extended obscured phase \citep{motta_2021, miller_2025}. Similarly, Swift J1858.6$-$0814 is omitted, owing to strong and variable line-of-sight absorption, which renders the source's state evolution and intrinsic X-ray luminosity uncertain \citep{hare_2020, van_den_eijnden_2020, rhodes_2022}. Circinus X-1 is also excluded, as the nature of the high-cadence flaring observed for this source is not well understood (Gasealahwe et al. in preparation).

\subsection{Pairing}

As a first approach, following many previous studies of the $L_R$--$L_X$ plane, for each source, we pair radio and X-ray data taken within a day of each other. Specifically, for each radio observation, we compute a weighted average of all X-ray measurements within $\pm$1 day bins, using the reciprocal of the variance (computed from the average of the upper and lower uncertainties) as the weight. This approach increases the effective SNR of the X-ray measurements, but may smooth over possible short-timescale X-ray variability. Alternatively, each radio observation could simply be paired with the nearest X-ray observation. However, since multiple X-ray measurements within a single time bin are rare, the difference in results between the two schemes is negligible. We also note that the non-simultaneity of our data points introduces some intrinsic scatter to our correlation, but we expect this to be largely accounted for by the systematic uncertainty added to each X-ray point. This pairing process results in 274 quasi-simultaneous data points on the $L_R$--$L_X$ plane -- which we refer to as our \emph{`paired' plane}. Of these, 169 correspond to the HS/QS, including 147 HS/QS points from the BHs and BHCs.

\subsection{Interpolation}\label{subsec: interp}

However, many of the MeerKAT and \emph{Swift}/XRT observations were not taken within a day of each other. To minimise data loss due to non-simultaneous X-ray and radio observations, we also constructed an \emph{`interpolated' plane}, by interpolating the X-ray measurements to the radio observation dates. This choice of interpolating to the radio dates is motivated by the fact that we are a radio-led programme, and the X-ray data are typically more densely sampled and therefore better suited for interpolation. Furthermore, this produces $L_R$:$L_X$ pairs spaced approximately weekly for each source -- rather than clusters of temporally concentrated, similarly-valued pairs that could arise from interpolating to the X-ray epochs, which could bias subsequent statistical analyses.

Since our aim is to capture general trends in the $L_R$--$L_X$ plane, we investigated using simple interpolation schemes from the \texttt{scipy.interpolate} package, in which the resultant interpolated curves pass through each data point by construction. After visually inspecting the results of various schemes, we selected Akima interpolation (\citealt{akima_1970}; implemented via \texttt{scipy.interpolate.Akima1DInterpolator}\footnote{\href{https://docs.scipy.org/doc/scipy/reference/generated/scipy.interpolate.Akima1DInterpolator.html\#scipy.interpolate.Akima1DInterpolator}{\nolinkurl{scipy.interpolate.Akima1DInterpolator}}}) -- which constructs a continuously differentiable sub-spline built from piecewise cubic polynomials -- as it yields smooth curves across all data sets, avoiding spurious high-amplitude oscillations that are not justified by the sparse sampling (i.e., it mitigates Runge's phenomenon). We chose to interpolate the logarithmic (base-10) X-ray fluxes, as this transformation reduces the dynamic range and yields smoother behaviour, while avoiding the unphysical negative values that can arise when interpolating in linear flux space.

More sophisticated approaches were considered but not adopted. In particular, smoothed splines and generalised additive models (GAMs) require explicit assumptions (such as the curve smoothness), which are difficult to justify given our sparse sampling. We also explored Gaussian Processes (GPs; see \citealt{aigrain_2023} for a review of GPs in astronomical time-series analysis), which incorporate measurement uncertainties in a self-consistent manner and can, in principle, leverage multi-band data to better constrain the interpolation. However, GP modelling introduces additional complexities, such as kernel selection and hyperparameter tuning.

To estimate the uncertainties on the interpolated points derived from the adopted Akima scheme, we performed a Monte Carlo (MC) simulation. In each of 1,000 iterations, before running the Akima interpolation, we resampled every observed X-ray flux value, $y$, by drawing from a Gaussian distribution in log space, with a mean $\log_{10}(y)$ and standard deviation max[$\log_{10}(y + \Delta y_{u}) - \log_{10}(y), \log_{10}(y) - \log_{10}(y - \Delta y_{l}) $], where $\Delta y_{u}$ and $\Delta y_{l}$ are respectively the upper and lower uncertainties on $y$. For each radio data point, we then calculated the corresponding X-ray flux as the median of the set of interpolated results at that time, and the upper and lower uncertainty bounds as the 16th and 84th percentiles, respectively. 

As a baseline comparison, we also employed a simple linear interpolation in log space, which yielded very similar results. In particular, this scheme allows uncertainties to be propagated directly, but may produce less reliable results near turning points in the light curves.

Before performing further analyses, we applied additional filtering to the interpolated X-ray data points, as the reliability of the interpolation decreases with increasing temporal separation from the nearest observed data. Each interpolated result at time $t$ is bracketed by two observed X-ray data points -- let $t_1$ denote the time of the observed point closest to $t$, and $t_2$ be the time of the other.  We only included the interpolated point if: (1) $|t-t_1| \leq 1$ day, or (2) $|t-t_1| \leq T_1$ day and $|t-t_2| \leq T_2$. After experimentation, we settled on using $T_1 = 3$ days and $T_2 = 10$ days, and while this choice is arbitrary, we find that varying these parameters slightly does not have any noticeable effect on the final results. When $t$ is outside the temporal range of the X-ray observations (i.e., extrapolation), we include only those with $|t-t_1| \leq 1$ day, adopting the X-ray flux at $t_1$.

\subsubsection{Limitations}

We note that the uncertainty on the interpolated result extracted from the MC simulations is purely driven by the uncertainty on the observed data points. As such, we may wish to add an additional uncertainty due to the interpolation, dependent on the values of $|t-t_1|$ and/or $|t-t_2|$, where the uncertainty is an increasing function of these values. For example, one could estimate this uncertainty by calculating the average slope in log space (i.e., d$\log(y$)/d$t$) for each source separately or all the sources together. However, given that the data are so sparsely sampled and that the behaviour of the light curves of each XRB varies over time, any scheme we implement would be relatively ad hoc. Therefore, we opt to not add any additional uncertainty, and simply note the limitations of the method. This approach is sufficient for our purposes in this paper, since the uncertainty on each X-ray point is already conservative, and individual sources span many orders of magnitude in luminosity, dwarfing any epoch-localised uncertainties arising from interpolation choices.

\begin{figure}
    \centering
    \includegraphics[width=\columnwidth]{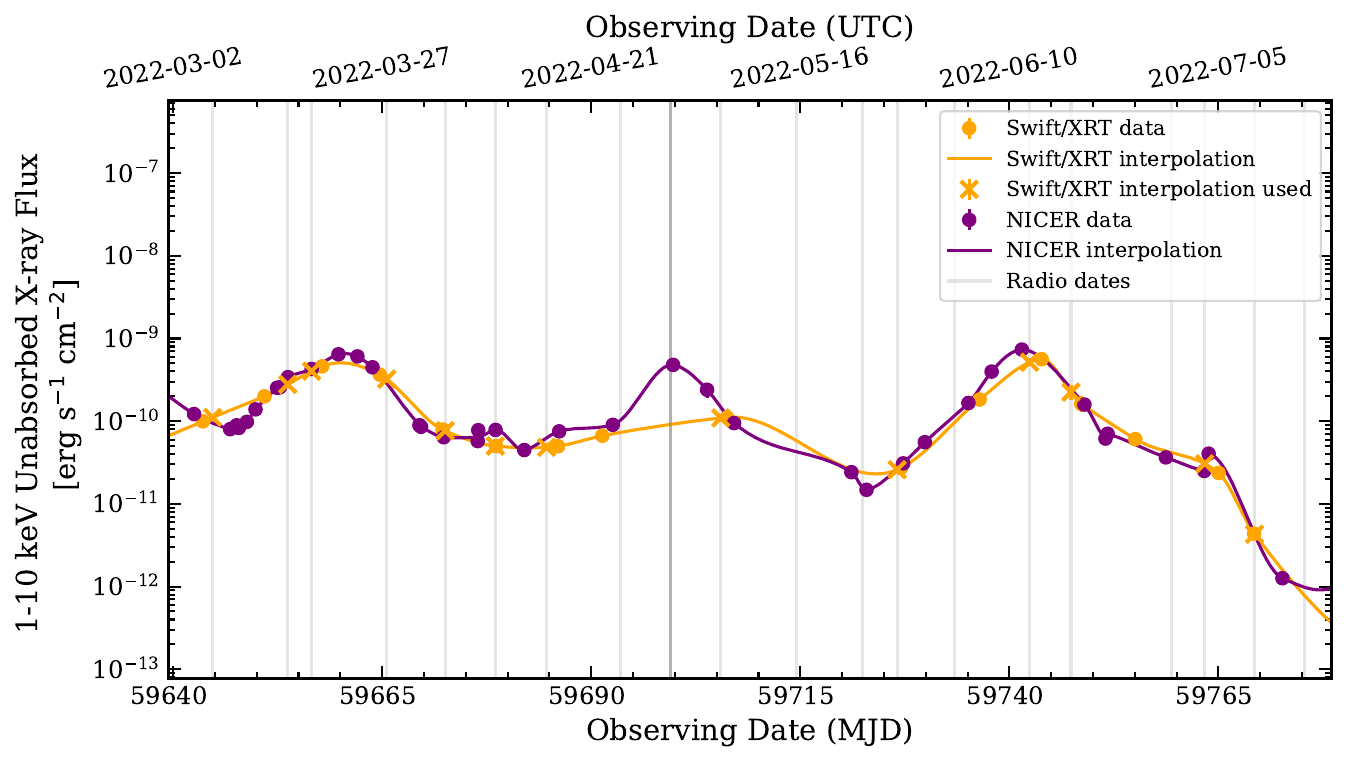}
    \caption{A subset of the X-ray results for 4U 1543$-$47 during a phase of HS-only reflares following its canonical outburst. The orange and purple data points represent the unabsorbed 1–10 keV fluxes from \emph{Swift}/XRT and NICER, respectively. The curves indicate the Akima interpolations of these data (see Section \ref{subsec: interp} for details). The grey vertical lines show the dates when MeerKAT radio observations were taken. The orange crosses indicate the interpolated data retained after filtering, as described in the text.}
    \label{fig: 4U1543_swift_nicer_comparison}
\end{figure}

Another primary limitation of our interpolation scheme stems from the relatively sparse \emph{Swift}/XRT sampling. XRBs can display short-timescale flares between observational epochs, which may only be revealed by higher-cadence monitoring. For example, in Figure \ref{fig: 4U1543_swift_nicer_comparison}, we show a segment of the 1--10 keV X-ray light curves of 4U 1543$-$47 during a phase of HS-only reflares (see also the third panel in Figure 1 of \citealt{zhang_2025a}), using data from \emph{Swift}/XRT (orange) and from the Neutron star Interior Composition ExploreR (NICER, \citealt{gendreau_2016}; purple). When comparing the interpolations obtained using the two data sets, it is clear that the former missed the HS flare at MJD $\sim$59700 (shown as the bold grey vertical line), leading the interpolated curve to underestimate the flux at that time. However, X-ray flaring is expected to be less common in the HS (which is of primary interest to us for the $L_R$--$L_X$ analysis) compared to the SS and IMS, and given that we filtered our interpolated results to only include those within three days of the nearest X-ray data point, these missed flares are generally excluded from our analysis. We assume that errors due to missed flares and dips are few, and that they are centred around zero (i.e., that they are relatively unbiased in any particular direction), such that our overall results are not affected. In all subsequent analyses in this paper, comparisons with the `paired' plane serve as an additional check that the interpolation does not introduce significant bias into our results.

\subsubsection{Expanded X-ray Data Set}\label{subsec: enhanced_plane}

Several of our sources were monitored -- sometimes at higher cadence -- with NICER and other X-ray telescopes. In upcoming work, we will construct an \emph{`enhanced'} $L_R$--$L_X$ plane by incorporating additional quasi-simultaneous X-ray observations, thereby increasing the number of data points. However, as illustrated in Figure \ref{fig: 4U1543_swift_nicer_comparison}, even quasi-simultaneous NICER and \emph{Swift}/XRT observations can yield fluxes that are inconsistent within their quoted uncertainties, potentially due to instrumental and calibration effects. Consequently, as previously described, combining data from multiple telescopes introduces heterogeneity into the sample and therefore requires a careful consideration of systematic uncertainties. 

This expanded X-ray data set will also allow a more direct assessment of uncertainties introduced by our interpolation scheme, particularly those arising from short-timescale flaring. Using our radio measurements together with the enlarged X-ray sample, we aim to investigate the HS variability of the XRBs in our sample (e.g., following a similar approach to \citealt{plotkin_2019} in their analysis of quiescent observations of V404 Cygni).

\begin{figure*}
    \centering
    \includegraphics[width=\textwidth]{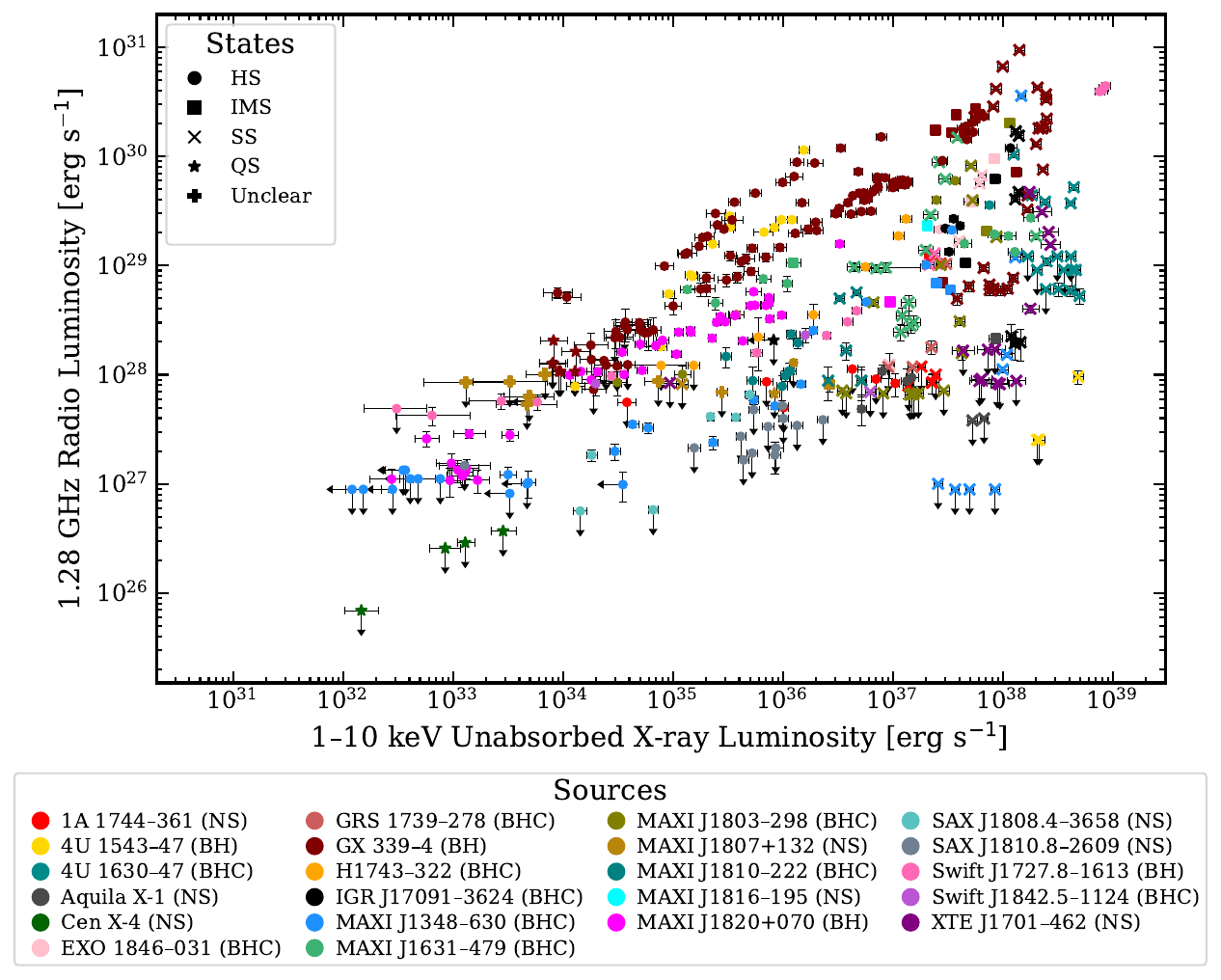}
    \caption{A version of our $L_R$--$L_X$ plane, constructed by interpolating our X-ray light curves, as described in Section \ref{subsec: interp}, and using the nominal source distances in Table \ref{tab: system_properties}. The error bars on the data do not include distance uncertainties. The colours for each source and markers for the spectral states are the same as in Figure \ref{fig: all_radio_luminosities}. Sources of all types are shown: black holes (BHs), black hole candidates (BHCs), and neutron stars (NSs). We include data for all the spectral states -- hard states (HS), intermediate states (IMS), soft states (SS), quiescent states (QS) -- and observations which we cannot label are called `Unclear'. This figure can be seen in greater detail on our \href{https://thunderkat.physics.ox.ac.uk/lrlx}{website}.}
    \label{fig: LrLx_plane_interp_sources}
\end{figure*}

\begin{figure}
    \centering
    \includegraphics[width=\columnwidth]{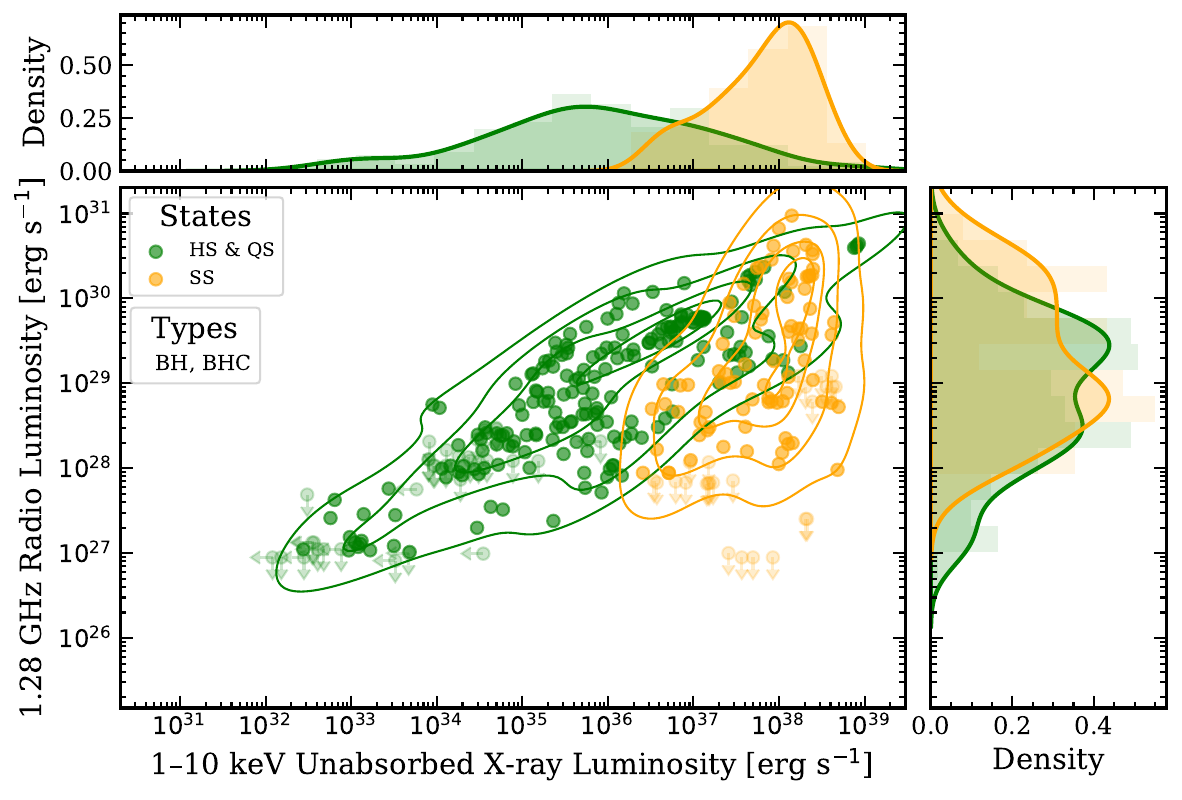}
    \caption{The `interpolated' $L_R$--$L_X$ plane for the black holes and black hole candidates in our sample, for the hard/quiescent (green) and soft (orange) spectral states. Error bars have been omitted for clarity. To aid visualisation, the data are overlaid with kernel density estimates (KDEs) computed using the detections; contours show the \{5, 25, 60, 90\} per cent highest-density regions. Upper limits were not used for the histograms or KDEs, but are displayed with lighter shading and arrows. }
    \label{fig: LrLx_plane_interp_HS_vs_SS}
\end{figure}

\subsection{The Radio:X-ray Plane}\label{subsec: lrlx_plane}

\begin{figure}
    \centering
    \includegraphics[width=\columnwidth]{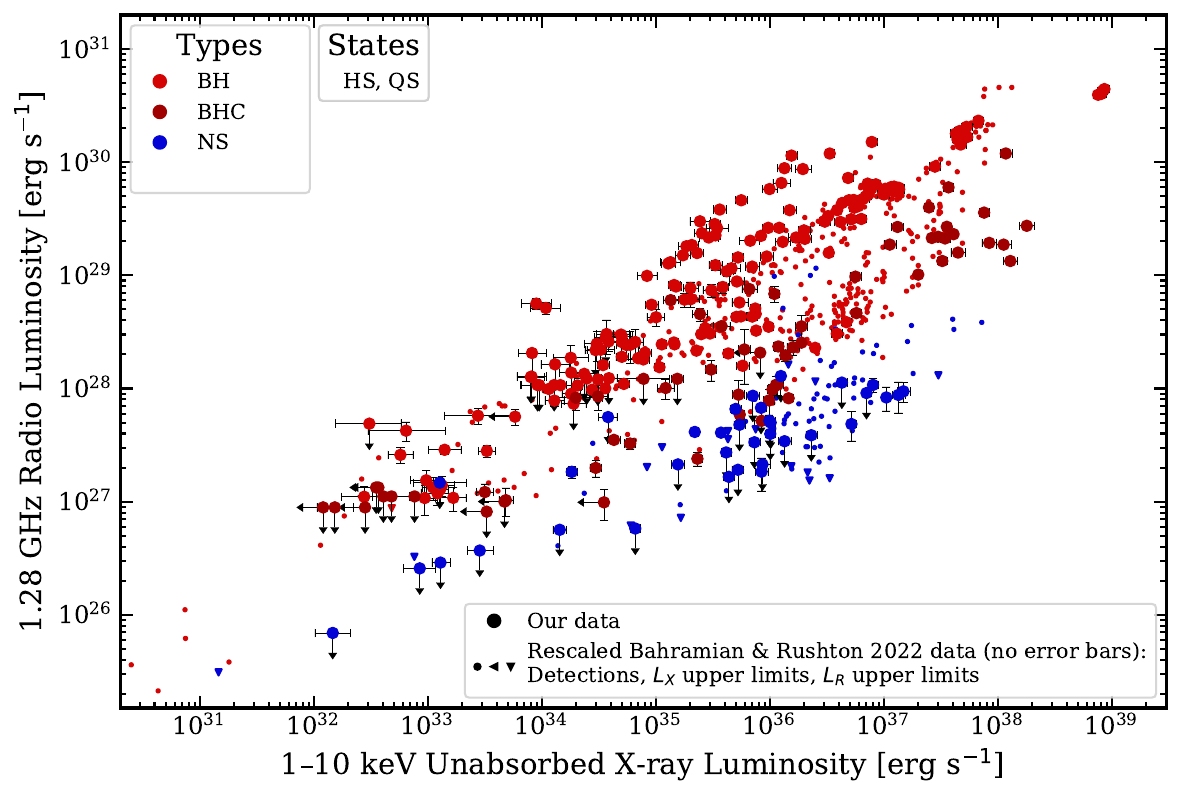}
    \caption{The same as Figure \ref{fig: LrLx_plane_interp_sources}, but showing only the data points at times when the sources were in the hard and quiescent states. The ThunderKAT/X-KAT data are shown by the large markers, where the colours indicate the various accretor types (light and dark red are respectively the black hole and black hole candidate systems, while blue indicates neutron stars). The smaller markers are the data from \citet{bahramian_2022} (error bars are omitted for visual clarity), where the radio data has been rescaled to 1.28 GHz by assuming flat spectral indices, for comparison with our sample.}
    \label{fig: LrLx_plane_interp_states}
\end{figure}
%

\begin{figure}
    \centering
    \includegraphics[width=\columnwidth]{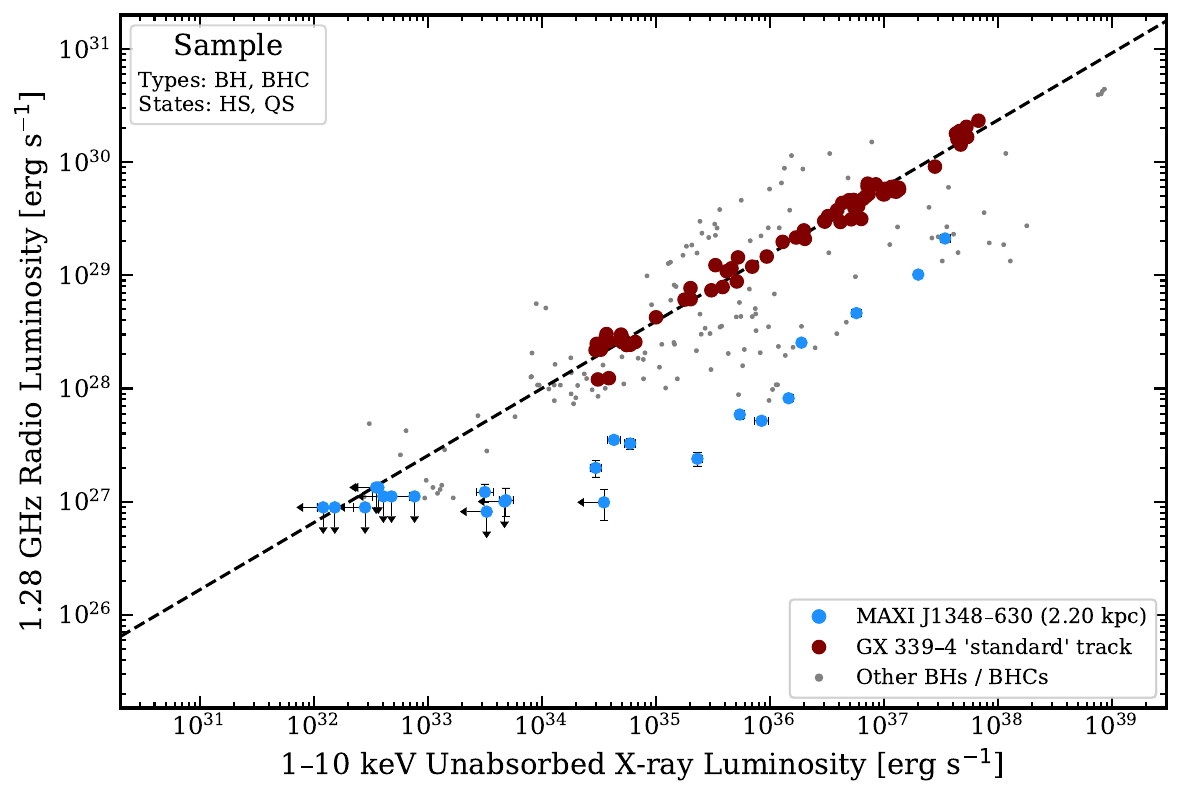}
    \caption{The same as Figure \ref{fig: LrLx_plane_interp_sources}, but showing only the data points for the black holes and black hole candidates during their hard and quiescent states. The dashed line is our defined `standard' track, obtained using a subset of GX 339$-$4 data (shown in maroon; see Section \ref{subsec: standard_track}). In blue, we highlight the `hybrid' behaviour of the black hole candidate system MAXI J1348$-$630.}
    \label{fig: LrLx_plane_interp_maxij1348}
\end{figure}

In Figure \ref{fig: LrLx_plane_interp_sources}, we show the `interpolated' $L_R$--$L_X$ plane for observations during all spectral states, where the luminosities were calculated using the nominal source distance estimates listed in Table \ref{tab: system_properties}. The corresponding `paired' plane can be seen on our \href{https://thunderkat.physics.ox.ac.uk/lrlx}{website}.

Furthermore, in Figure \ref{fig: LrLx_plane_interp_HS_vs_SS}, for the BHs and BHCs, we compare the HS/QS (green) and SS (orange) data points for the `interpolated' $L_R$--$L_X$ plane. As described in Section \ref{subsec: SS_emission}, core radio emission during the SS is generally interpreted as arising from ejecta (that are spatially disconnected from the core) rather than the compact jets, and is therefore not expected to correlate with the core X-ray emission. Two-sample (1D) Kolmogorov-Smirnov (KS) tests comparing the HS/QS and SS samples (detections only) yield $p$-values of $1.96 \times 10^{-2}$ for $L_R$ and $1.54 \times 10^{-28}$ for $L_X$. However, since these tests are performed on different properties of the same observed data points, they are not statistically independent and individual KS tests formally constitute multiple hypothesis tests and thus need to be corrected. To address this, we combine the two $p$-values using the harmonic mean $p$-value (HMP; e.g., \citealt{wilson_2019}) method with equal weights, which remains valid under positive dependence between tests. This yields a combined $p$-value of $3.08 \times 10^{-28}$, again indicating that the HS/QS and SS samples are inconsistent with being drawn from the same parent distributions, and therefore occupy different regions of the $L_R$--$L_X$ plane -- although we note that this combined result is strongly driven by the X-ray data. Repeating these tests after excluding the additional GX 339--4 branch (attributed to ejecta contamination) yields identical conclusions. It is important to emphasize that these tests compare the distributions of individual luminosity \emph{data points}, and therefore do not explicitly account for the hierarchical structure of the data (i.e., multiple data points per source).

We further apply a two-dimensional KS test \citep{peacock_1983, fasano_1987} to the $L_R$--$L_X$ detections (e.g., using \texttt{ndtest}\footnote{\url{https://github.com/syrte/ndtest}}), which yields $p=3.70 \times 10^{-19}$, likewise indicating that the HS/QS and SS data points occupy distinct regions in the $L_R$--$L_X$ plane. We caution, however, that 2D KS tests are approximate and not strictly distribution-free, and should therefore be regarded only as qualitative indicators of separation (see, e.g., Section 10.3 of \citealt{rizzo_2019} for a discussion of several more modern approaches).

Furthermore, the large number of SS $L_R$ upper limits at high $L_X$ seen in Figure \ref{fig: LrLx_plane_interp_HS_vs_SS} also supports the interpretation that the dominant radio-emission mechanism in the SS differs from that operating during the HS/QS. Additionally, the SS distribution shows no evidence for a coupling between $L_R$ and $L_X$. Consequently, for the remainder of this paper, in plots of the $L_R$--$L_X$ plane, we exclude SS data points. We also omit the few data points in the IMS, as radio emission during this state is likely also associated with discrete ejecta rather than compact jet emission. 

Figure \ref{fig: LrLx_plane_interp_states} shows the HS and QS data from our sample (large markers), colour-coded by accretor type. We also show the HS and QS data from \citet{bahramian_2022} (smaller markers) rescaled from 5 GHz to 1.28 GHz by assuming flat spectral indices. For this sample, we exclude objects classified as `candidateBH', as these were identified using different criteria from those applied to our sources -- primarily their $L_R$--$L_X$ behaviour rather than the outburst characteristics. We also exclude white dwarf systems (`WD') and tMSPs, since our sample does not include these subclasses.

\subsubsection{The `Standard Track'}\label{subsec: standard_track}

It is clear from Figure \ref{fig: LrLx_plane_interp_sources} that there is considerable scatter in the $L_R$--$L_X$ plane, making it difficult to discern the behaviour of each source. One of the most noticeable features is the distinct HS branch of GX 339$-$4 that lies at higher radio luminosities (for fixed X-ray luminosities) than its low-scatter canonical track identified in previous studies (e.g., \citealt{corbel_2003a}). This behaviour is attributed to contamination from unresolved ejecta components \citep{tremou_2026}. 

For the remainder of the paper, we define the \emph{`standard track'} as the correlation obtained by fitting a subset of the GX~339$-$4 HS data that excludes the low-luminosity cluster at $L_X\sim10^{34}$~erg s$^{-1}$, as well as epochs likely contaminated by ejecta emission -- specifically, the observations obtained at MJDs in the range 58956--59080 and 59496--59505. The first interval corresponds to the start of the 2020 HS decay, during which time ATCA observations confirmed the presence of an unresolved ejection. The second interval marks the beginning of the 2021 HS decay, when the core emission may again have been contaminated by ejecta components, as suggested by shifts in the measured SS emission in the days immediately preceding. 

The result of a linear regression fit to the standard track (see Section \ref{sec: individual_sources} for details) is shown as the grey line in Figure \ref{fig: LrLx_plane_interp_maxij1348}, along with the data used in the fit highlighted in maroon. In blue, we also show the HS and QS data for MAXI J1348$-$630, which clearly exhibits `hybrid' behaviour \citep{carotenuto_2021b}.


\section{Clustering Analysis}\label{sec: clustering}

In this section, we perform clustering on the $L_R$–$L_X$ plane, as a comparison with previous global/population-level analyses by \citet{gallo_2012, gallo_2014, gallo_2018}. Specifically, using a sample of 18 HS and QS BH XRBs and various clustering techniques, \citet{gallo_2012} reported statistical evidence for two tracks in the $L_R$–$L_X$ plane, although a subsequent analysis with an expanded sample and full uncertainty treatment by \citet{gallo_2018} found no robust partitioning of the population-level data into distinct luminosity tracks. 

To directly compare our results, we employ a Python\footnotetext{\href{https://scikit-learn.org/stable/modules/generated/sklearn.cluster.AffinityPropagation.html}{\texttt{sklearn.cluster.AffinityPropagation}}} implementation of the `affinity propagation' (APCLUST; \citealt{frey_2007}) clustering algorithm used by these authors. The method uses measures of pairwise similarity (typically the negative squared Euclidean distance) as input, and identifies a set of high-quality cluster centres (exemplars) through an iterative `message-passing' process in which data points exchange and update messages until convergence. A damping factor ($\lambda$) controls how strongly the current message update depends on the previous one, while the preference parameter ($p$) controls the number of clusters found, with higher values increasing the likelihood that more exemplars are chosen. The algorithm is limited to detections, so we remove the $L_R$ and $L_X$ upper limits, and use only the HS/QS data for the BHs and BHCs in our sample. Before performing the clustering analysis, we follow the data preprocessing scheme outlined in \cite{gallo_2018}. Specifically, we preprocess the $L_R$ and $L_X$ vectors by taking base-10 logarithms and standardising each vector to zero mean and unit variance to account for their differing dynamic ranges (i.e., $L_R$ and $L_X$ likely trace different fractions of the bolometric luminosity, and the data are more extended along the $L_X$ axis). Additionally, as the Euclidean distance metric performs poorly on highly correlated data, we apply principal component analysis (PCA; e.g. \citealt{jolliffe_2012}) with whitening, which rotates the data into an uncorrelated principal-component basis with unit variance. These transformations ensure comparable sensitivity (in terms of the ability to identify clusters) both parallel and perpendicular to the standard track. 

\begin{figure}
    \centering
    \includegraphics[width=\columnwidth]{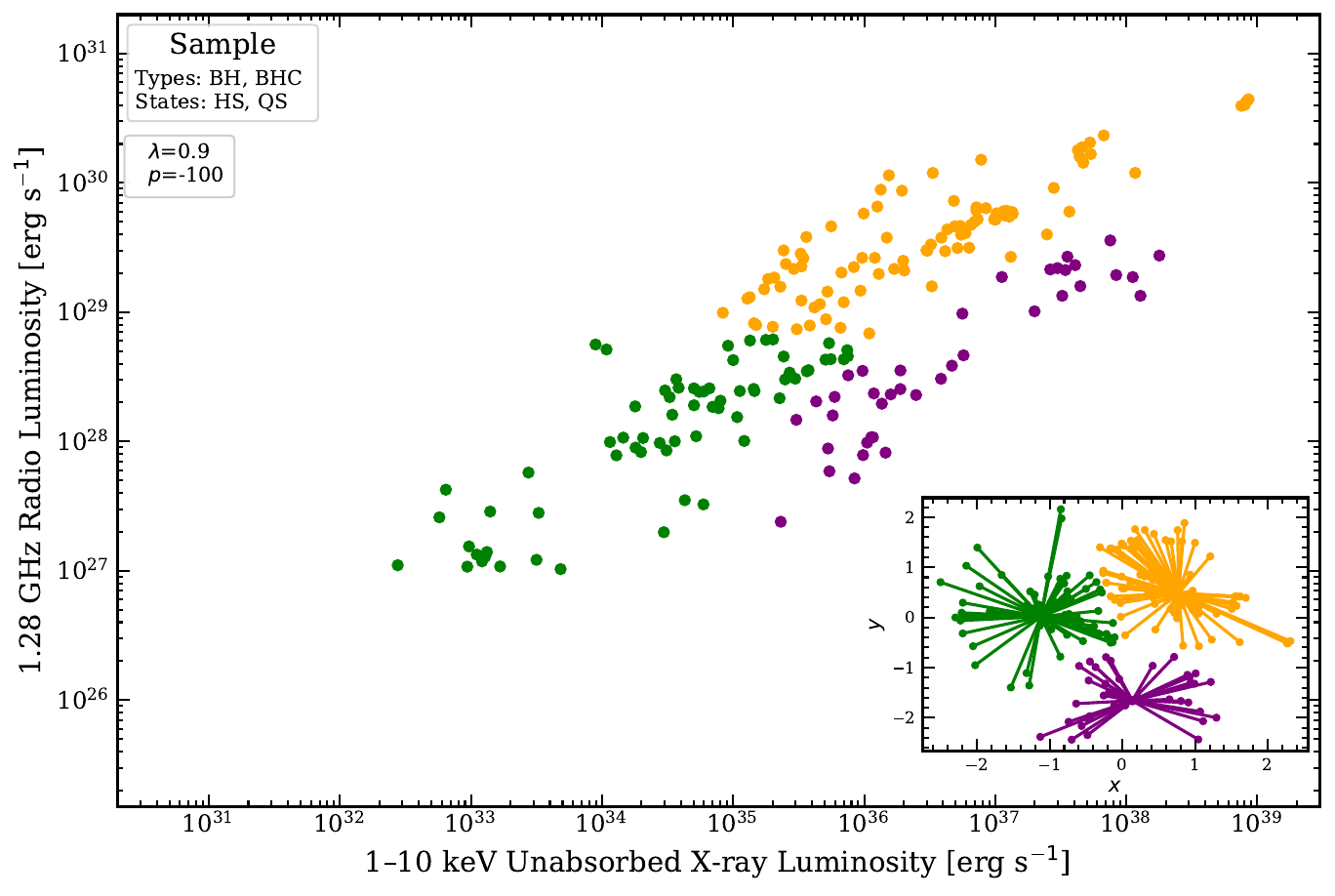} 
    \caption{Results of the affinity propagation clustering analysis applied to the `interpolated' $L_R$--$L_X$ plane (constructed as described in Section \ref{subsec: interp}, using the nominal source distances in Table \ref{tab: system_properties}), using the clustering hyperparameters $\lambda = 0.9$ and $p = -100$ for comparison with \citet{gallo_2012}. Only hard- and quiescent-state data for black hole and black hole candidate systems are included, and all $L_R$ and $L_X$ upper limits are excluded. The error bars have been omitted for clarity. The insets show the transformed data ($x$ and $y$) used in the clustering analysis (see Section \ref{sec: clustering}), and the clusters identified are shown in different colours. }
    \label{fig: clustering_comparison}
\end{figure}

We explored a range of values for the affinity propagation hyperparameters, but for direct comparison with the work of \citet{gallo_2012}, Figure \ref{fig: clustering_comparison} shows the result for $\lambda = 0.9$ and $p =-100$, applied to the `interpolated' $L_R$--$L_X$ plane constructed using the nominal source distance estimates (Table \ref{tab: system_properties}), with the inset showing the corresponding results in the transformed space. As expected, the clustering results are highly sensitive to the choice of hyperparameters. For example, increasing $\lambda$ to 0.95 leads to the formation of two clusters instead of three. Although a lower branch is identified in most cases, the algorithm does not yield a robust or stable partition of the data. Similar behaviour is observed when applying other clustering algorithms: the inferred clusters vary substantially depending on the specific method and parameter choices.

This clustering approach has several other limitations. The algorithms used consider only detections, ignoring upper limits (censored data) and measurement uncertainties, and although methods exist that address either (e.g., \citealt{shy_2022, lee_2012}), none were found that satisfactorily handle both. In addition, the algorithms ignore the hierarchical nature of the data, where multiple observations correspond to the same system and are linked by shared source properties and temporal evolution. As a result, heavily sampled systems such as GX~339–4 dominate clustering outcomes. For example, daily monitoring of a source would create dense clusters in the $L_R$–$L_X$ plane (especially if it is slowly varying) that could disproportionately weight its trajectory in the plane, and bias the inferred cluster structure. Although this effect is mitigated in our analysis by the approximately weekly sampling of each source, simple clustering methods remain insufficient to capture the full complexity of the data, and are included here primarily for comparison with previous studies.

Nevertheless, we emphasise that the inability of the clustering to robustly identify distinct tracks simply reflects that this type of analysis smooths over the source-specific trends that are clearly evident when individual systems are examined. The results also support the conclusion that these systems exhibit a continuum of behaviour, rather than a strictly bimodal distribution in the $L_R$--$L_X$ plane.


\section{Linear Regression}\label{sec: linreg}

\subsection{Methods}\label{subsec: linreg_method}

We estimate the slope and normalisation for our compiled $L_R$--$L_X$ plane by conducting a linear fit in log-log (base 10) space, using a Python implementation\footnotemark \footnotetext{\url{https://github.com/jmeyers314/linmix}} of \texttt{linmix} \citep{kelly_2007}, which was first applied in this context by \citet{gallo_2014}. In this method, Markov Chain Monte Carlo (MCMC) simulations are performed to fit a linear model of the form $Y = \alpha + \beta X + \epsilon$, accounting for uncertainties in both $X$ and $Y$ (which are assumed to be normally distributed and independent), and $Y$ upper limits (modelled via a Gaussian cumulative distribution function in the likelihood). Additionally, the algorithm includes an intrinsic random \emph{scatter} about the regression line (measured directly from the data), of the form $\epsilon \sim \mathcal{N}(0, \sigma_\epsilon)$. It adopts a Gaussian-mixture prior for the independent variable -- which is a well-justified assumption if the number of Gaussians is large enough (we use three) -- and performs a fit of the parameters $\alpha$, $\beta$, and $\sigma_\epsilon$. 

We emphasize that we specifically chose a regression scheme that allows for a scatter term, as this prevents data points with small errors but large offsets from the relation from being overweighted. This is a particularly important consideration when fitting all sources simultaneously, given the substantial intrinsic scatter. Likely contributors to this scatter include the non-simultaneity of luminosity measurements, relativistic beaming, variations in accretion rate at fixed X-ray luminosity, source-dependent environments, and the fact that (owing to differences in system properties, such as disc size and temperature) a given radio or X-ray waveband may probe different physical regions of the jet and accretion flow in different sources.

In our context, we linearise the model $L_R / L_{R,0} = \xi (L_X/L_{X,0})^\beta$ to obtain
\begin{equation} \label{eq: lin_reg}
\begin{aligned}
\log_{10}\left(\frac{L_R}{L_{R,0}}\right) 
&= \log_{10}(\xi) + \beta \log_{10}\left(\frac{L_X}{L_{X,0}}\right) + \epsilon \\
\Leftrightarrow\quad 
Y & = \alpha + \beta X + \epsilon
\end{aligned}
\end{equation}
where $\beta$ is the correlation coefficient (i.e., slope), and $\alpha$ is the normalisation. The luminosities are scaled by $L_{R,0}$ and $L_{X,0}$ for numerical stability, which we set to the median of the HS/QS radio and X-ray luminosity detections of the full data set, respectively: $L_{R,0} = 4.54 \times 10^{28}$ erg s$^{-1}$ and $L_{X,0} = 7.51 \times 10^{35}$ erg s$^{-1}$. Since \texttt{linmix} assumes normally distributed uncertainties for each data point, we conservatively adopt the larger of the upper and lower uncertainties -- i.e., max[$\log_{10}(L + \Delta_{L,u}) - \log_{10}(L), \log_{10}(L) - \log_{10}(L - \Delta_{L,l}) $], where $L$ is either the radio or X-ray luminosity, and $\Delta_{L,u}$ and $\Delta_{L,l}$ are respectively the upper and lower uncertainties. Parameter estimates ($\alpha$, $\beta$, and $\sigma_\epsilon$) and their uncertainties are taken as the median, 16th/84th percentiles of $10^4$ posterior samples, although consistent results are obtained when using the mean instead. 

To propagate distance uncertainties, we follow a similar approach to \citet{gusinskaia_2020} and \citet{van_den_eijnden_2022b, van_den_eijnden_2022c}. We perform $10^3$ Monte Carlo \emph{repeats}, each time drawing a distance for each source from its assumed probability distribution (either Gaussian or uniform; see Table~\ref{tab: system_properties}) and rescaling the luminosities by $(D_i / D)^2$, where $D_i$ is the sampled distance and $D$ is the nominal distance. We report the fitted parameters and their 1$\sigma$ uncertainties as the median and 16th/84th percentiles of the posterior samples from all the repeats. By combining all samples, we are essentially conflating distinct posterior shapes, thereby producing broader posteriors compared to a run using a single distance estimate. This provides more conservative and realistic uncertainty estimates compared to simply averaging the uncertainties from each repeat. Therefore, this approach is preferable, particularly when estimating the normalisation of a source, whose fit is dominated by distance uncertainties -- specifically, varying the assumed distance for a source may significantly affect its normalisation, while having only a minimal impact on its slope, as the correlation is effectively shifted diagonally in the $L_R$--$L_X$ plane. 

\subsection{Results: Black Hole X-ray Binaries}

\begin{figure}
    \centering
    \includegraphics[width=\columnwidth]{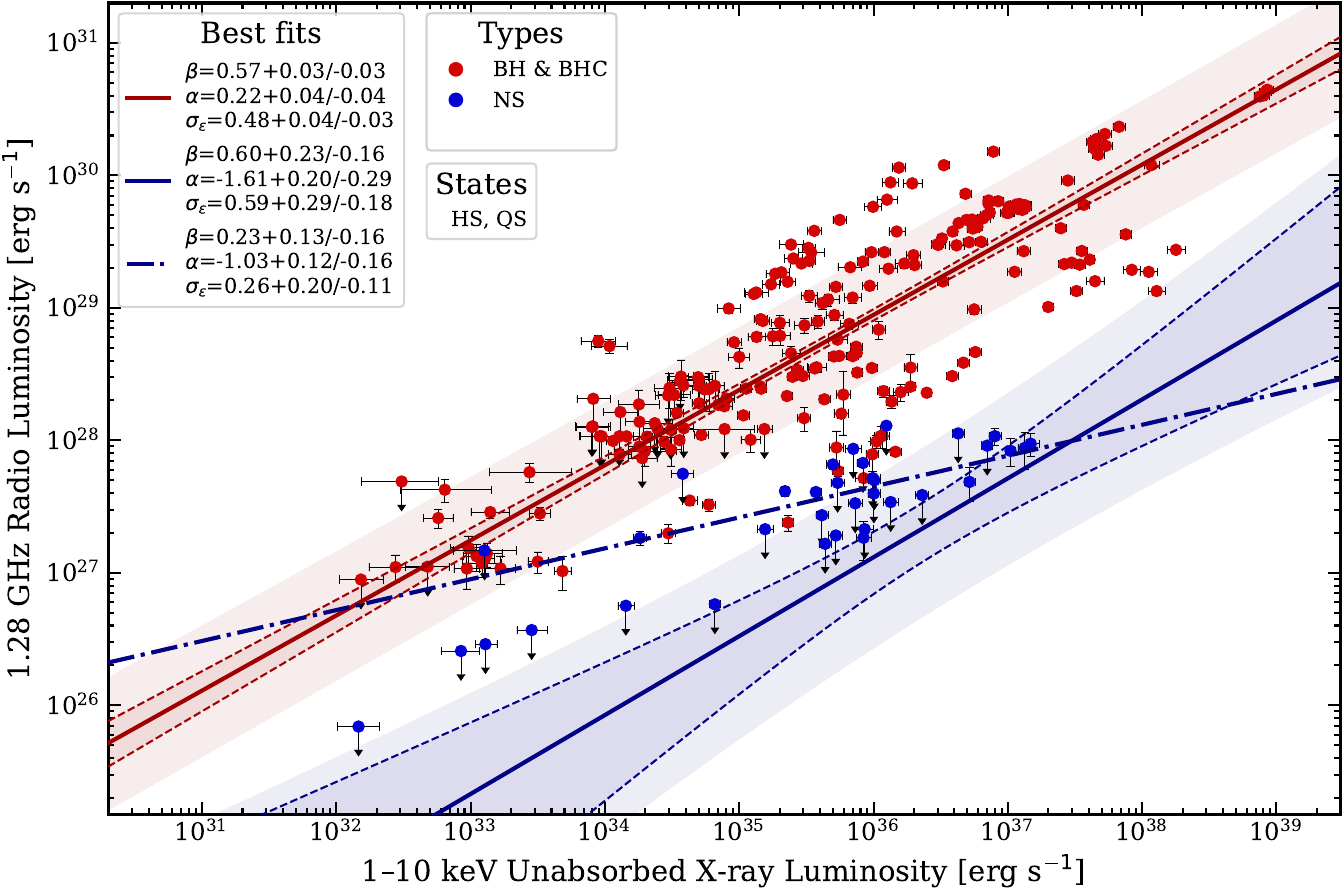} 
    \caption{Results of the \texttt{linmix} linear regression routine (Section \ref{subsec: linreg_method}) applied to the `interpolated' $L_R$--$L_X$ plane (Section \ref{subsec: interp}) for black hole and black hole candidate systems (red) and neutron star systems (blue) in the hard and quiescent spectral states. Best-fit values and their 1$\sigma$ uncertainties ($\beta$: slope; $\alpha$: normalisation; $\sigma_\epsilon$: intrinsic scatter) are calculated as the median and 16th/84th percentiles from $10^3$ fits in which the source distances were resampled from their assumed probability distributions (see Table \ref{tab: system_properties}). Dashed lines indicate the uncertainty bounds, while the lightly shaded regions represent the uncertainty including intrinsic scatter. The dot-dashed blue line shows the fit when only the NS detections are used (i.e., the upper limits are excluded). The error bars on the data do not include distance uncertainties, and the data are plotted for the nominal source distances. }
    \label{fig: interp_BH_NS_regression}
\end{figure}

For the \emph{BH and BHC} systems in our sample, a \emph{single} fit to the `interpolated' plane constructed using the nominal source distances yields $\beta = 0.57^{+0.03}_{-0.02}$, $\alpha = 0.22^{+0.03}_{-0.03}$, and $\sigma_\epsilon=0.48^{+0.03}_{-0.02}$. Using the `paired' plane instead produces consistent results within 1$\sigma$ uncertainties: $\beta = 0.58^{+0.03}_{-0.03}$, $\alpha = 0.22^{+0.04}_{-0.04}$, and $\sigma_\epsilon=0.46^{+0.03}_{-0.03}$. In all cases, the parameters are well constrained and the posterior distributions are approximately Gaussian.

We note that adopting a more conservative treatment of the uncertainties by adding an additional 0.15 dex in quadrature (in logarithmic flux-density space) to each $L_R$ and $L_X$ measurement on the `interpolated' plane -- following \citet{gallo_2018} -- results in an inferred slope and normalisation that agrees within uncertainties. Additionally, although the intrinsic scatter decreases slightly in this case, it remains consistent within uncertainties.

Conducting \emph{repeats} (i.e., varying the assumed source distances) using the `interpolated' plane produces similar results that agree within 1$\sigma$ uncertainties: $\beta = 0.57^{+0.03}_{-0.03}$, $\alpha = 0.22^{+0.04}_{-0.04}$, and $\sigma_\epsilon=0.48^{+0.04}_{-0.03}$ (see the fit in Figure \ref{fig: interp_BH_NS_regression}). This indicates that distance uncertainties have a relatively minor impact on the inferred parameters. The normalisation is expected to be affected more drastically than the slope because, as previously described, our sample is dominated by GX 339$-$4, and varying its distance shifts its correlation diagonally in the plane -- although this difference is not very apparent in our data.

For comparison, we also fit the `interpolated' plane using the restricted GX 339--4 subset defined in Section~\ref{subsec: standard_track} -- which excludes HS points potentially contaminated by ejecta emission -- while retaining all available HS data for the other sources. With this approach, we obtain slightly different best-fit parameters, most notably a lower normalisation: $\beta = 0.57^{+0.03}_{-0.03}$, $\alpha = 0.14^{+0.04}_{-0.04}$, and $\sigma_\epsilon = 0.42^{+0.04}_{-0.03}$.

\subsection{Results: Neutron Star Low-mass X-ray Binary Comparison}\label{sec: NS_comparison}

\begin{figure}\label{NS_comparison}
    \centering
    \includegraphics[width=\columnwidth]{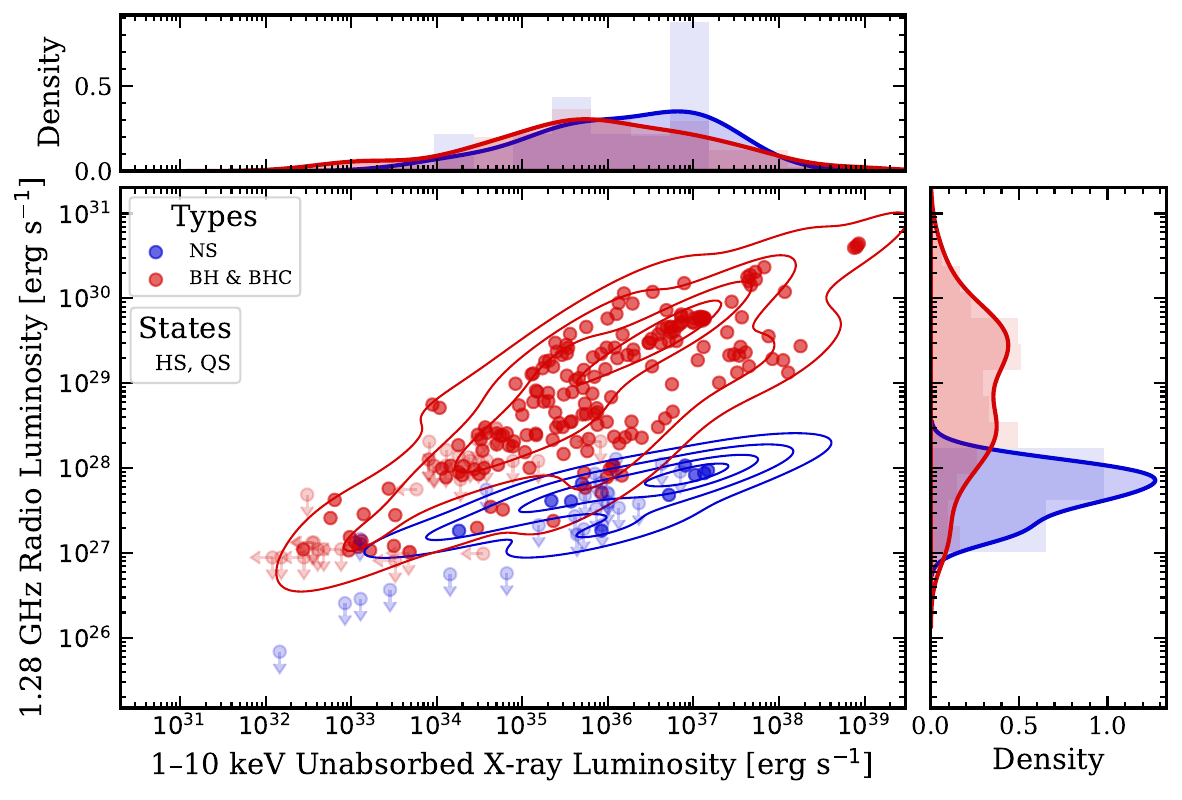}
    \caption{The `interpolated' $L_R$-$L_X$ plane (see Section \ref{subsec: interp}) for the data in the hard and quiescent spectral states, where the black hole and neutron star populations are coloured in red and blue, respectively. Error bars have been omitted for clarity. To aid visualisation, the data are overlaid with kernel density estimates (KDEs) of their distributions. Upper limits were not used for the histograms or KDEs, but are displayed with lighter shading and arrows.}
    \label{fig: BH_NS_2D_comparison}
\end{figure}

For comparison with previous studies, we examine population-level differences between the BHs and NS systems in our sample, focusing on their relative radio-loudness. This analysis is inherently limited by the small NS sample size and prevalence of radio upper limits among the NS data. Figure \ref{fig: BH_NS_2D_comparison} shows the `interpolated' $L_R$--$L_X$ plane, with BHs and NSs shown in red and blue, respectively. In this plot, kernel density estimates (using detections only) are overlaid to aid visualisation. 

A two-sample (1D) KS test of $L_X$ (detections only) for the two samples yields a $p$-value of 0.65, providing no evidence for a difference in their marginal distributions. Thus, the BH and NS data appear to occupy a comparable region of the X-ray parameter space, allowing a more meaningful comparison of their behaviour in the $L_R$--$L_X$ plane. Furthermore, a KS test applied to $L_R$ (detections only) for the two samples yields a $p$-value of 1.93$\times 10^{-7}$, suggesting a statistically significant difference in the distributions, with the BH sources generally exhibiting higher radio luminosities than the NS sources. However, as previously noted, we caution that the tests formally constitute multiple hypothesis tests on the same data. Combining the two $p$-values using the HMP method with equal weights yields a $p$-value of $3.87 \times 10^{-7}$, again suggesting that the BH and NS samples are not drawn from the same parent distributions. Repeating these tests without the additional branch of GX 339--4 (arising from ejecta contamination) leads to identical conclusions. We reiterate that this test compares the distributions of individual luminosity \emph{data points} rather than the distributions of the underlying BH and NS \emph{populations}, as the latter would require a more sophisticated approach that handles multiple data points per source. We do not conduct a 2D KS test as the number of data points is insufficient. 

Using the `interpolated' plane for the NSs, we perform a linear regression analysis with repeats, using the methods outlined above, and obtain: $\beta = 0.60^{+0.23}_{-0.16}$, $\alpha = -1.61^{+0.20}_{-0.29}$, and $\sigma_\epsilon=0.59^{+0.29}_{-0.18}$ (see the solid blue line in Figure \ref{fig: interp_BH_NS_regression}). 

These results indicate that the BH and NS slopes are consistent within uncertainties, although the NS constraints are weak due to limited data. In contrast, the best-estimates of the normalisations between the BH and NS populations differ significantly, suggesting that the BH sample is likely on average $\gtrsim$10–100 times more radio-loud than the NS one. Additionally, we note that sensitivity limits restrict us to the brightest portion of the NS $L_R$--$L_X$ plane. Therefore, if NSs exhibit an intrinsic scatter comparable to the BHs, they may be even fainter on average than inferred here.

We emphasize, however, that the NS fitting result is largely determined by the \texttt{linmix} method of handling upper limits. In fact, if we include only the detections (of which there are very few), we obtain a lower slope and higher normalisation: $\beta = 0.23^{+0.13}_{-0.16}$, $\alpha = -1.03^{+0.12}_{-0.16}$, and $\sigma_\epsilon=0.26^{+0.20}_{-0.11}$ (dot-dash line in Figure \ref{fig: interp_BH_NS_regression}). The disparity between these results and those obtained when upper limits are included hints at the diverse behaviour exhibited by NS XRBs in the $L_R$--$L_X$ plane (see Section~\ref{subsec: ns_comparison} for further discussion).

We also reiterate that the NS results are limited by the small sample size and are highly sensitive to the specific data points included. Notably, we chose to exclude the bright detection of the AMXP MAXI J1816--195, since the observation likely occurred during the HIMS rather than the HS, so the radio emission may not have been dominated by the compact jet. However, this classification remains uncertain, and opting to include this point significantly steepens the inferred slope and substantially affects the results. Clearly, additional observations of NS systems are required before any firm conclusions can be drawn.

\subsection{Limitations}

The analysis presented has several important limitations. Heavily monitored systems that have multiple detected outbursts -- particularly GX~339–4 -- dominate the global fits, introducing biases similar to those discussed in the clustering section. Although weighting fits by the number of data points per source or adopting a bootstrap approach could mitigate this effect, such methods would still smooth over the source-specific tracks that are evident in the $L_R$–$L_X$ plane when examining sources individually. 

We therefore advocate fitting \emph{each source individually} and examining the resulting distributions of slopes and normalisations, rather than focusing on population-level analyses. However, this approach is limited by two factors. First, many sources have few data points, and robust slope measurements require sampling over at least two orders of magnitude in X-ray luminosity \citep{corbel_2013}, particularly for identifying outlier tracks. Second, some sources (e.g., MAXI~J1348–630) are better described by broken power laws (i.e., a broken straight line in log-log space) or more complex functions (e.g., an exponential in log-log space), whereas \texttt{linmix} only supports single linear fits. To address these issues and enable a more detailed comparison between our sources, we are preparing a paper that expands the current data set using additional telescopes, and implements a custom regression framework that is capable of incorporating upper limits, treating distance uncertainties hierarchically, and fitting user-defined functional forms. To this end, we have recently extended the original functionality of the regression package \texttt{roxy}\footnote{\url{https://roxy.readthedocs.io/en/latest/}} \citep{bartlett_2023} to support the inclusion of upper limits. 

\subsection{Results: Individual Sources}\label{sec: individual_sources}

\begin{figure}
    \centering
    \includegraphics[width=\columnwidth]{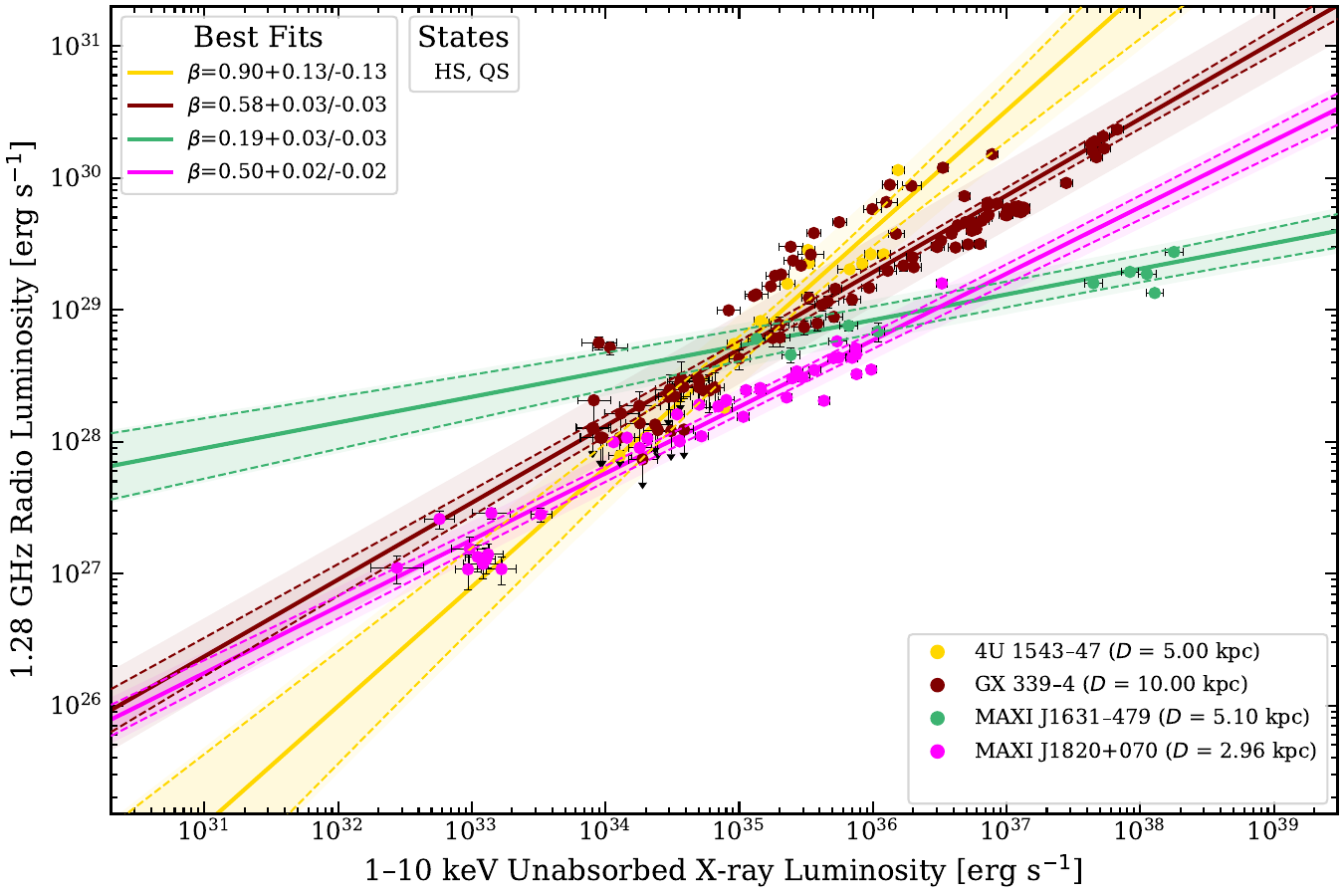} 
    \caption{Linear regression results for a few well-sampled sources, using the `interpolated' plane. The parameter $\beta$ is the slope extracted from a \texttt{linmix} fit, while accounting for distance uncertainties (Section \ref{subsec: linreg_method}). Dashed lines indicate the 1$\sigma$ uncertainty bounds, while the lightly shaded regions represent the uncertainty with intrinsic scatter included. The error bars on the data do not include distance uncertainties, and the data are plotted using the nominal source distances. }
    \label{fig: individual_lin_reg}
\end{figure}

\begin{figure}
    \centering
    \includegraphics[width=\columnwidth]{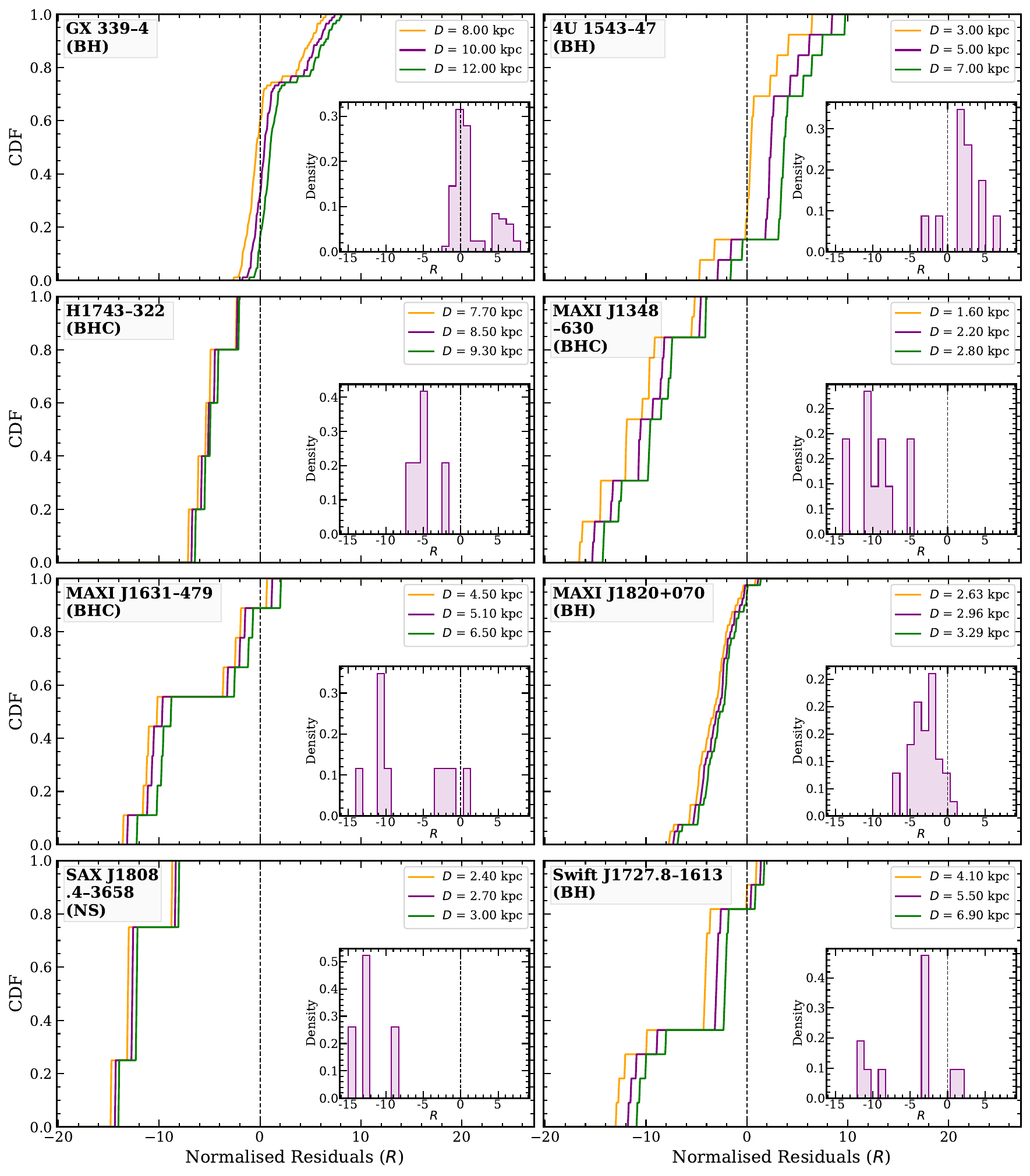} 
    \caption{The cumulative distribution functions (CDFs) of the normalised residuals ($R$) of the radio detections for several sources relative to the `standard' track defined by GX 339$-$4 (vertical dotted line; see Section \ref{sec: individual_sources}). For each source, we include only hard- and quiescent-state detections (in both $L_X$ and $L_R$), obtained from the `interpolated' plane. Data points to the right of the vertical dashed line are more `radio-bright' than the standard track, while those to the left are more `radio-faint'. The different colours indicate the source distances used, with purple corresponding to the best-estimate distance. The inset in each panel shows the histogram of the normalised residuals computed using the nominal distance for that source. The additional branch of GX 339$-$4 above the standard track (arising from unresolved ejecta emission) is clearly visible. Most sources in the sample are found to be radio-quiet, and Swift J1727.8$-$1613 is one example of a source showing both radio-quiet data points and other points lying close to the standard track.}
    \label{fig: normalised_residuals}
\end{figure}

In Figure~\ref{fig: individual_lin_reg}, we present the results of the linear regression fits (accounting for distance uncertainties, as described in Section \ref{subsec: linreg_method}) for our few well-sampled sources that are adequately described in log-log space by a single linear relation. The fits clearly illustrate the range of slopes and normalisations spanned by the sources in our sample.

In Section~\ref{subsec: standard_track}, we defined the `standard track' as the correlation obtained by fitting the subset of GX~339–4 HS data that excludes the potentially contaminated epochs, as well as the low-luminosity cluster at $L_X \sim 10^{34}$ erg s$^{-1}$. Accounting for distance uncertainties, a fit to this track yields: $\beta = 0.59^{+0.01}_{-0.01}$, $\alpha = 0.46^{+0.05}_{-0.05}$, and $\sigma_\epsilon = 0.06^{+0.01}_{-0.01}$, consistent with the canonical $\sim$0.6 slope reported in the literature (e.g., \citealt{corbel_2013}).

For the analysis that follows, we define the \emph{`normalised residual'} of each radio \emph{detection} relative to the standard track as: $R = [l_r - (\beta \times l_x + \alpha)]/\sigma_t$, where $l_r\equiv\log_{10}\left(L_R/L_{R,0}\right)$, and $l_x\equiv\log_{10}\left(L_X/L_{X,0}\right)$, and $\sigma_t$ is the total uncertainty of the numerator of $R$ (propagated using the intrinsic scatter, measurement errors, and uncertainties in slope and intercept). In Figure \ref{fig: normalised_residuals}, cumulative density functions (CDFs) of the results are shown for a few well-sampled sources, computed using several possible source distances. In these plots, data points to the right of the vertical dashed line are more `radio-bright' than the standard track, while those to the left are more `radio-faint', and the value of $R$ at CDF = 0.5 represents the median offset from the standard track. The insets show the corresponding histogram constructed using only the nominal source distance. 

Although simple, the metric $R$ provides a clear visual comparison of source behaviour, and a quantitative measure of `radio-loudness'. In particular, we see that GX~339–4 exhibits an additional `radio-loud' branch above the standard track (resulting from ejecta contamination, as previously described). 4U 1543–47 also has a number of data points that lie above the standard track. In contrast, the detections for the outliers H1743–322 and MAXI~J1348–630 lie below the standard track, with the former located closer to it. MAXI J1631–479 and MAXI~J1820+070 similarly fall predominantly below the track, with the latter lying closer to it. Swift J1727.8–1613 exhibits data points below the standard track, as well as a few that fall slightly above it. Lastly, all data for the NS system SAX~J1808.4–3658 lie far below the standard track. 

In what follows, we summarise the HS/QS $L_R$--$L_X$ behaviour of each source in our sample. 
\begin{itemize}[itemsep=0.8em]

\item \textbf{GX 339–4}: As noted previously, the source exhibits an additional branch above the standard track during the decay phases of its 2020 and 2021 `full' outbursts. This excess is likely attributable to emission from ejecta that remained unresolved from the core at the angular resolution of MeerKAT -- a conclusion supported by four dual-frequency observations of resolved ejecta with ATCA in mid-2020 \citep{tremou_2026}. This branch has a similar slope ($\sim$0.6), suggesting that the transient ejection is fading at a constant rate. Previous work has reported that GX~339–4 exhibits variations in both the normalisation and slope of its $L_R$–$L_X$ correlation, depending on whether it is in the rising or decaying HS and whether the outburst is failed or successful, suggesting an evolution in its jet-accretion coupling over time (e.g., \citealt{corbel_2013, islam_2018, koljonen_2019, de_haas_2021}; see Section~\ref{sec: evolution} for a discussion). A detailed multi-outburst analysis of the source using ThunderKAT and X-KAT data will be presented in a forthcoming paper.

\item \textbf{MAXI J1820+070}: Our MeerKAT data covered the decaying HS of a canonical outburst and three HS-only rebrightenings. During this time, the source followed a relatively constant $L_R$--$L_X$ slope that is shallower than the standard track, as also noted by \citet{bright_2025}. We obtain a 90\% upper limit of $\sim$0.53 for the slope, consistent with the result reported by \citet{shaw_2021} using data from a previous outburst.

\item \textbf{MAXI J1631–472}: It exhibits a shallower correlation coefficient than the standard track, as shown in \citet{monageng_2021}, with a 90\% upper limit on the slope of $\sim$0.24.

\item \textbf{H1743–322}: It was one of the first well-sampled `outlier' sources \citep{coriat_2011}, found to lie below the standard track and to follow a curved trajectory in the (logarithmic) $L_R$–$L_X$ plane, with a shallower slope at low luminosities and a steeper one at high luminosities. Our MeerKAT observations of a single HS-only outburst show similar behaviour \citep{williams_2020}, suggesting that the trajectory may be largely independent of whether the outburst is successful. Although, we note that during our monitoring, the source does not fully return to the standard track at either extreme, unlike earlier reports.

\item \textbf{MAXI J1348–630}: Our monitoring covers a single full outburst and several HS-only reflares. Like H1743–322, it follows a curved track in the plane (but at lower radio luminosities), rejoining the standard track as it approaches quiescence, as discussed in detail in \citet{carotenuto_2021b}.

\item \textbf{Swift J1727.8–1613}: It displays outlier behaviour similar to MAXI J1348–630, which becomes even more apparent when additional ATCA and VLA radio data are included \citep{hughes_2025_lrlx}.

\item \textbf{MAXI J1803–298}: Although our data include too few points to constrain the $L_R$–$L_X$ behaviour, Espinasse et al. (submitted) additionally use NICER, ATCA, and VLA observations and suggest that the source exhibits hybrid behaviour.

\item \textbf{4U 1543–47}: It exhibits a steep slope and large scatter in the $L_R$–$L_X$ plane, with a 90\% lower limit on the slope of 0.74 from our `interpolated' plane, consistent with the result obtained by \citet{zhang_2025a} using additional NICER data. In this study, the radio excess above the standard track is attributed to variable Doppler boosting rather than ejecta contamination, likely due to the low jet inclination ($\lesssim$19.7$^\circ$; \citealt{zhang_2025a}) and possibly a variable Lorentz factor or jet precession. It is also relevant to note that its HS points on our $L_R$–$L_X$ plane were obtained during a series of HS-only reflares, rather than the main canonical outburst, which may alternatively be the cause of its unexpected behaviour. 

\item \textbf{MAXI J1810–222}: It displays an unusual, almost vertical track in the $L_R$--$L_X$ plane (spanning a very narrow range in $L_X$), as previously found by \citet{russell_2022} using ATCA data. 

\item \textbf{IGR J17091–3624}: It has several HS data points in the $L_R$--$L_X$ plane, exhibiting a steep best-fit slope. However, these data span a narrow range in $L_X$, and therefore may not reflect the full evolution of the source in the plane. In Russell et al. (in preparation), the $L_R$--$L_X$ correlation for the source will be studied using additional ATCA data.

\item \textbf{Other BH systems}: The remaining BH and BHC XRBs in our sample are too sparsely sampled in the `interpolated' HS/QS $L_R$--$L_X$ plane to draw firm conclusions. Both \textbf{4U 1630–47} and \textbf{GRS 1739–278} contribute only a single HS point each in the plane. \textbf{Swift J1842.5–1124} has two HS points, and likely lies below the standard track with a flatter slope, although its distance is poorly constrained \citep{zhang_2022}. For \textbf{EXO 1846–03}, \citet{williams_2022} additionally used VLA and AMI-LA data and found that the source lies below the standard correlation for the range of distances assumed.

\item \textbf{SAX J1808.4–3658}: It is our best-sampled NS system, and is found to lie completely below the standard correlation in the $L_R$–$L_X$ plane. An in-depth analysis of this source, using additional high-cadence quasi-simultaneous NICER data from the 2019 outburst, together with new data from the 2025 outburst, will be presented in Gasealahwe et al. (in preparation). This source was previously studied in detail by \citet{tudor_2017}, who found no clear $L_R$–$L_X$ correlation, potentially due to sparse sampling during the reflares, a large time delay between the radio and X-ray emission, or strong propeller-driven ejections.

\item \textbf{Other NS systems}: The remaining NS XRBs are relatively poorly sampled in the `interpolated' HS/QS $L_R$–$L_X$ plane. \textbf{1A 1744–361} has five HS upper limits and a single detection (although the state classification of the latter is not certain). This places it among the faintest radio NS LMXBs, though we note that its distance is poorly constrained \citep{ng_2024}.  \textbf{Aquila X-1} contributes four HS detections to the $L_R$--$L_X$ plane over a narrow range in $L_X$, occupying a similar region identified in more extensive studies \citep{gusinskaia_2020}. \textbf{Cen X-4} is undetected in our MeerKAT observations, and provides deep constraints on the $L_R$–$L_X$ correlation at low X-ray luminosities ($L_X \lesssim 10^{33}$ erg s$^{-1}$) \citep{van_den_eijnden_2022b}. \textbf{MAXI J1807+132} has only two HS/QS upper limits. \textbf{MAXI J1816–195} contributes no HS points to the plane. However, it has a very bright observation that is currently classified as being in the IMS, although this remains uncertain. If this data point is reclassified as being in the HS, it would represent the brightest NS data point in the plane: for a source distance of 6 kpc, it lies at ($L_X$, $L_R$)$\sim$ ($2 \times 10^{37}, 2\times 10^{29}$) erg s$^{-1}$, although we note that the distance is poorly constrained. \textbf{SAX J1810.8–2609} has many HS observations, but most are upper limits (due to its emission being confused with an unrelated AGN), making its regression results highly uncertain. \textbf{XTE J1701–462} is observed only in the SS, where it follows a steep, vertical track in the plane (see also \citealt{gasealahwe_2024}).

\end{itemize}


\section{Discussion}\label{sec: discussion}

In this section, we briefly review the physical mechanisms proposed to explain the diverse behaviour of BH LMXBs in the $L_R$--$L_X$ plane, while recognising that multiple mechanisms may operate simultaneously with evolving relative importance throughout outbursts. In particular, we highlight insights that can be gleaned from the sources in our sample. We also discuss comparisons with the $L_R$--$L_X$ plane for NS LMXBs. In addition, we address why HMXBs have been excluded from our compiled $L_R$--$L_X$ plane, owing to the effects of wind-fed accretion and dense environments that complicate the interpretation of radio and X-ray emission. We review parallel studies of supermassive BHs, highlighting how they complement investigations of XRBs. We also discuss evidence that $L_R$--$L_X$ correlations may evolve in quiescence, and note how optical–X-ray correlations can further constrain the jet–accretion coupling in these systems.

\subsection{Bolometric Corrections}\label{sec: bolometric}

Most studies of the $L_R$--$L_X$ plane use narrow X-ray energy bands (typically 1--10 keV or 3--9 keV) as proxies for the \emph{bolometric} X-ray luminosity (and hence the accretion rate), despite the fact that the majority of the HS luminosity is emitted at $\sim$100 keV. In fact, a study of Cygnus X-1 showed that the power-law indices measured in narrow X-ray bands depend strongly on the selected energy range \citep{zdziarski_2011b, zdziarski_2012}. Moreover, \citet{koljonen_2019} found that the bolometric correction for narrow-band X-ray fluxes is not constant, but instead decreases at higher fluxes due to an evolving hard X-ray spectrum. Therefore, the choice of energy range to use for the $L_R$--$L_X$ plane can significantly affect the inferred properties. Notably, analyses of the plane using bolometric luminosities found that GX 339$-$4 exhibits a slightly higher correlation index ($\beta \approx 0.81$ when all the data are fitted together; \citealt{islam_2018}) compared to studies using narrow-band fluxes, with its correlation steepening at higher X-ray fluxes (see also \citealt{koljonen_2019}). In other words, when including higher-energy bands, GX 339$-$4 exhibits a type of hybrid behaviour. This raises the question of whether all sources might display similar behaviour when bolometric X-ray luminosities are considered. Nevertheless, \citet{koljonen_2019} concluded that bolometric corrections alone cannot account for the observed dichotomy between sources occupying different tracks in the $L_R$--$L_X$ plane, indicating that additional factors must be tuning the jet–accretion coupling.

\subsection{Evolution of Radio:X-ray Correlations}\label{sec: evolution}

It is well established that X-ray luminosities are typically higher during the rising HS than the decaying HS (when the compact jet is being re-established), forming a characteristic hysteresis cycle \citep{maccarone_2003_hysterisis}. Furthermore, \cite{corbel_2013} showed that the decaying HS of GX 339$-$4 tends to be more radio-bright \emph{for a given X-ray luminosity} (or more X-ray-faint for a given radio luminosity) than the rise, implying different normalisations in the $L_R$--$L_X$ plane; note that in this case, contamination from discrete ejecta can be ruled out. These authors also suggested that the $L_R$--$L_X$ correlation of GX 339$-$4 may steepen at high $L_X$, during the rising phase. Similar conclusions were reached by \cite{islam_2018} and \cite{koljonen_2019} using bolometric X-ray fluxes (see also \citealt{barnier_2022}). Specifically, \citet{islam_2018} found that GX 339$-$4 exhibits a steep $L_R$--$L_X$ correlation during its HS rise (with $\beta \approx 1.05$, compared to $\beta \approx 0.62$ for the decay), comparable to the steep branch of the outlier source H1743$-$322. In our sample, tentative evidence for an evolving $L_R$--$L_X$ correlation is seen during the outburst of Swift J1727.8$-$1613, where fitting the steep outlier track while excluding data points from the rising HS yields a steeper correlation (\citealt{hughes_2025_lrlx}, Figure 1). 

Additional complexity is evident between outbursts. For example, \citet{islam_2018} found that different outbursts of GX 339$-$4 exhibit varied behaviour, and \citet{de_haas_2021} reported a shallower correlation during its HS-only outbursts ($\beta \approx 0.46$) compared to the successful ones ($\beta \approx 0.62$). In our sample, we find no strong evidence for this behaviour. For example, although our MAXI J1820+070 data set is dominated by a series of HS-only (i.e., failed) reflares and followed a shallow correlation index of $\sim$0.5 in the $L_R$--$L_X$ plane, MAXI J1631$-$479 exhibited an even shallower slope during a successful outburst. 

These studies point to a complex coupling between the jet and accretion flow -- likely governed by multiple factors -- that evolve during outbursts. In this picture, \emph{all} sources may traverse different tracks in the plane during various stages of an outburst, driven by changes in the parameters tuning the jet--accretion coupling (see further discussion below), and sources that exhibit only a single track may simply never experience the conditions required to transition onto a steeper branch. In upcoming work, we will investigate whether any patterns emerge when comparing data from the rising HS, decaying HS, and HS-only outbursts/reflares across our sample.   

\subsection{X-ray-quiet vs. X-ray-loud Paradigm}

Assuming that the X-ray emission is dominated by the accretion flow, a possible explanation for the multiple tracks in the $L_R$--$L_X$ plane is that they reflect differences in the radiative efficiency of the accretion flow, giving rise to \emph{`X-ray-quiet'} and \emph{`X-ray-loud'} behaviour. In this framework, \citet{coriat_2011} proposed that sources on the steep outlier track host \emph{radiatively efficient} accretion flows –- such as luminous hot accretion flows (LHAF; \citealt{yuan_2004}) –- which convert a large fraction of the accreted mass into radiation and are therefore brighter in X-rays than standard-track sources. In contrast, the standard-track sources might be powered by \emph{radiatively inefficient} accretion flows, such as advection-dominated accretion flows (ADAFs; \citealt{narayan_1994}), in which energy is advected inward rather than radiated efficiently; in BH systems, it may cross the event horizon, while in NSs, it may be transferred to the stellar surface. Extensions to this scenario include strong mass loss through winds (advection-dominated inflow-outflow solution; ADIOS; \citealt{narayan_1995, blandford_1999}). 

Simple scaling arguments support this interpretation: if jet power scales with mass accretion rate ($L_{\text{jet}} \propto \dot{M}$; e.g., \citealt{heinz_2003}), and the radio and total jet luminosity scale as $L_R \propto L_{\text{jet}}^{\sim1.4}$ (for an optically thick jet; e.g. \citealt{heinz_2003, coriat_2011}), then radiatively inefficient flows (for which we expect $L_X \propto \dot{M}^{2-3}$; \citealt{merloni_2003}) produce $L_R \propto L_X^{0.5-0.7}$, consistent with the standard track. Conversely, radiatively efficient flows (for which $L_X \propto \dot{M}$) predict $L_R \propto L_X^{1.4}$, as observed for the steep branch of the outlier track in H1743$-$322 \citep{coriat_2011}. We note that this slope for the outlier track is slightly steeper than those measured for MAXI J1348$-$630 \citep{carotenuto_2021b} and Swift J1727.8$-$1613 \citep{hughes_2025_lrlx}, although a slope of $\sim$1.4 is obtained when fitting only the decay phase of the outlier track in the latter.


In this interpretation, one possibility is that hybrid sources transition from radiatively inefficient ADAFs to efficient LHAFs as the mass accretion rate increases above a critical value of $\dot{M}_C \propto 5\theta_e^{3/2} \alpha_\nu^2 \dot{M}_\text{Edd}$ (\citealt{xie_2012, xie_2016}), where $\dot{M}_\text{Edd} \equiv 10 L_\text{Edd}/c^2$ is the Eddington mass accretion rate, $\theta_e = k_B T_e/m_e c^2$ is the dimensionless electron temperature (e.g., $kT_e \sim 10{-}500$ keV; \citealt{koljonen_2019, yan_2020}), and $\alpha_\nu$ is the disc viscosity parameter (e.g., $\alpha_\nu \sim 0.1{-}1$; \citealt{tetarenko_2018}). Specifically, close to $\dot{M}_C$, the radiative efficiency can rise sharply while the mass accretion rate changes very little, producing the shallow outlier branch (i.e., transitional track) of the $L_R$--$L_X$ plane. After crossing $\dot{M}_C$, a two-phase accretion LHAF may develop: dense cold clumps form within the hot flow, supplying additional seed photons for Comptonisation in the corona, thereby enhancing the X-ray emission and shifting the source to the steep outlier track. By contrast, standard sources may have higher $\dot{M}_C$ and therefore remain radiatively inefficient throughout an outburst. Observationally, $\dot{M}_C$ (and thus the corresponding critical luminosity, $L_C$) spans many orders of magnitude across different sources (e.g., \citealt{islam_2018, koljonen_2019, carotenuto_2021b, hughes_2025_lrlx}). An assessment regarding the viability of the scenario presented requires better constraints on $\theta_e$ and $\alpha_\nu$ for the hybrid sources in our sample.


Assuming instead that the sources remain in an ADAF throughout, \citet{meyer_hofmeister_2014} proposed an alternative geometry that can arise at sufficiently high accretion rates ($L_X \gtrsim 10^{-3}{-}10^{-4} L_\text{Edd}$), in which a weak, cool disc forms from the re-condensation of coronal matter, providing similar seed photons that lead to enhanced X-ray emission and therefore shift sources to the steep outlier track. At lower values of $L_X$, the re-condensation model predicts that no inner disc can exist, explaining the transition back to the standard track. Furthermore, at high luminosities near the SS transition, the additional luminosity becomes unimportant and the two tracks merge.  

Observational support for the emergence of a cold component in hybrid sources includes the detections of high-energy X-ray spectral cut-offs (at tens of keV), implying an effective cooling of the Comptonising electrons and a corresponding morphological change in the accretion flow \citep{koljonen_2019}. Consistent with this picture, \citet{dincer_2014} found that outlier sources exhibit lower X-ray RMS variability than standard-track sources, as expected if a cool disc (or clumps) forms within the hot flow and suppresses variability (see \citealt{koljonen_2019} for further discussion). Note, however, that \cite{motta_2018} instead suggested lower \emph{hard-line} (i.e., X-ray count rate versus RMS) slopes for the outliers compared to standard-track sources, and thus a higher RMS at a given count rate.

Further evidence that the different tracks reflect distinct accretion properties was presented by \citet{cao_2014}, who showed that $L_R$--$L_X$ correlations are closely linked to X-ray spectral evolution. In particular, data exhibiting an anti-correlation between the HS photon spectral index ($\Gamma$) and X-ray flux ($F_X$) were found to follow the standard or shallow outlier tracks. On the other hand, a positive $\Gamma$--$F_X$ correlation was associated with the steep branch of the outlier track. The transition between these regimes across a critical flux suggests a switch in the radiation mechanism at a certain accretion rate. The authors showed that this behaviour is especially clear in multi-outburst observations of H1743--322, and a similar steepening of the $L_R$--$L_X$ correlation (to $\beta \sim 1.1$) is observed in GX 339$-$4 for data points with a positive $\Gamma$--$F_X$ correlation. Support for this interpretation is found in our sample: for Swift J1727.8$-$613, its photon index reaches a minimum at the transition between its steep and shallow $L_R$--$L_X$ tracks, consistent with a genuine physical change in the accretion flow \citep{hughes_2025_lrlx}. We will investigate whether a similar pattern emerges for GX 339$-$4 in an upcoming paper focused on the detailed evolution of its X-ray properties.  


\subsection{Radio-loud vs. Radio-quiet Paradigm}

An alternative explanation is that the tracks in the $L_R$--$L_X$ plane reflect differences in jet properties, leading to \emph{`radio-loud'} and \emph{`radio-quiet'} behaviour. Evidence for this scenario was reported by \citet{espinasses_2018}, who found that (on average) radio-quiet BH XRBs have steeper radio spectral indices ($\bar{\alpha} \sim -0.3$) than radio-loud ones ($\bar{\alpha} \sim +0.2$), implying different emission characteristics. These authors proposed that, assuming inclination effects are negligible, this could imply that radio-loud sources host jets that are optically thick (self-absorbed), whereas radio-quiet sources may be dominated by unresolved discrete ejecta. However, this scenario does not explain the presence of the shallow outlier track. An alternative, although similarly unlikely, explanation is that radio-quiet sources have spectral breaks (from optically thick to thin) that lie within the radio band, implying a change in jet structure when sources transition between tracks. In our sample, we find no clear evidence for a connection between the tracks and spectral indices, although this may reflect observational limitations, as intra-band spectral indices were not calculated for all epochs and are known to have biases at modest detection significances \citep{heywood_2016}. 


Additionally, in this context, we also caution that at L-band, modest low-frequency absorption (either internal synchrotron self-absorption or external free-free absorption) may in principle reduce the observed radio flux in some sources \citep{van_der_laan_1966}, potentially contributing to scatter in the $L_R$--$L_X$ relation.

The $L_R$--$L_X$ behaviour of a source may alternatively depend on how efficiently accretion energy is channelled into the jet as a function of accretion rate \citep{coriat_2011}. Jet emission may also strongly be influenced by the magnetic field strength: strong internal fields could suppress observable radio emission, producing radio-quiet behaviour \citep{casella_2009, peer_2009}. In future work, we will extract the polarisation properties of our entire LMXB sample, as a means to probe the magnetic fields and test this hypothesis.

\subsection{System Properties}\label{sec: doppler}

Many studies have attempted to explain the range of normalisations observed among sources in the $L_R$--$L_X$ plane by differences in their intrinsic system parameters -- although no clear trends have been found to date. 

For example, a naive correction for accretor mass using the Fundamental Plane (FP) relation ($L_{R,\text{corr}} \approx L_R/ M^{0.78}$; \citealt{merloni_2003}) cannot fully account for the observed scatter, since the mass range for XRB BHs spans a factor of $\sim$4. 

Furthermore, \citet{fender_2010} (see also \citealt{espinasses_2018}) found no correlation between BH spin and jet power, suggesting that spin does not drive the presence of the tracks. Although these studies are inherently uncertain due to the difficulty of robustly constraining spin measurements, this consensus is supported by the fact that a source's spin is not expected to vary on the timescales of a BH outburst, and therefore cannot account for the different tracks seen in hybrid sources \citep{gallo_2012}. 

Finally, \citet{soleri_2011} showed that the existence of multiple tracks does not reflect a trend in any other known physical parameters, such as the system's disc size or orbital period. However, see \cite{garcia_2003} for a possible indication that long-period transient BH XRBs produce larger jets than short-period systems. 

Another property that affects a source's location in the $L_R$--$L_X$ plane is its inclination relative to the observer. Specifically, \emph{Doppler boosting} (or relativistic beaming) is the enhancement or suppression of the observed luminosity of a moving source due to Doppler and relativistic aberration effects. X-ray emission is often assumed to be isotropic and therefore unaffected, although growing evidence suggests that source inclination may affect the X-ray spectral and timing properties of XRBs (e.g., \citealt{munoz-darias_2013, motta_2018, koljonen_2019}), so this assumption may not always hold. By contrast, radio emission from jets is directional, so the observed and intrinsic radio luminosities are related by $L_R = L_\text{int} \delta^{p}$, where $\delta$ is the Doppler factor defined as $\delta_{\text{rec/app}} = \Gamma_{\text{jet}}^{-1} [1 \pm \beta_{\text{jet}} \cos\theta]^{-1}$ for the receding and approaching jet components, $\Gamma_{\text{jet}} = (1 - \beta_{\text{jet}}^2)^{-1/2}$ is the bulk Lorentz factor, $\beta_{\text{jet}} = v/c$ is the intrinsic jet speed normalised by the speed of light, and $\theta$ is the jet angle with respect to the line of sight. The exponent is $p \equiv k - \alpha_s$, with $k = 2$ for a compact, continuously-replenished jet (e.g., \citealt{fender_2006}), and the radio spectral index is $\alpha_s\sim 0$. In most cases, the approaching jet dominates the observed emission, but when both components are considered, the effective Doppler factor for unresolved radio emission can be approximated (for $p=2$) as $\delta_{\text{eff,R}} = [(\delta_{\text{app}}^2 + \delta_{\text{rec}}^2)/2]^{1/2}$ \citep{gallo_2003}.

Several studies have examined the effects of Doppler boosting in the $L_R$--$L_X$ plane. Notably, \citet{motta_2018} proposed that the tracks may reflect \emph{inclination} effects, with radio-quiet sources being high-inclination (Doppler de-boosted) objects. \emph{Variable} Doppler boosting has also been invoked to explain some features in the plane: \citet{soleri_2011} showed that an increasing $\Gamma_{\text{jet}}$ above $10^{-3}$ L$_{\text{Edd}}$ could explain the enhanced scatter at high $L_X$, while \citet{russell_2015} attributed the steep $L_R$--$L_X$ slope of MAXI J1836$-$194 to an evolution of the form $\log(\Gamma_{\text{jet}}) \propto L_X$ that is consistent with simulations \citep{peault_2019}. Similarly, the peculiar behaviour of 4U 1543$-$47 (namely, its steep slope and large scatter) has been attributed to its low jet inclination angle and to variable Doppler boosting driven by changes in the jet's Lorentz factor or inclination angle (\citealt{zhang_2025a}). However, other studies argue against Doppler effects as the primary driver of the observed tracks. For example, \citet{espinasses_2018} found no correlation between whether a source is radio-loud/radio-quiet and its inclination, and \citet{gallo_2014} showed that Doppler boosting alone cannot account for the normalisation differences observed between V404 Cyg and XTE J1118+480.

A major limitation in assessing Doppler effects is the uncertainty in jet inclination estimates and the fact that the compact jet speeds are usually not known. While compact jet speeds are often assumed to be only mildly relativistic ($\Gamma_{\text{jet}} \lesssim 2$; e.g., \citealt{gallo_2003}), and the few resolved HS jets generally support relatively low Lorentz factors (e.g., \citealt{wood_2024}), some studies have reported significantly higher values for a few systems (e.g., \citealt{casella_2010, saikia_2019, tetarenko_2019, tetarenko_2021}). Constraints on jet inclinations are similarly uncertain: direct measurements are rare and precession may affect the inferred value, while indirect estimates based on orbital or inner-disc inclinations may not hold due to disc-jet misalignments (e.g., \citealt{martin_2008, miller_jones_2019_v404}). 

In our sample, GX 339$-$4 shows little scatter about its best-fit $L_R$--$L_X$ relation, suggesting that its Lorentz factor and inclination angle remain approximately constant during outbursts. By contrast, if the standard correlation is indeed `universal', hybrid sources (e.g., MAXI J1348$-$630) would require enhanced Doppler de-boosting near $L_C$ (relative to both higher and lower $L_X$), implying Lorentz factors that are larger than typically assumed (or precessing jets). Furthermore, no consistent behaviour could explain the different tracks seen for various sources. Therefore, while Doppler (de-)boosting undoubtedly contributes to scatter in the plane, it is unlikely to be the dominant cause of the distinct tracks. However, we reiterate that Doppler effects are difficult to isolate (particularly if it influences both radio and X-ray emission), and merit reconsideration as improved constraints on jet Lorentz factors and inclinations become available. 

\subsection{Neutron Star Comparison}\label{subsec: ns_comparison}

Comparisons between BH and NS LMXBs provide key insights into the roles of the event horizon, gravitational potential, and stellar magnetic field in jet production. However, they are limited by observational biases, as NSs are typically radio fainter and undergo faster state transitions. The first detailed comparison in the $L_R$--$L_X$ plane \citep{migliari_2006} found that NSs exhibit a steeper slope than BHs on average (respectively $\beta \gtrsim 1.4$ versus $\beta \sim 0.7$). This difference was interpreted as evidence for radiatively efficient accretion in NSs (due to the presence of a solid surface), in contrast to radiatively inefficient flows in BHs. The study also suggested that, on average, BHs are $\sim$30 times more radio-loud than NSs at a given X-ray luminosity. Using a larger NS sample, \citet{gallo_2018} found a shallower NS slope ($\beta \approx 0.45$), broadly consistent with that of BHs within 2.5$\sigma$ uncertainties, but confirmed that BHs remain significantly more radio-loud by a factor of $\sim$22. 

In earlier sections, we compared the radio-loudness of our BH and NS samples, although this analysis is limited by the small number of NS detections and the prevalence of upper limits. Despite these limitations, we found that BHs are likely $\gtrsim$10--100 times more radio-loud than NSs, on average (i.e., radio brighter at a given $L_X$, or X-ray fainter at a given $L_R$). Several effects may contribute to this offset \citep{migliari_2006, gallo_2018}: differences in the accretor mass (\citealt{merloni_2003}), bolometric corrections to $L_X$ \citep{zdziarski_2004, galloway_2008, gallo_2018}, and additional X-ray emission from the NS boundary layer \citep{burke_2017}. However, even when correcting for these effects, the NS population remains systematically less radio-loud than BHs at a given $L_X$. The origin for this difference remains uncertain, but may potentially be due to BHs advecting a portion of their energy across their event horizons \citep{narayan_1995}, or producing more powerful jets.

However, we note that caution is required when examining the NS $L_R$--$L_X$ plane, as different types of NS systems exhibit distinct relations, and substantial differences have also been observed between individual systems within the same class (e.g., \citealt{tetarenko_2016_exo, tudor_2017, qiao_2019, van_den_eijden_2021}). For example, AMXPs appear to represent extreme systems: some are among the most radio-loud NS LMXB systems and may approach the standard BH $L_R$--$L_X$ track (e.g., \citealt{migliari_2011, russell_2018}), while others are significantly more radio-faint (e.g., \citealt{tetarenko_2018_igr_j1659}). Although our sample does not include any tMSPs, we note that these systems have similarly been found to follow distinct correlations compared to all other LMXBs, possibly due to being jet-dominated accretion systems operating in a propeller mode \citep{deller_2015, bogdanov_2018}. Additionally, some NS XRBs show drastic changes in their radio jet emission despite small changes in their X-ray luminosity \citep{panurach_2021, panurach_2023}. Such varied behaviour between the NS populations could reflect differences in jet-launching mechanisms, magnetic field strengths, geometry, spin, compact object mass, orbital period, or the onset of strong propeller modes which could additionally potentially introduce longer delays between the radio and X-ray bands compared to BH systems (e.g., \citealt{migliari_2011, deller_2015, tudor_2017, van_den_eijden_2021}). Given this diversity, detailed source-specific studies spanning wide luminosity ranges are needed for robust conclusions to be drawn (e.g., \citealt{gusinskaia_2020}).

We emphasize again that, given that many of our NS radio measurements are upper limits and that the sample is very small, our results are likely biased and may not be measuring the true typical slope of NSs in the $L_R$--$L_X$ plane. Ongoing X-KAT observations will help to address these issues by substantially expanding our data set, enabling a more comprehensive comparison with BH systems and between different types of NS systems.

\subsection{High-Mass X-ray Binaries}

For completeness, we briefly consider HMXBs, which provide an instructive contrast to the Roche-lobe-fed LMXBs discussed above. Many HMXBs are wind-fed and often host strongly magnetised NSs, although some contain BHs, and a substantial fraction are Be/XRBs accreting from decretion discs (see \citealt{kretschmar_2019, martinez_2017}). In wind-fed systems, dense and clumpy stellar winds can absorb and reprocess X-ray emission, and obscure or contaminate radio emission via free-free absorption, thermal wind emission, or wind-jet interactions (see, e.g., Section 4 of \citealt{van_den_eijden_2021}). As a result, the observed X-ray luminosity no longer traces the instantaneous accretion power as they do in Roche-lobe-fed LMXBs, and the radio luminosity is sometimes decoupled from the intrinsic jet power. As a result, HMXBs generally lie offset from the canonical LMXB $L_R$--$L_X$ relation (e.g., \citealt{zdziarski_2016}). 

Our sample includes a single data point for the NS HMXB Vela X-1, which lies below our NS LMXB $L_R$--$L_X$ correlation, at $(L_X, L_R)\sim(1\times10^{36},6\times10^{26})$ erg s$^{-1}$, assuming a distance of 1.99 kpc. We note that since the radio emission from this source may include a substantial contribution from the stellar wind \citep{van_den_eijden_2021}, the measured $L_R$ likely overestimates the intrinsic jet luminosity and should therefore be treated as an upper limit.

\subsection{Extension to Active Galactic Nuclei}

Studies of stellar-mass and supermassive BHs are highly complementary: BH XRBs probe accretion and jet physics on much shorter timescales but are few in number \citep{corral-santana_2016}, whereas AGN evolve slowly yet are abundant \citep{flesch_2023}, enabling population-level studies across a wide range of BH masses. However, Fundamental Plane (FP) studies are limited by uncertainties in AGN BH mass estimates (\citealt{gultekin_2009}), the timescales over which radio and X-ray observations are paired (e.g., \citealt{miller_2010, king_2011}), and a strong sensitivity to sample selection. For example, restricting FP analyses to sub-Eddington, low-accretion-rate AGN (low-luminosity AGN; LLAGN) -- which appear to resemble BH XRBs in the HS (e.g., \citealt{kording_2006_agn_vs_xrb, ho_2005}) -- significantly reduces scatter (e.g., \citealt{kording_2006_fp, plotkin_2012, saikia_2018}) compared to mixed samples (e.g., \citealt{merloni_2003,falcke_2004}). By contrast, AGN accreting at a few per cent of the Eddington rate may resemble SS XRBs \citep{maccarone_2003_agn}, while some AGN accreting at $\sim$10 per cent Eddington display distinct radio:X-ray behaviour that suggests a different mode of disc–jet coupling \citep{king_2011}. Separate FP relations are also observed for radio-loud\footnote{This is not directly analogous to the upper track of the $L_R$–$L_X$ plane for LMXBs. In AGN, radio loudness is typically defined using the ratio of radio to optical flux \citep{kellermann_1989}, although alternative X-ray-based definitions have also been used \citep{terashima_2003}.} and radio-quiet AGN, with an additional dependence on the Eddington-ratio of X-ray luminosity (e.g., \citealt{wang_2006, li_2008, bariuan_2022, wang_2024_FP}). Notably, \citet{dong_2014} showed that bright ($L_{\text{bol}}/L_{\text{Edd}} \gtrsim 1\%$), radio-quiet AGN exhibit steep correlation coefficients similar to outlier LMXBs, consistent with radiatively efficient accretion. These results highlight the complexity of FP studies and their utility as powerful diagnostics of the origin of X-ray emission, distinguishing jet-dominated from accretion-flow-dominated regimes.

\subsection{Quiescence}

Quiescent systems are challenging to study due to their faintness, particularly in the radio band, and it remains unclear whether the $L_R$--$L_X$ correlations evolve in this regime. While quiescence has traditionally been viewed as a low-luminosity extension of the HS, some studies suggest it may instead represent a distinct state, and it remains unclear whether compact jets are always produced in this regime. In particular, once the accretion rate drops below a critical value, the structure of the accretion flow is expected to change, and several studies have reported a softening of the X-ray power-law spectral component (photon index of $\Gamma \approx 2.1$) relative to the canonical HS (e.g., \citealt{corbel_2006, homan_2013, plotkin_2013, plotkin_2015}). During this state, X-ray emission is thought to arise from radiatively inefficient processes in the accretion flow (e.g., ADAFs) or jet. Alternatively, a hybrid jet/ADAF model has been proposed, which notably predicts jet-dominated X-ray emission in quiescence, leading to a steeper $L_R$--$L_X$ correlation \citep{yuan_2005_model, yuan_2005}.

Observationally, detections of BH XRBs suggest that they broadly continue to follow the standard $L_R$--$L_X$ correlation down to the QS (e.g., \citealt{gallo_2003, gallo_2006, corbel_2008, corbel_2013, plotkin_2017a, tremou_2020, shaw_2021}), and hybrid systems appear to rejoin this track at low $L_X$ (e.g., \citealt{carotenuto_2022_quiescence}). However, some studies report tentative evidence for changes to the correlation slope at low $L_X$ (e.g., \citealt{pszota_2008, rodriguez_2020}), though the statistical significance remains uncertain. Results in AGN are similarly mixed: while some studies find evidence for a steepening at low $L_X$ (e.g., \citealt{wu_2007, wrobel_2008, yuan_2009}), others report contrary results and/or highlight potential biases due to sample selection \citep{plotkin_2012, dong_2015}.

Even fewer observations exist for NS systems in quiescence. Early studies found that they are systematically more X-ray luminous than BH systems at comparable accretion rates \citep{garcia_2001, mcClintock_2003}, initially interpreted as evidence for advection across BH event horizons, although this may instead reflect selection effects or the possibility that BH XRBs are jet-dominated (e.g., \citealt{fender_2003, jonker_2006}). The behaviour of quiescent NSs in the $L_R$--$L_X$ plane remains uncertain (e.g., \citealt{van_den_eijnden_2022b, pattie_2026}). Several sources transition from radio detections above $L_X \approx 10^{36}~\text{erg}~\text{s}^{-1}$ to predominantly non-detections at lower luminosities -- possibly indicating a break in the correlation due to quenched or fainter radio jets (\citealt{van_den_eijnden_2022b}) -- although there is no consistent luminosity threshold across sources (e.g., \citealt{tudor_2017, gusinskaia_2020}).

Deeper radio and X-ray observations of both BH and NS systems are therefore required to constrain the properties of faint jets and clarify the jet–accretion coupling in the low-luminosity regime.

\subsection{Coupled Radio, Near-Infrared and X-ray Correlation}

Attempts to physically interpret the jet–accretion coupling of XRBs are greatly aided by studies of optical:X-ray and near-infrared (NIR):X-ray correlations. In particular, NIR emission originates close to the jet base and is therefore more directly connected to the X-ray-emitting flow than radio emission which arises further downstream \citep{russell_2006, russell_2007_tracks}. Using our MeerKAT data, a radio:optical:X-ray plane for MAXI J1820+070 has already been constructed \citep{bright_2025}, and similar analyses will soon be conducted for several other BH XRBs in our sample, including Swift J1727.8–1613.


\section{Conclusions}\label{sec: conclu}

\begin{itemize}

\item The primary aim of this paper was to present the results of five years of coordinated MeerKAT and \emph{Swift}/XRT monitoring of X-ray binaries, conducted as part of the ThunderKAT programme. \emph{These data are publicly released on our project website}: \\ \url{https://thunderkat.physics.ox.ac.uk/}.

\item We presented light curves for all sources in our sample, and discussed the nature of the observed \emph{soft-state core radio emission}. This is likely dominated by unresolved jet ejecta components, with occasional reflaring arising from interactions of ejecta with the interstellar medium, internal shocks, or the launch of additional ejecta during brief transitions to the intermediate state. Continued radio monitoring in the soft state is essential to determine how common such reflares are and to better understand their origin. 

\item Using the quasi-simultaneous radio and X-ray data from ThunderKAT, we constructed the \emph{largest homogeneous radio:X-ray plane for X-ray binaries to date}, and performed linear regressions to characterise the observed correlations. Consistent with previous studies, we found evidence for sources having a range of behaviour, indicating a complex and potentially evolving coupling between core accretion and jet production in different systems. We also discussed the possibility that all sources may transition between different radio:X-ray tracks as a function of X-ray luminosity.
\end{itemize}

\noindent\emph{Ongoing and future monitoring} of outbursts for the sources in our sample will enable tighter constraints on their behaviour in the radio:X-ray plane. Particularly important is the expansion of the currently sparse NS sample. This goal will be facilitated by the X-KAT programme, which will continue to monitor bright, active X-ray binaries in the southern hemisphere using MeerKAT and \emph{Swift}/XRT. Together with the existing ThunderKAT data set, these observations will form one of the most comprehensive databases of X-ray binaries to date, enabling more detailed studies of the radio:X-ray correlation and thereby providing crucial insights into the disc–jet coupling in these systems.  \\

\noindent\emph{Future Work:}

\noindent In forthcoming papers, we aim to further exploit the ThunderKAT data set by constructing an updated radio:X-ray plane that incorporates additional quasi-simultaneous X-ray observations -- while carefully treating cross-instrument systematics -- and new X-KAT data. The additional X-ray data will enable a more robust quantification of the uncertainties introduced by non-simultaneity when pairing radio and X-ray observations for placement on the plane. Using a custom regression scheme, we will extract and systematically compare the slopes and intercepts of the radio:X-ray correlations for our best-sampled sources. Alongside this enhanced data set, we will develop an updated prescription linking radio luminosity to the kinetic power of hard-state jets. We will further exploit the full polarimetric capability of MeerKAT by measuring total linear polarisations across all ThunderKAT and X-KAT observations, enabling new constraints on the magneto-ionic properties of jets. By combining ThunderKAT and X-KAT measurements with analogous samples of active galactic nuclei, we will construct an updated Fundamental Plane of Black Hole Activity spanning stellar-mass and supermassive accretors, thereby helping to advance our understanding of accretion and jet physics across mass scales.

\section*{Acknowledgements}

We thank Elena Gallo for the comments. We acknowledge Fabio Pintore for providing scripts used to perform \emph{Swift}/XRT data reductions for a few of the sources in our sample. 

JCM acknowledges support from the Rhodes Trust at the University of Oxford. RF acknowledges support from STFC, The ERC and the Hintze Foundation charitable trust. MDS acknowledges ASI-INAF program I/004/11/6 (Swift) and INAF Large Grant 2023 BLOSSOM O.F. 1.05.23.01.13. LR acknowledges support from the Trottier Space Institute Fellowship and from the Canada Excellence Research Chair in Transient Astrophysics (CERC-2022-
00009).

The MeerKAT telescope is operated by the South African Radio Astronomy Observatory, which is a facility of the National Research Foundation, an agency of the Department of Science and Innovation. We acknowledge the use of the Inter-University Institute for Data Intensive Astronomy (IDIA) data intensive research cloud for data processing. IDIA is a South African university partnership involving the University of Cape Town, the University of Pretoria and the University of the Western Cape.

\section*{Data Availability}

The data used in this paper is available to download in machine-readable format on our website: \url{https://thunderkat.physics.ox.ac.uk/}. Data from MeerKAT are available through the SARAO data archive (Proposal IDs: SCI-20180516-PW-01 and SCI-20230907-RF-01): \url{https://archive.sarao.ac.za/}. Data from \emph{Swift}/XRT are publicly available through the \emph{Swift} archive: \url{https://www.swift.ac.uk/swift_portal}.

The code used to conduct the analysis and generate the plots in this paper are available at: \url{https://github.com/JustineCrook/Radio_X-ray_Plane_Paper_2026}; any changes to the data and/or results will be listed here.



\bibliographystyle{mnras}
\bibliography{references} 


\appendix
\renewcommand{\sectionautorefname}{Appendix}

\section{Overview of Data Reduction} \label{app: data_reduction}

In this section, we highlight key aspects of the MeerKAT and \emph{Swift}/XRT data reduction processes, with the aim of incorporating lessons learnt into future analyses, and working toward a standardised data reduction approach across our collaboration.

\subsection{MeerKAT Data Reduction}

For the radio data that had not been reduced, we made use of the semi-automated routine \texttt{oxkat}\footnote{\url{https://github.com/IanHeywood/oxkat}} \citep{heywood_oxkat_2020}  to perform flagging, calibration, and imaging. When using this pipeline, since we are primarily interested in continuum emission, the recorded visibilities (taken in the 32K correlator mode) are frequency-averaged by a factor of 32 prior to calibration, resulting in 1024 channels. Initial flagging and reference calibration are performed using \texttt{CASA} \citep{casa_2022}. The target data are further flagged with \texttt{tricolor} \citep{hugo_tricolour_2022}, and an initial image of the source field is created with \texttt{WSCLEAN} \citep{offringa_wsclean_2014}. A deconvolution mask is then produced from this unmasked image, and applied for an initial masked deconvolution. The resulting model image is then used for direction-independent self-calibration with \texttt{CUBICAL} \citep{kenyon_cubical_2018}. Thereafter, a second round of masked deconvolution is conducted using the self-calibrated visibilities, producing a set of channelised images evenly spaced in frequency, along with a frequency-averaged (i.e., a multi-frequency synthesis, MFS) image that is representative of the sky at the central frequency of 1.284 GHz. The \texttt{oxkat} pipeline also has the functionality to solve for direction-dependent self-calibration solutions, although this was not required for any of the new data we reduced. For some sources, \texttt{polkat}\footnote{\url{https://github.com/AKHughes1994/polkat}} \citep{hughes_2025_polkat} was used instead, which is a modified version of \texttt{oxkat}, extended to enable full polarisation calibration and imaging of Stokes I, Q, U, and V. 

In each MFS image for every source, the root-mean-square noise (RMS; $\equiv \sigma_\text{rms}$) was measured in a source-free region near the target, using the \texttt{CASA} task \texttt{imstat}. The source was considered to be detected when the peak flux density at the source position was $>$3$\sigma_\text{rms}$; else, we report 3$\sigma_\text{rms}$ as an upper limit. All the new targets considered were found to be unresolved at every epoch, so flux densities were obtained with \texttt{imfit} by fitting an elliptical Gaussian whose shape was fixed to match the synthesized beam (i.e., a point-source model), using a small region centred on the source. The quoted 1$\sigma$ uncertainties are those obtained from the fits; however, for at least one source in our sample, $\sigma_\text{rms}$ was used as the uncertainty instead, which should be comparable and potentially slightly more conservative. 

The quoted time ($t$) for each radio data point is the midpoint of the scan in which the target was observed (typically $\sim$15 minutes long). For targets observed in multiple scans, $t$ is the midpoint between the start of the first scan and the end of the last scan.

\subsection{\emph{Swift}/XRT Data Reduction}

Each \emph{Swift}/XRT observation is taken in either photon counting (PC) mode, which provides full imaging and spectral information with 2.5 s resolution, or windowed timing (WT) mode, which sacrifices one imaging dimension to achieve a higher time resolution of 1.7 ms for bright sources. 

For data sets that had not been reduced, spectra were extracted using the \texttt{SWIFTTOOLS} \emph{product generator}\footnote{\url{https://www.swift.ac.uk/user_objects/}} \citep{evans_2007, evans_2009}, which applies the latest software and calibration files, and is the recommended tool for this purpose (UK \emph{Swift} team, private communication). Notably, it accounts for \emph{pile-up}, which occurs when multiple photons hit the same detector pixel in rapid succession and are recorded as a single higher-energy event, distorting the spectrum, underestimating the true flux, and potentially introducing artifacts in images of bright sources. For \emph{grade filtering} when extracting spectra, we used either the `All valid' (grades 0--12 in PC mode, and 0--2 in WT mode) or `Restricted' (0--4 in PC mode, and 0 in WT mode) options (see the grade definitions on page 10 of the reduction guide\footnote{\url{https://www.swift.ac.uk/analysis/xrt/files/xrt_swguide_v1_2.pdf }}). `Zero only' filtering is no longer recommended (for mitigating pile-up) because radiation damage and the build-up of charge traps in the charge-coupled device (CCD) can lead to `grade migration' -- whereby some of the trapped charge `trails' behind as the CCD is read out, meaning that grade 0 PC events appear as grade 1 PC -- which is an effect that is not currently accounted for in the instrument response calibration files (UK \emph{Swift} team, private communication). In addition, to minimise calibration uncertainties, we used the same grade selection for each source. 

Spectral fitting was performed using \texttt{XSPEC} \citep{arnaud_1996}, ignoring channels that had been flagged as `bad'. We fit over the \emph{energy range} 0.5 / 0.6 --10 keV, varying the lower-energy bound in particular, as this choice is important for heavily absorbed sources in WT mode which may exhibit artificial low-energy features due to redistribution and charge-trapping effects\footnote{\url{https://www.swift.ac.uk/analysis/xrt/digest_cal.php}}. For spectral \emph{binning}, many of our collaboration members grouped spectra with sufficient counts (generally $\gtrsim$300) into 20-count bins and fit using $\chi^2$ statistics, while lower-count spectra were fitted using single-count binning and Cash statistics (\texttt{cstat}). However, since studies suggest that $\chi^2$ fitting may be biased even at high counts (e.g., \citealt{humphrey_2009}), for the newly reduced data sets, we also fitted all spectra with single-count binning and \texttt{cstat}. Given the large dynamical range of the data, the difference between the two approaches is comparatively negligible. For bright WT epochs, when fitting with $\chi^2$ statistics, a 3 per cent systematic uncertainty was generally added to the response (note that this has no effect when fitting using \texttt{cstat})\footnotemark \footnotetext{\url{https://www.swift.ac.uk/analysis/xrt/files/SWIFT-XRT-CALDB-09_v16.pdf}}.

In all cases, interstellar absorption was modelled with \texttt{tbabs}, parametrised by $N_H$ (the equivalent hydrogen column density along the line of sight), with \texttt{wilm} abundances \citep{wilms_2000} and \texttt{vern} cross-sections \citep{verner_1996}. Since our goal was to estimate unabsorbed fluxes rather than detailed spectral parameters, simple \emph{spectral models} were sufficient. We employed power-law
models (\texttt{tbabs*pegpwrlw} or \texttt{tbabs*powerlaw}) to model the Comptonised coronal emission, the disc blackbody model (\texttt{tbabs*diskbb}) to model a multi-colour accretion disc, or a combination of both models (\texttt{tbabs*(powerlaw+diskbb)}), as appropriate. One collaboration member who conducted a data reduction opted to use the thermally Comptonised continuum model \texttt{nthcomp} instead of \texttt{powerlaw}. However, we verified that the differences in the flux results between these two models are negligible; they would likely only be significant if we were estimating bolometric fluxes (from $\sim$0.01 keV), since a de-absorbed \texttt{powerlaw} can diverge at low energies and overestimate fluxes. Additionally, for some neutron star (NS) systems, a blackbody component (\texttt{bbodyrad} or \texttt{bbody}) was added to model thermal emission from the surface or boundary layer. 

The unabsorbed 1--10 keV flux was extracted directly for the \texttt{tbabs*pegpwrlw} model, whereas for other models, it was estimated by applying the \texttt{cflux} convolution after initial fitting (e.g., $\texttt{tbabs*cflux*(powerlaw+diskbb)}$). Furthermore, for spectra with very few counts (usually $\lesssim$50), where spectral parameters cannot be well constrained, we generally used only single-component models and fixed all parameters (except the flux) to appropriate values. Uncertainties were estimated using the \texttt{error} command, and are reported at the 1$\sigma$ level for all newly reduced data.

Within our sample, different approaches were used to handle $N_H$. In general, it is often assumed to be constant for a given source and therefore fixed during fitting, either to a weighted average from unconstrained \emph{Swift}/XRT fits, or to a value measured with another instrument. The former approach may be unreliable due to parameter degeneracies, model simplicity, and weak constraints on absorption when fitting \emph{Swift}/XRT spectra above $\sim$1 keV. Such estimates should therefore be compared with literature values -- noting that reliable constraints typically require high-resolution soft X-ray spectroscopic instruments, such as the Reflection Grating Spectrometer (RGS) aboard XMM-Newton \citep{rgs_newton_2001} -- though cross-instrument calibration differences may introduce discrepancies. Apparent epoch-to-epoch variations in $N_H$ may reflect instrumental systematics or modelling simplifications, but could also indicate genuine changes in the intrinsic absorption. Therefore, for some sources, we allowed $N_H$ to vary between epochs -- as this yielded statistically better fits -- while verifying that it remains consistent within an acceptable error tolerance.

Another effect to consider is X-ray scattering by interstellar dust in our Galaxy. For detectors with sufficient imaging resolution, such as \emph{Swift}, this produces \emph{dust-scattering halos} around sources with $N_H \gtrsim 10^{22} ~\text{cm}^{-2}$ \citep{corrales_2016}. Even when too faint to detect directly, these halos can redistribute source photons and bias flux estimates if not modelled (see, e.g., \citealt{beardmore_2016, heinz_2016} for modelling approaches in V404 Cygni). Although we do not model these effects, the resulting uncertainties are negligible compared to the large dynamical ranges of our data sets, and are accounted for by our conservative 10 per cent systematic uncertainty on each flux measurement.  

For each newly reduced data point, the quoted time ($t$) is the midpoint between the \texttt{TSTART} and \texttt{TSTOP} values in the spectral header -- which are respectively the start and end times of the extracted data, and include spacecraft clock-drift corrections -- converted to Modified Julian Date (MJD) using \texttt{swifttime}.

\section{Overview of source properties}\label{app: source_properties}

In Appendix \ref{app: sources}, for each source in our sample, we provide a brief overview of its accretor type, distance, and the jet inclination angle estimates found in the literature. However, many of these properties remain poorly constrained, and when discrepancies exist, we adopt values based on our judgement. In this section, we briefly describe how these properties are determined, highlighting the most reliable methods. 

Each system in our sample is assigned an \emph{accretor type} -- black hole (BH), BH candidate (BHC), or NS -- referring to the nature of the compact object. A NS can be definitively identified by the detection of Type I X-ray bursts, which indicate thermonuclear burning of accreted material on a solid surface \citep[see][]{galloway_2021}. A compact object is typically identified as a BH if its dynamical mass estimates (from radial velocity curves) exceed the theoretical NS limit ($\sim$3$M_\odot$; \citealt{kalogera_1996, ozel_freire_2016}). In the absence of such constraints, a source is classified as a BHC based on its spectral and timing behaviour, such as the absence of surface phenomena (e.g., pulsations or bursts), strong variability, and state transitions similar to those observed in confirmed BH systems. We note that it is extremely rare for a BHC to later be reclassified as a NS. 

The most reliable estimates of X-ray binary (XRB) \emph{distances} are via parallax measurements from the Gaia Data Release 3 (DR3; \citealt{gaia_2021}) or very long baseline interferometry (VLBI; e.g., \citealt{atri_2020}). For Gaia parallaxes, distances are generally inferred within a Bayesian framework that incorporates appropriate priors (e.g., for Galactic XRBs from \citealt{atri_2019}), after applying the relevant zero-point corrections. Estimates via kinematic methods -- using measured proper motions -- are also generally reliable (e.g., \citealt{reid_2023}). Other approaches include using observations of the X-ray source or donor star (e.g., assumed state transition luminosities; \citealt{vahdat_2019, abdulghani_2024}), jet proper motions (e.g., \citealt{mirabel_1994}), X-ray dust scattering (e.g., \citealt{lamer_2021}), line-of-sight (LOS) HI absorption (e.g., \citealt{chauhan_2021}), and estimates of the hydrogen column density (e.g., \citealt{balakrishnan_2023}). In some NS systems, Type I bursts exhibit photospheric radius expansion (PRE), allowing distance estimates via their peak flux (e.g., \citealt{galloway_2020}), although systematic uncertainties arising from the assumed system parameters must be considered. 

We emphasize that there is heterogeneity in the literature when reporting distance estimates, as some authors include additional systematic uncertainties while others do not. In cases where systematic uncertainties are not provided, we adopt the quoted statistical uncertainties without introducing any ad hoc additions. When the inferred source distance is modelled as an asymmetric Gaussian, we use the larger of the upper and lower uncertainties in our Monte Carlo simulations when varying the distance. Any bias resulting from this approach is negligible for our purposes, especially since the upper and lower uncertainties are similar or equal for all the assumed distance distributions. When a source distance is poorly constrained, we adopt a uniform probability distribution over the range permitted by constraints in the literature. We note, however, that an alternative approach would be to use the distance distributions derived from the Galactic mass density model of low-mass XRBs (LMXBs; \citealt{grimm_2002}), as implemented in \cite{atri_2019}. 

Jet \emph{inclination angles} (relative to the LOS) are estimated using three main approaches. The most direct (preferred) method uses spatially resolved radio and/or X-ray ejecta (typically assumed to be symmetric), and then adopts the same inclination for the compact jets even though this may differ if the jets precess (e.g., \citealt{margon_1984, miller-jones_2019}). When direct measurements are unavailable, we adopt the binary orbital or inner-disc inclinations as a proxy, assuming that the jet is perpendicular to these planes, although it may differ from this due to factors such as disc tearing or the Bardeen-Petterson effect \citep{bardeen_1975, nealon_2015, liska_2021, maccarone_2002}. Orbital inclinations are typically inferred from optical modelling of ellipsoidal variations of the tidally distorted companion star in quiescence, while disc inclinations are derived from X-ray data. In particular, the presence of X-ray dips (caused by partial obscuration at the interaction point between the accretion stream and disc) without eclipses from the companion star implies inclinations of $\sim$60--75$^\circ$ (e.g., \citealt{frank_1987}). Furthermore, the presence of equatorial disc winds in the soft state (SS; seen as highly ionised absorption X-ray spectral features) also suggests a high inclination \citep{ponti_2012}. Inclinations can also be estimated from spectral modelling of the inner-disc X-ray reflection component, though this remains the least reliable method due to the strong dependence on model assumptions, despite efforts to standardise approaches (e.g., \citealt{draghis_2024}). 

\section{Details Regarding Sources} \label{app: sources}

In the following, for each source in our sample, we describe the data reduction process, referring to published work where available, and summarise the evolution of its MeerKAT light curve, with a particular focus on any launched ejecta.

\subsection*{1A 1744–361}

\href{https://thunderkat.physics.ox.ac.uk/source/1A%201744-361}{1A 1744–361} was first discovered by the Ariel 5 satellite in 1976 \citep{davison_1976, carpenter_1977}, and Type I X-ray bursts have confirmed a NS accretor (e.g., \citealt{bhattacharyya_2006}). Although the source distance remains poorly constrained, the lack of PRE during bursts implies an upper limit of $\sim$9 kpc \citep{bhattacharyya_2006}. We adopt 8 kpc as a nominal distance, with an allowed range 1--9 kpc. Its observed X-ray dips and absence of eclipses suggest an orbital inclination angle of $\sim$$60^\circ$–$75^\circ$ \citep{ng_2024}. 

In 2022, 1A 1744–361 entered its longest recorded outburst to date \citep{kobayashi_2022}. Our MeerKAT campaign began with a rapid-response observation on 2022 May 31 (MJD 59730; \citealt{hughes_atel_2022}), $\sim$2 days after its detection, followed by regular $\sim$weekly monitoring from 2022 June 3 to August 27 (MJDs 59733--59818), for a total of 14 epochs. Details of the MeerKAT data reduction, as well as the X-ray spectral and timing analyses used for state classification, are given in \citet{ng_2024}. During the outburst, the source was predominantly in an atoll-like state, though it exhibited Z-source-like behaviour near the peak. After an initial MeerKAT observation in the intermediate state (IMS), 1A 1744–361 transitioned to the SS and exhibited a steepening in-band radio spectral index, suggesting that the subsequent SS detections are due to unresolved transient ejecta launched during this transition. The final three MeerKAT observations, obtained after a return to the hard state (HS), yielded non-detections. 

In 2024, during X-KAT, 1A 1744–361 was reported to have entered a new outburst \citep{zu_2024ATel}. We obtained eight $\sim$weekly MeerKAT observations of the source during this period, between 2024 July 19 and September 1 (MJDs 60510--60554). The MeerKAT data were reduced in the same way as described in \cite{ng_2024}. Additional data from the Monitor of All-sky X-ray Image (MAXI) indicate that the duration of this outburst was considerably shorter than the 2022 one. Using data from the Neutron Star Interior Composition Explorer Mission (NICER) and the Hard X-ray Modulation Telescope (Insight-HXMT) for this outburst, \cite{wang_2025_1A} suggested that the source was in the HS until MJD 60516, transitioning to the SS thereafter, and remaining there until the end of their campaign on MJD 60522 -- although they note that their classification is uncertain. Therefore, it is likely our first two MeerKAT observations (MJDs 60510 and 60512) occurred during the rising HS, with the latter resulting in a detection. Thereafter, from MJDs 60516--60522, the source was likely in the SS, consistent with the observed drop in the radio emission (specifically, the upper limit on MJD 60518) as the compact jet quenched. The \emph{Swift}/XRT observation on MJD 60525 marginally favours a SS classification, although this is uncertain. Shortly afterwards, the source likely returned to the HS during the decaying phase, as suggested by the low-SNR radio detection on MJD 60527, which may indicate a reactivation of the compact jet. However, it is also possible that this was a SS reflare. \emph{Swift}/XRT observations from MJD 60532 onward suggest that the source had returned to the HS. During our final five MeerKAT observations in the decaying HS, the source emission had dropped below the detection threshold.

For the \emph{Swift}/XRT data reduction, spectra with ${\geq}\,500$ counts were binned into 25-count intervals (${\geq}\,20$ bins) and fit using $\chi^2$ statistics. Motivated by the models of \citet{lin_2007}, we adopted a state-dependent approach: $\texttt{tbabs*(pegpwrlw+bbody)}$ for HS epochs and $\texttt{tbabs*(diskbb+bbody)}$ for SS epochs. For the IMS epochs on MJDs 59731 and 60525, both models yielded statistically indistinguishable fit statistics and fluxes (${\ll}$$1\sigma$), so we adopted the SS model due to its slightly lower $\chi^2$. The low-count spectra were instead fit with single-count bins and Cash statistics, using a simplified $\texttt{tbabs*pegpwrlw}$ model. In all cases, spectra were fit over the range 0.5--10\,keV, $N_{\rm H}$ was fixed at $4.4\,{\times}\,10^{21}\,{\rm cm^{-2}}$ \citep{ng_2024}, and unabsorbed 1--10\,keV fluxes were extracted using \texttt{cflux} after convergence.

\subsection*{4U 1543–47}

\href{https://thunderkat.physics.ox.ac.uk/source/4U%201543-47}{4U 1543–47} was discovered during an X-ray outburst by the Uhuru telescope in 1971 \citep{matilsky_1972}. Dynamical studies of the companion star indicate a compact object mass exceeding $\sim$3$M_\odot$, consistent with a BH accretor \citep{orosz_1998}. While optical studies have estimated a source distance of $\sim$7.5 kpc \citep{orosz_2002}, we adopt the independent Gaia DR3 parallax-based estimate of 5.0$^{+2.0}_{-1.2}$ kpc \citep{zhang_2025b}. The system's orbital inclination angle is reported to be 20.7$^{\circ} \pm$1.5$^{\circ}$ \citep{orosz_1998, orosz_2003}. 

In 2021, following $\sim$19 years of quiescence, 4U 1543–47 went into outburst \citep{negoro_2021}, and was subsequently monitored with MeerKAT from 2021 June 19 to 2023 March 9 (MJDs 59384--60012), for a total of 72 epochs. Details of the MeerKAT and \emph{Swift}/XRT data reductions are presented in \citet{zhang_2025b, zhang_2025a}. At the start of our MeerKAT monitoring, the source was in the SS, where it remained until late 2021, aside from a possible brief excursion to the IMS during a flaring episode around MJD $\sim$59453 \citep{zhang_2025a}. In the following weeks, two transient ejecta components were resolved, having inclinations $<$30$^\circ$ \citep{zhang_2025b}. The first was estimated to have been launched near the HS$\rightarrow$SS transition prior to our monitoring (MJD $\sim$59376). The second was likely launched during a possible short IMS excursion around MJD $\sim$59505, and although no flaring was detected at this time, it may have been missed due to the monitoring cadence. Thereafter, the source briefly entered the HS before fading to quiescence. Later, the source exhibited a series of HS-only reflares during the remainder of our campaign.

\subsection*{4U 1630–47}

\href{https://thunderkat.physics.ox.ac.uk/source/4U%201630-47}{4U 1630–47} is an extremely active LMXB that was initially discovered by Uhuru in 1969 \citep{giacconi_1972, priedhorsky_1986}, and has since undergone recurrent outbursts every $\sim$2--3 years. Despite lacking a dynamical mass confirmation, its accretor is classified as a BHC based on spectral and timing properties (e.g., \citealt{parmar_1986, klein-wolt_2004}). It is located near the Galactic plane and its high LOS extinction (e.g., \citealt{gatuzz_2019}) make the determination of its binary properties challenging. In particular, no optical counterpart has been identified, likely due to its high reddening and crowded source field. Infrared observations place the source in the direction of a giant molecular cloud, and modelling of its dust-scattering halo yields a most likely source distance of 11.5 kpc, with a systematic plus statistical error of 1 kpc (dominated by uncertainties in determining the cloud distance; \citealt{kalemci_2018, kalemci_2025}). A high source inclination (60--75$^\circ$) is inferred due to the presence of X-ray flux dips \citep{tomsick_1998, kuulkers_1998}, the lack of full eclipses, the detection of highly ionised absorption spectral lines (e.g., \citealt{kubota_2007}), and the presence of powerful equatorial X-ray disc winds in the SS (e.g., \citealt{king_2014, miller_2015_disc_winds}).

4U 1630–47 was observed with MeerKAT during three discrete periods between 2020 March 21 and 2023 June 17 (MJDs 58929--60112), yielding 39 epochs. Details of the MeerKAT data reduction and spectral state classifications are given in \citet{zhang_2026}. Our 2020 MeerKAT observations detected the source during the SS, with the emission likely arising from residual ejecta. During the 2021 outburst, the source was observed once in the HS, and then it returned to the SS, where optically thin radio in-band spectral indices again suggested emission from unresolved transient ejecta. The SS of 2022--2023 showed much weaker radio emission, again likely arising from ejecta interactions. 

\emph{Swift}/XRT spectra that were quasi-simultaneous with the MeerKAT observations were extracted, although we excluded a few epochs where the source was too faint or near the edge of the CCD. To derive the unabsorbed 1--10 keV fluxes, all spectra were fit with \texttt{tbabs*cflux*diskbb}. The source was in the SS for all epochs, except for two in the HS which were nevertheless well fit by this simple model. $N_H$ was allowed to vary, yielding values consistent with those reported in the literature (e.g., \citealt{Capitanio15, Kuulkers98}). Our 1--10 keV fluxes were found to agree within uncertainties (including systematics) with the results obtained from a more detailed X-ray analysis using data from NICER \citep{zhang_2026}. A more detailed analysis of the ThunderKAT and X-KAT $L_R$--$L_X$ behaviour of 4U 1630–47 will be presented in Mariani et al. (in preparation). 

\subsection*{Aquila X-1}

\href{https://thunderkat.physics.ox.ac.uk/source/Aquila%20X-1}{Aquila X-1} (Aql X-1) / V1333 Aql was one of the first NS LMXBs to be discovered \citep{kunte_1973}, and has been studied in great detail due to its frequent $\sim$yearly outbursts. Its nature as a NS was confirmed through the detection of Type-I X-ray bursts (e.g., \citealt{galloway_2008, mandal_2025}). The source is identified as an intermittent accreting millisecond X-ray pulsar (AMXP) due to the observation of X-ray pulsations \citep{casella_2008}. Its distance is estimated to be $6\pm 2$ kpc from near-infrared photometry, using spectroscopic constraints on the donor star and veiling, and assuming an evolved Roche-lobe-filling companion \citep{mata_sanchez_2017}. This is in agreement with the parallax distance using the Gaia DR3, although this is not well constrained due to the large uncertainties on the parallax. \cite{mata_sanchez_2017} also estimated the inclination to be $36^\circ<i<47^\circ$. 

As part of ThunderKAT, we obtained two MeerKAT observations of Aql X-1 during its August 2020 outburst (MJDs 59083 and 59091), shortly after renewed activity was reported for the source \citep{saikia_2020_atel}. Additionally, after a new outburst was reported in September 2024 \citep{liu_2024_atel}, we monitored the source on a weekly basis as part of X-KAT, from 21 September to 30 November (MJDs 60574--60644), obtaining 11 epochs. Each observing block consisted of 15 minutes of on-target time, bracketed by $\sim$2-minute scans of the phase calibrator J2011--0644, along with a 5--10 minute observation of the flux/bandpass calibrator J1939--6342. The data were reduced using \texttt{oxkat} with the default parameters. For all the detections, the source at the core position was observed to be unresolved, so it was fit with a point source model.

For the \emph{Swift}/XRT data, spectra were extracted using the online product generator. The data were fit using \texttt{cstat} over the range 0.6--10 keV, using a \texttt{powerlaw} or \texttt{diskbb} model, or a combination of one of these models with \texttt{bbodyrad}, as appropriate. $N_H$ was fixed to 0.5$\times 10^{22}$ cm$^{-2}$, following previous works (e.g., \citealt{tudose_2009a, marino_2025}).

The spectral state designations for each observation were determined using a combination of \emph{Swift}/XRT spectral fits, the MAXI hardness ratio (HR), as well as the results presented in \cite{marino_2025}. The two 2020 MeerKAT epochs were likely obtained during a rising HS. Our first MeerKAT observation in 2024 was also likely taken during the rising HS, after which the source transitioned to the IMS on MJDs 60581.5--60582 \citep{marino_2025}. This was followed by several radio non-detections during the SS, consistent with a quenched compact jet. On MJD 60609, the source was detected again with MeerKAT (at $\sim$0.13 mJy), which may suggest a re-activation of the compact jet and return to the HS. VLA observations taken on MJDs 60611, 60614, 60616, and 60619 show a flat radio spectrum, further supporting this interpretation (S. Fijma, private communication). However, \emph{Swift}/XRT and NICER observations between MJDs 60613 and 60617 suggest that the source may still be in the HIMS, although this classification is not definitive, and it is possible that the source may have rapidly switched between the HS and IMS during this period (A. Marino, private communication). We tentatively label the MeerKAT observations on MJDs 60609 and 60615 as being in the HS, although this classification is uncertain. By MJD $\sim$60619, the source was definitively in the HS, where it remained for the rest of the outburst as it faded toward the quiescent state (QS).  

\subsection*{Centaurus X-4}


\href{https://thunderkat.physics.ox.ac.uk/source/Cen%20X-4}{Centaurus X-4} (Cen X-4) was discovered in outburst by the Vela 5B satellite in 1969 \citep{conner_1969}, with a second outburst observed in 1979 \citep{kaluzienski_1980}. The detection of a Type I X-ray burst confirmed the accretor to be a NS \citep{matsuoka_1980}. We adopt the recent Gaia parallax-based distance estimate of $1.87^{+0.75}_{-0.42}$ kpc \citep{van_den_eijnden_2022b}. The system's orbital inclination was estimated to be $32^\circ$$^{+8^\circ}_{-2^\circ}$ using XRB model fits to the secondary star’s Roche lobe-distorted absorption lines \citep{shahbaz_2014}, consistent with independent estimates from ellipsoidal light curve modelling \citep{hammerstein_2018}.  

Details regarding the \emph{Swift}/XRT and four MeerKAT observations of Cen X-4 are given in \citet{van_den_eijnden_2022b}. The first MeerKAT observation (2020 September 26; MJD 59118) was taken during the QS with $\sim$4 hours on target, while the remaining three (each with $\sim$15 minutes on target) were taken during a `failed' outburst in January 2021 (MJDs 59221--59230). No significant radio emission from Cen X-4 was detected in these epochs, although the resulting upper limits provide valuable constraints in a poorly explored regime, as NS detections below $L_X = 10^{34}$ erg s$^{-1}$ are rare.

\subsection*{Circinus X-1}

\href{https://thunderkat.physics.ox.ac.uk/source/Cir%20X-1}{Circinus X-1} (Cir X-1), discovered in 1971 \citep{margon_1971}, is the youngest-known LMXB, embedded within its natal supernova remnant. The source exhibits Type I bursts \citep{linares_2010} -- confirming a NS accretor -- and appears to show characteristics of both Z- and atoll sources. A kinematic source distance of $9.4^{+0.8}_{-1.0}$ kpc (1$\sigma$ uncertainties, including the effects of random cloud-to-cloud and streaming motions) has been estimated using its X-ray dust scattering light echoes \citep{heinz_2015}. Its jet's orientation and morphology are complex, and it is suggested to precess on year-long timescales \citep{coriat_2019, cowie_2025}. 

During ThunderKAT, Cir X-1 was observed with MeerKAT for one epoch on 2018 October 27 (MJD 58418) and for 36 additional epochs with $\sim$daily monitoring between 2021 August 3 and September 5 (MJDs 59429--59462). As described in detail in \cite{gasealahwe_2025}, the data were reduced and imaged using the \texttt{oxkat} pipeline with modified imaging scripts, and 3GC peeling was performed. Flux densities were subsequently extracted using \texttt{CASA imfit}, using a point source model. In \cite{gasealahwe_2025}, these observations were additionally stacked, producing the deepest radio image of the field to date. Given the system’s complexity and poorly understood flaring behaviour, we exclude it from our $L_R$--$L_X$ analysis in this paper.

\subsection*{EXO 1846–031}

\href{https://thunderkat.physics.ox.ac.uk/source/EXO%201846-031}{EXO 1846–031} was discovered as an `ultra-soft X-ray transient' by the European X-ray Observatory Satellite (EXOSAT) in 1985 \citep{parmar_1985}. It was subsequently classified as a BHC due to its X-ray properties \citep{parmar_1993, miller_jones_2019_exo}. Given the lack of an optical counterpart, its distance is poorly constrained. An initial estimate of $\sim$7 kpc was derived using the EXOSAT peak flux \citep{parmar_1993}, while a later constraint of 2.4--7.5 kpc was obtained by assuming that BH XRBs reach 0.3--3 per cent of the Eddington luminosity during the soft-to-hard state transition \citep{kalemci_2013, williams_2022}. We adopt the preferred distance of $\sim$4.5 kpc, based on the source’s position on the $L_R$--$L_X$ plane \citep{williams_2022}. Using X-ray data, the source's inclination is inferred to be relatively high \citep{draghis_2020,draghis_2024,nath_2024}.

After more than 30 years since its discovery, EXO 1846-031 entered a new outburst phase in July 2019 \citep{negoro_2019}. During ThunderKAT, the source was monitored with MeerKAT at approximately weekly cadence from 2019 August 4 to 2020 January 3 (MJDs 58699--58851), with two additional observations on 2020 February 21 and April 10 (MJDs 58900 and 58949), for a total of 25 epochs. \citet{williams_2022} provide details regarding the MeerKAT and \emph{Swift}/XRT data, along with quasi-simultaneous observations using the Arcminute Microkelvin Imager Large-Array (AMI-LA) and Karl G. Jansky Very Large Array (VLA). Spectral states were determined from the X-ray spectral fits, the MAXI hardness-intensity diagram (HID), and the results of \citet{liu_2021}. 

The MeerKAT data reveal two radio flares: one during the IMS (HS$\rightarrow$SS transition) around MJD $\sim$58705 -- just prior to a VLA observation (on MJD 58709) that revealed an ejection component -- and a broader flare during the SS between MJDs 58731–58739. A later VLA observation on MJD 58776 resolved an ejection component, which may be a rebrightening of the initial ejection or a second ejection event.

\subsection*{GRS 1739–278}

\href{https://thunderkat.physics.ox.ac.uk/source/GRS%201739-278}{GRS 1739–278} was discovered during an outburst in 1996 by the SIGMA telescope onboard the GRANAT space observatory \citep{paul_1991}. It has been classified as a BHC due to its X-ray spectral properties and the presence of a very strong 5 Hz quasi-periodic oscillation (QPO) in the soft-intermediate state \citep{borozdin_1998, borozdin_2000}. Based on early measurements of $N_H$, the source distance was initially estimated to be 6–8.5 kpc, and it was suggested to lie in the Galactic bulge \citep{greiner_1996}. Although later studies noted this $N_H$ exceeds recent LOS estimates (\citealt{mereminskiy_2019}), we adopt a nominal distance of 8 kpc, consistent with its projected position in the bulge. The inclination has so far only been constrained through reflection spectroscopy, although these results show significant discrepancies \citep{miller_2015, mereminskiy_2019, draghis_2024}. 

In June 2023, the source exhibited renewed X-ray activity during a SS, though the outburst may have begun earlier \citep{negoro_2023, kennea_2023_grs_1739, li_2023a}. ThunderKAT radio monitoring commenced on 2023 June 30 (MJD 60125; \citealt{hughes_2023}) and continued weekly until 18 August (MJD 60174), followed by two additional epochs on 4 and 8 September (MJDs 60191 and 60195), for a total of 10 epochs. During each observation, the target was observed for 15 minutes, and PKS J1939–6342 and J1833–2103 were used as the flux/bandpass and complex gain calibrators, respectively. The data were reduced using \texttt{polkat}, with all the default parameters, except that some of the (aggressive) flagging was removed. When detected, a point source was fit at the position of GRS 1739$-$278 using \texttt{CASA} \texttt{imfit}, while for non-detections, 3$\sigma_\text{rms}$ upper limits were obtained from the RMS measured in a nearby off-source region. Spectral states were inferred using the \emph{Swift}/XRT spectral fits and MAXI HID, as no detailed X-ray analysis of the 2023 outburst has been published. The source was detected in the first two MeerKAT epochs, followed by three non-detections, then a $\sim$300 $\upmu$Jy detection on MJD 60160 when it was likely in the HS, and finally a return to quiescence. 

The \emph{Swift}/XRT spectra were well-fit with a \texttt{tbabs*(diskbb+powerlaw)} model, which adequately reproduced the observed continuum in all cases. $N_{\rm H}$ was left free, except for the final two low-count observations where it was fixed to the best-fit value from earlier data -- specifically, $N_{\rm H} \simeq 2.3 \times 10^{22}$~cm$^{-2}$, consistent with literature values (e.g., \citealt{miller_2015}; \citealt{Furst2016}). The final two observations are classified as HS, as their spectra are dominated by a power-law component with a photon index $\Gamma<2$, while the remaining epochs are labelled as SS, since the thermal disk component contributes more than 80\% of the total flux. We note that a dust-scattering halo has been reported for this source, whose effect cannot be easily isolated and may contribute to the reported fluxes \citep{greiner_1996}. 

\subsection*{GRS 1915+105}

\href{https://thunderkat.physics.ox.ac.uk/source/GRS%201915+105}{GRS 1915+105} is a well-studied XRB containing a BH (e.g., \citealt{reid_2014}), which was discovered in 1992 by the WATCH all-sky monitor on GRANAT as it started a decades-long outburst \citep{castro-tirado_1992, castro-tirado_1994}. It is particularly renowned for its quasi-periodic `heartbeat' variability, characterised by repetitive, structured X-ray flares, thought to be driven by radiation-pressure instabilities in the inner accretion disc (e.g., \citealt{belloni_2000, neilsen_2011}). It was also the first Galactic source observed to exhibit apparent superluminal motion of its jet components \citep{mirabel_1994}. Using Galactic proper motions and LOS radial velocities (and assuming low peculiar velocities), a three-dimensional kinematic distance of 9.4 $\pm$ 0.6 (statistical) $\pm$ 0.8 (systematic) kpc has been estimated \citep{reid_2023}, in agreement with an independent radio parallax estimate \citep{reid_2014}. Using this distance and the proper motion measurements of the source's radio jet \citep{miller_jones_2007}, its inclination was inferred to be $64^\circ \pm 4^\circ$ \citep{reid_2023}. 

GRS 1915+105 was extensively observed by MeerKAT during ThunderKAT. We release 72 epochs of core radio flux density results spanning 2018 December 8 to 2023 October 6 (MJDs 58460--60223). The radio reduction is described in \citet{motta_2021, motta_2025}. The heavy dust obscuration of the source \citep{motta_2021, miller_2025} renders the \emph{Swift}/XRT results and spectral state classifications uncertain, so we opt to exclude it from our $L_R$–$L_X$ analysis.

\subsection*{GX 339–4}\label{sec:gx339} 

\href{https://thunderkat.physics.ox.ac.uk/source/GX%20339-4}{GX 339–4} is a well-known LMXB, discovered in 1972 by the MIT X-ray detector on board the Orbiting Solar Observatory 7 satellite \citep{markert_observations_1973}. Dynamical studies indicate that its compact object is a BH \citep{hynes_2003}. Distance estimates for this source are uncertain, due to its companion star being very faint. Using Nuclear Spectroscopic Telescope Array (NuSTAR) and \emph{Swift} data, a distance of $8.4 \pm 0.9$ kpc was derived \citep{parker_nustar_2016}, while a more recent estimate of $\approx$ 8--12 kpc was obtained from evolutionary modelling of the donor \citep{zdziarski_x-ray_2019}. We adopt a nominal distance of 10 kpc, with a uniform probability distribution of 8--12 kpc. The source's orbital inclination has been estimated to be $\sim$37–78$^\circ$ using a variety of methods (e.g., \citealt{kolehmainen_limits_2010, shidatsu_x-ray_2011, heida_mass_2017, zdziarski_x-ray_2019}), with the upper limit set by the absence of X-ray eclipses, and the lower limit set by mass constraints on the donor and compact object \citep{munoz-darias_masses_2008, heida_mass_2017}.  

GX 339–4 comprises the largest subset of our sample. We report on 261 MeerKAT observations, taken between 2018 April 15 and 2023 September 23 (MJDs 58223--60210). The data reduction procedures for both MeerKAT, as well as additional 2020 observations with the Australia Telescope Compact Array (ATCA; taken at 5.5 GHz and 9 GHz), are described in \citet{tremou_2020, tremou_2026}. 

\emph{Swift}/XRT spectra were extracted using the online product generator using the restricted grade selection (to mitigate the effects of pile-up). We also verified that results obtained using all grades are consistent within 1$\sigma$ uncertainties. Spectral fitting was performed over 0.6--10 keV using Cash statistics, adopting three models as appropriate: a power-law, disc blackbody, or a combination of both. We excluded epochs flagged as `upper limits' by the light curve product generator, and those with $<$15 counts in WT mode or $<$10 counts in PC mode over the fitted energy range. 

For spectra with $>$100 counts over the fitted energy range, $N_H$ was left free and found to be consistent (within $\sim$90 per cent confidence intervals) with the nominal value of $\sim$$0.6 \times 10^{22} \text{cm}^{-2}$ found in the literature (e.g., \citealt{zdziarski_1998}). Fixing $N_H$ at this value produces fluxes consistent within 1$\sigma$, confirming that our results are not sensitive to this assumption. For spectra with $<$100 counts, we fixed $N_H = 0.6 \times 10^{22} \text{cm}^{-2}$ , although we note again that varying $N_H$ within the range reported in the literature ($\sim$$0.5{-}0.9 \times 10^{22} \text{cm}^{-2}$) does not significantly affect the derived fluxes. For the subset of these spectra with $<$50 counts, we adopted a power-law model and additionally fixed the photon index to a nominal value of $\Gamma=1.6$ for epochs in the HS and $\Gamma=2.0$ for the few during quiescence (due to the expected spectral softening at low luminosities, as noted by \citealt{corbel_2006, tremou_2020}). Although the limited counts in our QS observations prevent meaningful constraints on $\Gamma$, adopting either value for $\Gamma$ yields fluxes consistent within 1$\sigma$, given that statistical uncertainties dominate. We note that for many of the epochs with $<$50 counts, the detected photons are confined to very low energies, resulting in highly uncertain 1--10 keV flux estimates; these epochs correspond to the cluster of low-luminosity points with large $L_X$ error bars in the $L_R$--$L_X$ plane for the source. 

It is also worth mentioning that, although we excluded spectra with very low counts from our sample (as previously noted), if we instead opt to include these data -- by fixing all spectral parameters except the flux -- the resulting slope and intercept for the $L_R$–$L_X$ correlation of the source remain consistent within uncertainties. 

For several of the \emph{Swift}/XRT epochs, the source was observed in PC mode during extremely bright periods, when WT mode would have been more appropriate, resulting in extreme pile-up, and therefore uncertain flux estimates. An example is the spectrum with ID 00032898238, which we examined in detail with the UK \emph{Swift}/XRT team, using both the product generator and manual extractions. Despite the extreme pile-up producing a hole in the point spread function (PSF; caused by events being rejected due to the excess of photons) that coincides with a bad column, the product generator's source centroiding was found to work reliably. Additionally, changing the annular extraction region had little impact on flux estimates for a given grade selection, confirming that the product generator’s pile-up correction performs adequately. Furthermore, although no dust-scattering rings are visible in the PC observations of GX 339$-$4, the PSF for this epoch appears slightly irregular (and historical WT data show slightly extended wings relative to the PSF model), suggesting possible low-level dust scattering; however, this is unlikely to significantly affect results. 

We also note that several \emph{Swift}/XRT observations were taken in very close succession or are interspersed, and in some of these cases, the inferred unabsorbed fluxes do not agree within 1$\sigma$ uncertainties. While this could arise from statistical fluctuations (i.e., measurement uncertainties), preliminary analysis of intra-observation light curves suggests that the source can exhibit significant variability (see, e.g., \citealt{gandhi_2010}). This behaviour will be explored in future work. 

The spectral state at each epoch was classified using the \emph{Swift}/XRT fits and HR, as well as the MAXI HID. When our monitoring began in mid-2018, GX 339–4 was in quiescence. By early November, shortly after an observed increase in its X-ray flux signalled the beginning of a new outburst, it became unobservable with \emph{Swift}/XRT due to Sun constraints (MJDs 58423--58504) -- although the source continued to be monitored with MeerKAT, during which time its emission was seen to rise. When \emph{Swift}/XRT observing resumed in late January 2019, the source was in the HS with declining X-ray and radio fluxes. 

Around June 2019 (MJD $\sim$58650), the X-ray and radio fluxes of GX 339–4 began to rise again, and it remained in the HS until late October, after which \emph{Swift}/XRT observations halted due to Sun constraints (from MJD 58788). When X-ray observations resumed in late January 2020 (MJD 58873), GX 339–4 was in a decaying SS. Although the exact HS$\rightarrow$SS transition cannot be constrained, it likely occurred near MJD 58837, coinciding with the quenching of the core radio emission. Shortly after, on MJD 58845, the SS MeerKAT core flux density peaked (88.8 mJy), likely from unresolved transient ejecta, and then began to fade. The persistent radio emission during the SS is likely due to ejecta, as suggested by the steep spectral indices ($\alpha \lesssim -0.5$). The source remained in the SS until MJD $\sim$58922, when a modest radio rebrightening suggested a possible reactivation of the HS jets and a transition to the IMS. Following this IMS, GX 339–4 stayed in the HS for the remainder of the outburst. During the IMS/HS phase between MJDs 58922 and 59034, positional shifts of the radio peak/centroid relative to the core were observed, likely from unresolved ejecta, and an ATCA observation on MJD 58970 confirmed the presence of a discrete ejection (which was subsequently monitored for 42 days). Finally, the source reached quiescence in late November 2020.

GX 339–4 entered a new outburst in early January 2021 (MJD $\sim$59218), though \emph{Swift}/XRT observations were Sun-constrained until late January (MJD 59235). Type-C QPOs detected between MJDs 59285–59301 \citep{zhang_2024} confirmed the source remained in the HS during this period. Around MJD 59301, a drop in the core radio flux indicated compact-jet quenching and a transition to the IMS, consistent with the detection of Type-B QPOs at MJD $\sim$59303 \citep{periano_2023}. The subsequent SS core radio flux density peaked (27.9 mJy) at MJD 59314, likely from unresolved transient ejecta, and then faded; a positional shift in the MeerKAT images between MJDs 59335–59484 further supports this interpretation. Around MJD 59480, the source began to transition back to the HS via the IMS, accompanied by renewed core radio emission from the reactivated compact jet. It remained in the HS for the rest of the outburst, although it was Sun-constrained between MJDs 59519--59600, and then returned to quiescence by late January 2022 (MJD $\sim$59608). 

Another outburst began around August 2022 (MJD $\sim$59797) and lasted until late January 2023 (MJD $\sim$59965), during which the source remained in the HS. Then, after a period of quiescence (comprising mostly radio non-detections, with a few detections close to the quiescent level), another outburst began in mid-August 2023 (MJD $\sim$60170).

As part of X-KAT, we have continued monitoring GX 339–4. In forthcoming work, we will present a comprehensive analysis of the source's X-ray properties, incorporating \emph{Swift}/XRT observations from both ThunderKAT and X-KAT, along with additional quasi-simultaneous X-ray data (which will allow better constraints on the transition dates between spectral states). We will also examine potential phase-dependent variations in its $L_R$–$L_X$ relation, as suggested by previous studies (e.g., \citealt{corbel_2013, islam_2018, koljonen_2019}).

\subsection*{H1743–322}

\href{https://thunderkat.physics.ox.ac.uk/source/H%201743-322}{H1743–322} was first detected by Ariel 5 in 1977 \citep{kaluzienski_1977}, and has undergone many complete and `failed' outbursts (e.g., \citealt{capitanio_2009, tetarenko_2016}). Owing to its similarities with the dynamically confirmed BH XRB XTE J1550–564, the source has been classified as a BHC \citep{mcClintock_2009}. We adopt a nominal distance of 8.5 $\pm$ 0.8 kpc (statistical uncertainty; \citealt{steiner_2012}). This was obtained by fitting a simple, symmetric kinematic model to the trajectories of the radio and X-ray ejecta during the 2003 outburst, which also yielded an estimate of $75^\circ\pm3^\circ$ for the radio jet inclination angle.

Our ThunderKAT MeerKAT monitoring began on 2018 September 5 (MJD 58366), just three days after a new outburst was reported \citep{grebenev_2018}, and continued until 2018 November 10 (MJD 58432), yielding a total of 11 epochs. Details of the MeerKAT and \emph{Swift}/XRT data reduction are provided in \citet{williams_2020}. All the radio detections show an unresolved point source at the core position, with no evidence for transient ejecta, and the X-ray spectral fits indicate that the source remained in the HS throughout.

\subsection*{IGR J17091–3624}

\href{https://thunderkat.physics.ox.ac.uk/source/IGR%20J17091-3624}{IGR J17091–3624} was discovered with the INTErnational Gamma-Ray Astrophysics Laboratory (INTEGRAL) in 2003, while entering outburst \citep{kuulkers_2003}. Subsequent observations suggested a BH accretor, based on its spectral behaviour and significant radio activity (e.g., \citealt{lutovinov_2003}), as well as the similarity of its X-ray properties to those of GRS 1915+105 \citep{altamirano_2011}. We adopt a nominal distance of 14 kpc, based on the $\sim$11--17 kpc range derived from the inferred luminosity at the HS transition, assuming a typical BH mass \citep{rodriguez_2011}. However, we note that a larger estimate of $\sim$20--25 kpc has also been proposed \citep{rao_2012}. The presence of X-ray dips (e.g., \citealt{ewing_2025}) and wind signatures in its spectra (e.g., \citealt{wang_2024}), combined with the absence of eclipses and similarities to the high-inclination system GRS 1915+105, suggest an inclination in the range $\sim$60--75$^\circ$ (e.g., \citealt{king_2012, rao_2012}). 

IGR J17091–3624 entered a new outburst phase in March 2022 \citep{miller_2022}. As part of ThunderKAT, we obtained 42 MeerKAT epochs between 2022 March 17 and 2023 February 11 (MJDs 59655--59986). Details of the MeerKAT and \emph{Swift}/XRT data reductions are given in Russell et al. (in preparation). We note that the \emph{Swift}/XRT data set includes one 3$\sigma$ upper limit, derived by converting the count rate to flux using \texttt{WebPIMMS}. To classify the spectral states, we relied on the \emph{Swift}/XRT fits, the radio behaviour, as well as the NICER-based classifications reported by \cite{wang_2024}. 

At the start of our monitoring, IGR J17091–3624 was in the late rising HS or hard IMS (HIMS), before transitioning to a soft IMS (SIMS) at MJD $\sim$59660. It remained in the SIMS and subsequently the SS for $\sim$120 days. However, during this period, it exhibited heartbeat-like variability in its X-ray light curve, reminiscent of GRS 1915+105, which \cite{wang_2025} classify as an `exotic' state. Around MJD $\sim$59785, the source transitioned back to the IMS, after which point the MeerKAT flux density increased significantly, suggesting that the compact jet may have been reactivated  during this period (although it is also possible that the flaring arose from discrete ejecta interacting with the surrounding environment). Finally, the source re-entered the HS, likely at approximately MJD $\sim$59850; however, due to a lack of X-ray data because of Sun constraints, the exact transition date is uncertain. Supporting ATCA monitoring shows that the radio spectrum became flatter around this time, implying the presence of a compact jet and, therefore, a return to the HS.

\subsection*{MAXI J1348–630}

\href{https://thunderkat.physics.ox.ac.uk/source/MAXI%20J1348-630}{MAXI J1348–630} was discovered as a bright X-ray transient in January 2019 \citep{yatabe_2019} by MAXI. Intense multi-wavelength follow-up observations have identified it as a BHC (e.g., \citealt{sanna_2019}, \citealt{zhang_2020}). We adopt a nominal distance of 2.2$^{+0.5}_{-0.6}$ kpc, derived from HI absorption measurements using the Australian Square Kilometre Array Pathfinder (ASKAP) and MeerKAT \citep{chauhan_2021}. However, we note that an alternative geometrical estimate of $3.39\pm0.34$ kpc was obtained by analysing the giant dust-scattering ring around the system \citep{lamer_2021}. Recently, using the maximum best-fit radio jet proper motions from \citet{carotenuto_2021a} and a distance of 2.2 kpc, an upper limit on the jet inclination of $i \lessapprox 71.4^\circ$ was derived \citep{cooper_2025}; tighter kinematic constraints are not possible due to the absence of a detected counter-jet (F. Carotenuto, private communication). 

MAXI J1348–630 was monitored with MeerKAT at $\sim$weekly cadence between 2019 January 29 and 2020 March 3 (MJDs 58512--58910), yielding 48 observations. We also report on an additional radio detection on 2020 September 20 (MJD 59112). Details of the MeerKAT and \emph{Swift}/XRT data reductions, as well as additional ATCA observations (at 5 and 9 GHz), are given in \citet{carotenuto_2021a, carotenuto_2021b}. Several 3$\sigma$ \emph{Swift}/XRT upper limits are reported; these were estimated by fitting a power-law model (with $\Gamma=2.2$) to a circular region centred on the source position in the background spectrum, and adopting three times the resulting flux as the upper limit (F. Carotenuto, private communication). Spectral states were classified based on the spectral and timing analysis of NICER data by \citet{zhang_2020}. 

At the start of our campaign, MAXI J1348–630 was in a rising HS, before transitioning through the IMS into the SS where it remained for $\sim$2 months. At the onset of the SS (MJD $\sim$58523), the radio spectrum steepened and a strong MeerKAT flare ($\sim$486 mJy) was observed, followed shortly by an additional ATCA flare at MJD $\sim$58530, after which the core radio emission faded. From MJD 58551, a discrete component (initially unresolved from the core) was detected with MeerKAT, and interpreted as an approaching ejection. It was monitored over the following months, exhibiting high proper motion ($\sim$100 mas d$^{-1}$) and subsequent deceleration (likely as it propagated through a low-density cavity in the interstellar medium, ISM), with proper motion fits suggesting an ejection on MJD $\sim$58522 during the IMS just prior to the first core flare \citep{carotenuto_2022}. The second ATCA flare may indicate a second ejection or represent the receding component of the first event. 

Later in the SS (MJDs 58573–58582), a reflare of optically thin core radio emission was observed. Then, on MJD 58589, a second radio knot was detected (with ATCA and later MeerKAT), moving in the same direction as the first (with no counter-jet), and was tracked for about a month. It was inferred to have been launched during the SS reflare (MJD $\sim$58579), contemporaneous with a drop in the fractional RMS variability, likely during a brief excursion to the IMS \citep{carotenuto_2025}. In addition, another ejection (partially resolved in the ATCA epoch on MJD 58589) may have been launched some time between MJDs 58578--58589 \citep{carotenuto_2025}. 

Around MJD 58597, the source re-entered the IMS and then the decaying HS. We note that MeerKAT core measurements on MJDs 58614, 58621, and 58628 may be contaminated by unresolved ejecta. Before returning to quiescence, the source subsequently exhibited several HS-only reflares: a main one in June 2019, a shorter one observed only in the optical and X-rays around mid-October, one in November–December 2019, and another in February 2020. A final MeerKAT detection occurred during another HS reflare on 2020 September 20 (MJD 59112).

\subsection*{MAXI J1631–479}

\href{https://thunderkat.physics.ox.ac.uk/source/MAXI%20J1631-479}{MAXI J1631–479} was first detected by MAXI in December 2018, and was initially mistaken for the X-ray pulsar AX MAXI J1631–4752 \citep{kobayashi_2018}. However, its spectral properties inferred through further observations by NuSTAR suggested that it is in fact a binary system with an accreting BH \citep{miyasaka_2018}. The source distance is poorly constrained. \citet{rout_2023} derived a lower limit of $\sim$4.5 kpc by comparing observed and theoretical optical fluxes, and noted that, although the high column density permits distances $\gtrsim$15 kpc, such values are disfavoured by the source's position on the $L_R$--$L_X$ plane. More recently, \cite{zdziarski_2025} estimated a distance of $D = 5.1^{1.1}_{-0.4}$ kpc (statistical uncertainty) through relativistic disc-continuum fitting of the SS X-ray spectrum (using \texttt{bhspec} with $\alpha=0.1$). We therefore adopt the range 4.5--6.5 kpc, with a nominal value of 5.1 kpc, although we emphasize that this estimate is very uncertain. The system's inclination has only been constrained through reflection spectroscopy: \citet{rout_2023} inferred a high inclination of $\sim$50$^\circ$–72$^\circ$, which they argue is supported by the presence of wind signatures, while other studies \citep{xu_2020, rout_2023, zdziarski_2025} have suggested low inclinations. 

We obtained 26 epochs of MeerKAT data for MAXI J1631–479 between 2019 January 12 and August 10 (MJDs 58495--58705). The MeerKAT and \emph{Swift}/XRT data reductions are described in \citet{monageng_2021}. The spectral states were classified using the \emph{Swift}/XRT spectral fits, MAXI HID, and X-ray analyses from the literature. At the start of the outburst, the source was in the SS. Following a brief excursion to the HS/IMS, a radio flare was observed (on $\sim$MJD 58565) as the source returned to the SS, likely due to the launch of ejecta (although none were resolved). The source then briefly re-entered the HS, during which time a contribution to the core radio emission from fading ejecta cannot be ruled out. Subsequently, as the source re-entered the SS, the core radio emission quenched before rising again, possibly due to an additional ejection.

\subsection*{MAXI J1803–298}

\href{https://thunderkat.physics.ox.ac.uk/source/MAXI%20J1803-298}{MAXI J1803–298} was first detected by MAXI in May 2021, during the onset of its only known outburst to date \citep{serino_2021}, and lies $\sim$4$^\circ$ from the Galactic centre on the plane of the sky. Mass estimates strongly support a BH accretor \citep{mata_sanchez_2022, chand_2022}. The source distance remains uncertain, but estimates using its inner disc radius in the SS and its peak luminosity suggest that either moderate ($\sim$6 kpc) or large ($\gtrsim$11 kpc) distances are favoured \citep{shidatsu_2022}. Following previous work (e.g., \citealt{chand_2022, wood_2023}), we adopt a nominal distance of 8 kpc, with an allowed range 6--12 kpc. Observations with multiple X-ray telescopes have revealed periodic absorption dips (e.g., \citealt{xu_2021}; \citealt{jana_2022}), and strong evidence for a disc wind \citep{zhang_2024}, both consistent with a high orbital inclination of $\sim$65--70$^\circ$ \citep{shidatsu_2022, sanchez_2022}.  

MAXI J1803–298 was first observed with MeerKAT on 2021 May 4 (MJD 59338), a few days after its X-ray discovery, and was subsequently monitored at $\sim$weekly cadence until 2021 December 18 (MJD 59566), yielding 30 epochs. Details of the MeerKAT and \emph{Swift}/XRT data reductions, along with quasi-simultaneous VLA and ATCA radio observations, are presented in Espinasse et al. (submitted). The spectral state transitions were determined using the spectral fit results and the literature \citep{steiner_2021, shidatsu_2022}. 

At the onset of the outburst, MAXI J1803–298 was in the late-HS phase, showing a rapid rise in radio emission with slightly positive spectral indices, consistent with compact jet activity. As the source transitioned into the IMS at MJD $\sim$59345, a core optically thin radio flare was observed, indicative of discrete ejecta. Although ejecta were not spatially resolved in MeerKAT, VLA, or ATCA data, observations by \citet{wood_2023} during the IMS using the Very Long Baseline Array (VLBA) revealed at least two components moving away from the core. As the source subsequently entered the SS, core radio emission dropped before further optically thin flaring -- with a MeerKAT-detected peak at MJD $\sim$59370 and a secondary ATCA peak $\sim$12 days later -- likely due to shocks produced as the previously-launched ejecta interacted with the ISM, or consecutive ejections as the source entered the SS (Espinasse et al. submitted). After nine MeerKAT non-detections, another optically thin flare was observed on MJD $\sim$59477, possibly due to a brief excursion to the IMS during which further discrete ejecta might have been launched (although this occurred relatively late in the SS compared to other sources; Espinasse et al. submitted). MAXI J1803–298 eventually returned to the HS, where positive spectral indices signalled the reappearance of compact jets.

\subsection*{MAXI J1807+132}

\href{https://thunderkat.physics.ox.ac.uk/source/MAXI%20J1807+132}{MAXI J1807+132} was discovered in 2017 by MAXI \citep{negoro_2017, shidatsu_2017}, although its position was found to be consistent with that of a long-term flaring event detected as early as 2011 \citep{kawamuro_2016}. The compact object in the system is known to be a NS, due to the detection of Type I thermonuclear bursts during its 2019 outburst \citep{albayati_2021}. In particular, the source's spectral and timing properties are found to resemble a typical atoll source \citep{rout_2025a}. It is known to have short outbursts and reflares, with unusually high amplitude X-ray and optical variability on day- to week- timescales \citep{jimenez_2019, albayati_2021}. To be consistent with a NS accretor, a low distance of $\sim$1--5 kpc has been suggested, with the upper limit set by the source's position on the optical:X-ray luminosity diagram and the lower limit by its peak luminosity \citep{jimenez_2019, rout_2025a}. However, more recently, \citet{saavedra_2025} derived a preferred distance of 6.3 $\pm$ 0.7 kpc (statistical) by extending the absolute magnitude versus orbital period correlation for BH transients to NS systems. We adopt 6.3 kpc as our nominal estimate, with an allowed range 4.2--8.4 kpc ($3\sigma$ CI, given the uncertainty of the method). The system inclination has been estimated to be $72^\circ \pm 5^\circ$ by modelling its optical light curve \citep{saavedra_2025}. 

In July 2023, MAXI J1807+132 was reported to be in outburst \citep{saikia_2023}. As part of ThunderKAT, we collected 15 MeerKAT epochs between 2023 July 7 and October 14 (MJDs 60132--60231), as detailed in \citet{rout_2025b}. Our spectral state classifications follow \citet{rout_2025a}, although we note that the processes driving the X-ray/optical/UV reflares after MJD $\sim$60155 remain uncertain, so the states during this period are labelled as `Unclear'. The radio morphology in the vicinity of MAXI J1807+132 exhibited temporal variability, which was confirmed (through additional VLA monitoring) to be due to three moderately variable but unassociated point sources within $\sim$10 arcsec of the source \citep{rout_2025b}. Our radio observation campaign yielded a single detection at the source position during the IMS, although the nature of this emission remains unclear.

\emph{Swift}/XRT spectra were fit over the range 0.5--10 keV, fixing $N_{\rm H} = 2.616\,{\times}\,10^{21}\,{\rm cm^{-2}}$, consistent with the value inferred by \citet{rout_2025a}. Low-count (${\lesssim}\,500$) spectra were fit using Cash statistics and an absorbed power-law, while the remainder were binned into 25-count intervals and fit using $\chi^2$ statistics. HS epochs were modelled with $\texttt{tbabs*(bbody+pegpwrlw)}$. In the SS and IMS epochs, a $\texttt{diskbb}$ component was added, following \citealt{rout_2025a}, which was required for convergence in several bright observations. The three-component model was additionally adopted during bright epochs in the late-time flaring phase, as it offered the most flexibility.

\subsection*{MAXI J1810–222}

\href{https://thunderkat.physics.ox.ac.uk/source/MAXI%20J1810-222}{MAXI J1810–222} was discovered in 2018 \citep{negoro_2018} by MAXI. The source's nature remains unconfirmed and it is classified as a `soft X-ray transient', though its X-ray and radio behaviour suggest that it is a BH XRB \citep{russell_2022}. Using a Galactic mass density model for LMXBs \citep{grimm_2002, atri_2019}, \citet{russell_2022} estimated that there is a $\sim$94 per cent probability that the source lies beyond 6 kpc (and 31 per cent at $>$10 kpc). Using this model, the source distance is additionally inferred to be $\lesssim$17 kpc at the $\sim$94 per cent confidence interval. We therefore adopt a nominal distance of 8 kpc, with an allowed range of 6--17 kpc, though we note that this estimate is very uncertain. Although there are no strong constraints on the source's inclination angle, the lack of X-ray dips or QPO detections suggest that it is low \citep{russell_2022}.

In April 2023, following reports of MAXI J1810–222 being in the HS  \citep{morrow_2023}, we began our ThunderKAT monitoring, obtaining 34 epochs of MeerKAT data at $\sim$weekly cadence between 2023 April 30 and 2024 January 29 (MJDs 60064--60338). Each observing block consisted of 15 minutes of on-target time, bracketed by $\sim$2-minute scans of the phase calibrator J1833–2103, and ended with a 5–10 minute observation of the flux/bandpass calibrator J1939–6342. The data were reduced with \texttt{oxkat}, using the default parameters. For all the epochs, the source at the core position was observed to be unresolved, so it was fit with a point source model.

\emph{Swift}/XRT spectra between April and October 2023 were extracted, although a few observations were excluded because the source was near the edge of the CCD. From November 2023 to February 2024, the source was not visible with \emph{Swift}/XRT due to visibility constraints. Spectral states were determined using the \emph{Swift}/XRT HR and spectral fits. To derive the unabsorbed 1–10 keV fluxes, HS epochs were fit with \texttt{tbabs*powerlaw}, while SS epochs were fit with \texttt{tbabs*(powerlaw+diskbb)} with the power-law photon index restricted to 1.2–3. We allowed N$_{\mathrm{H}}$ to vary within the range (0.6–1.0) $\times$ 10$^{22}$ cm$^{-2}$ since this interval encompasses the typical values reported for the source in the literature \citep{russell_2022,del_santo_2023}. For instance, analysis of data from RGS on board XMM-Newton suggests 0.73$\times$ 10$^{22}$ cm$^{-2}$ \citep{Pinto26}. 

\subsection*{MAXI J1816–195}

\href{https://thunderkat.physics.ox.ac.uk/source/MAXI%20J1816-195}{MAXI J1816–195} was discovered by MAXI/GSC during its outburst in June 2022 \citep{negoro_2022}, and the detection of 528 Hz pulsations and a thermonuclear X-ray burst indicated that the source is a NS AMXP \citep{bult_2022}. Distance estimates based on its X-ray bursts suggest upper limits of 6.3 kpc \citep{chen_2022}, 8.6 kpc owing to the lack of evidence for PRE \citep{bult_2022b}, and $\sim$8.7 kpc \citep{mandal_2023}. We adopt a nominal distance of 6 kpc (following \citealt{wang_p_2024}), allowing it to vary uniformly in the range 1--8.7 kpc. The inclination of the system has not been well constrained. When conducting our data reductions, we use the updated source position by \citet{kennea_2022}. 

We obtained five MeerKAT epochs for MAXI J1816–195 between 2022 June 8 and 2022 July 15 (MJDs 59738--59775). In each observation, the target was observed for approximately 15 minutes, with J1939–6342 and J1833–2103 as the flux/bandpass and phase calibrators, respectively. The data were reduced using \texttt{oxkat} with the default parameters. The RMS ($\equiv$$\sigma_\text{rms}$) was determined using a two-arcminute-wide square source-free region slightly offset from the target position. For the detections (i.e., peak source flux density $>$$3\sigma_\text{rms}$), a point source model was fitted, while for non-detections, $3\sigma_\text{rms}$ upper limits are reported. The source was detected at $\sim$4.2 mJy during the first epoch (MJD 59738.97), but was no longer detected four days later. It then reappeared at a lower flux density in early July, while two subsequent weekly observations yielded non-detections. For the first epoch, we imaged eight sub-bands spanning the observing frequency range and measured a spectral index of $\alpha = -0.1 \pm 0.08$, assuming a power-law spectrum ($F_\nu \propto \nu^\alpha$).

Studies examining the evolution of the source's X-ray spectral parameters around this time suggest that its disc-related parameters were in a transitional phase, and that the HR was decreasing (e.g., \citealt{li_p_2023, wang_p_2024a, wang_p_2024}). Furthermore, an analysis of the NICER power spectrum at MJD 59738.32 is suggestive of the source being in the HS, whereas those on MJDs 59738.77 and 59739.35 are more consistent with an IMS (P. Wang, private communication). Therefore, at the time of our first radio epoch (MJD 59738.97), the source was likely undergoing a transition from the late HS to IMS, so we classify it as being in an IMS. 

We report on a single \emph{Swift}/XRT WT epoch from 2022 June 8 (MJD 59738.20), as the only other quasi-simultaneous observation yielded an upper limit. We fitted this spectrum over the energy range 0.6--10 keV, using single-count binning with Cash statistics, leaving $N_H$ free. We note that a fit performed using 20 counts per bin and $\chi^2$ statistics yielded results that agree within uncertainties. The data are well fit by \texttt{tbabs*powerlaw} and \texttt{tbabs*(bbody+powerlaw)} models, with photon indices in the typical range for the HS/HIMS. We adopt the latter fit, although both models yield fluxes consistent within 10 per cent. The resulting $N_H$ is slightly higher than those determined using other telescopes (e.g., \citealt{chen_2022, mandal_2023, wang_p_2024}), although fixing it to these values changes the flux by $<$10 per cent. We note that, although \emph{Swift} observations in June/July 2022 revealed a dust-scattered halo at $\sim$2–6 arcmin from the source \citep{beardmore_2022}, the high count rate in our reported epoch means that its impact on the flux is likely negligible.

\subsection*{MAXI J1820+070}

\href{https://thunderkat.physics.ox.ac.uk/source/MAXI%20J1820+070}{MAXI J1820+070}/ASASSN-18ey was discovered as an optical transient in March 2018 by the All-Sky Automated Survey for SuperNovae (ASAS-SN; \citealt{tucker_2018}). It was subsequently identified as an XRB following MAXI observations \citep{kawamuro_2018,denisenko_2018}, and later dynamically confirmed to host a BH \citep{torres_2019}. Owing to its proximity and high luminosity, it has been an ideal target for multi-wavelength monitoring since its initial outburst. Using radio parallax measurements in the HS from the VLBA and the European Very Long Baseline Interferometry Network (EVN), the source distance was estimated to be 2.96 $\pm$ 0.33 kpc, consistent with values derived using the Gaia DR2 \citep{atri_2020} and DR3 \citep{wood_2021}. 

Bipolar relativistic ejecta -- likely launched during a HS$\rightarrow$SS transition at MJD $\sim$58306 -- were extensively monitored in the radio with the Multi-Element Radio Linked Interferometer Network (eMERLIN), VLBA, AMI-LA, VLA, and MeerKAT, and observed to propagate up to $\sim$10 arcsec from the core with high proper motion, with their rebrightenings likely caused by in-situ particle acceleration as they interacted with the ISM \citep{bright_2020, wood_2021}. These ejecta were also detected in X-rays up to $\sim$12 arcsec from the core, using Chandra observations between late 2018 and 2019 \citep{espinasse_2020}. By fitting the VLBA proper motion measurements and adopting the radio parallax distance, \citet{wood_2021} derived a jet inclination of 64 $\pm$ 5$^\circ$, consistent with estimates by \citet{atri_2020} using the fits from \citet{bright_2020}, as well as by \citet{carotenuto_2022} using a dynamical blast-wave model. 

MAXI J1820+070 was observed during ThunderKAT from 2018 September 28 (MJD 58389), during its SS$\rightarrow$HS transition \citep{shidatsu_2019}. We report on 55 MeerKAT observations until 2022 July 9 (MJD 59769), during which time the source underwent three HS-only rebrightenings \citep{bright_2025}. During the first 12 MeerKAT observations (up to and including MJD 58454), both the core and approaching ejecta were resolved and modelled with multiple point sources. From MJD 58460 onward, the source appeared unresolved and was fitted with a single component at the core position. However, between MJD 58460 and 58551, residual low-level emission from the receding ejecta may have slightly affected fits.

\subsection*{SAX J1808.4–3658}

\href{https://thunderkat.physics.ox.ac.uk/source/SAX%20J1808.4-3658}{SAX J1808.4–3658} was discovered in 1996 through the detection of a Type I X-ray burst using the BeppoSAX satellite \citep{zand_1998}. Shortly after, the detection of 401 Hz pulsations with the Rossi X-ray Timing Explorer (RXTE) established the source as the first known AMXP \citep{wijnands_1998}. It has since been extensively monitored due to its recurrent outbursts every 2--4 years. \cite{goodwin_2019} used a Bayesian approach for matching X-ray burst data with models to estimate a distance of $3.3 ^{+0.3}_{-0.2}$ kpc (statistical). We adopt the distance estimate by \citet{galloway_2024}, who obtained $2.7 \pm 0.3$ kpc (statistical), attributing the lower value greater burst anisotropy. The inclination was estimated to be $50^\circ$$^{+6^\circ}_{-5^\circ}$ using quiescent optical light curve modelling \citep{wang_2013}, while similar values were derived from iron line fitting \citep{cackett_2009} and X-ray burst modelling \citep{goodwin_2019}.

After SAX J1808.4–3658 was reported to have entered a new outburst in July 2019 \citep{russell_2019_j1808}, we obtained six MeerKAT observations between 2019 July 31 and August 31 (MJDs 58695--58726), and three additional ones during renewed activity \citep{sanna_2022} between 2022 August 27 and September 9 (MJDs 59818--59831). Details regarding the MeerKAT and \emph{Swift}/XRT data reductions can be found in \citet{gasealahwe_2023} and Gasealahwe et al. (in preparation). The spectral states were determined using \emph{Swift}/XRT spectral fit results and the MAXI HID. The source was also extensively monitored with NICER, revealing that some flaring was missed at the low cadence of the \emph{Swift}/XRT observations. 

In Gasealahwe et al. (in preparation), we compile an updated $L_R$--$L_X$ plane for the source, using both \emph{Swift}/XRT and NICER data, together with additional data points from X-KAT observations in 2025 that were obtained after the source entered a new outburst \citep{russell_2025_atel}.

\subsection*{SAX J1810.8–2609}

\href{https://thunderkat.physics.ox.ac.uk/source/SAX%20J1810.8-2609}{SAX J1810.8–2609} was discovered in 1998 by the BeppoSAX satellite \citep{ubertini_1998}, and a subsequent detection of a Type I X-ray burst has confirmed a NS accretor \citep{natalucci_2000}. The source is classified as a `soft X-ray transient', though its moderate peak X-ray luminosity ($L_X \leq 4 \times 10^{36}$ erg s$^{-1}$) and transient behaviour suggest it is an atoll. The PRE of the X-ray burst was used to estimate a source distance of 4.9 $\pm$ 0.3 kpc \citep{natalucci_2000}, although this uncertainty is purely statistical so likely underestimated. No constraints currently exist on the system’s inclination angle.

The source entered its fifth recorded outburst in May 2021 \citep{iwakiri_2021}, and as part of ThunderKAT, we monitored the full $\sim$5-month outburst (\citealt{hughes_2024}). We obtained 21 $\sim$weekly MeerKAT observations from 2021 May 22 to 2021 October 23 (MJDs 59356--59510), as well as two deep (1-hour) observations during X-ray quiescence on 2023 May 22 and 2023 August 16 (MJDs 60086 and 60172). The deep images strongly suggest the presence of an unresolved, persistent, steep-spectrum radio source spatially coincident with SAX J1810.8$-$2609 (within $\pm$3 arcsec), with a mean flux density $93 \pm 5\,\upmu\mathrm{Jy}$. Therefore, for each epoch, we fitted a point source at the core position and subtracted this mean background level. Detections were defined as those with a residual flux density that is $\geqslant$3$\sigma_\text{rms}$, where $\sigma_\text{rms}$ is the RMS in a nearby source-free region. For non-detections, upper limits are given as the residual flux density plus 3$\sigma_\text{rms}$. Based on the radio and X-ray spectral properties, all epochs were classified as being in the HS (although the epoch on MJD 59384 may be in the IMS), suggesting that SAX J1810.8$-$2609 underwent a HS-only outburst.

\subsection*{Swift J1727.8–1613}

\href{https://thunderkat.physics.ox.ac.uk/source/Swift%20J1727.8-1613}{Swift J1727.8–1613} was discovered by the \emph{Swift} Burst Alert Telescope (BAT) in August 2023 during its first recorded outburst \citep{page_2023, kennea_2023_j1727}, and hosts a dynamically confirmed BH accretor \citep{mata_sanchez_2025}. We adopt the most recent distance estimate of $5.5^{+1.4}_{-1.1}$ kpc, derived from its near-ultraviolet reddening and optical photometry, assuming a main-sequence K4($\pm1$)V companion \citep{burridge_2025}. However, we note that an alternative recent estimate places the distance at 3.4 $\pm$ 0.3 kpc \citep{mata_sanchez_2025}. X-ray spectral modelling (e.g., \citealt{peng_2024}, \citealt{svoboda_2024}, \citealt{liu_2024}) suggests a medium-to-low inclination of $\sim$30--50$^\circ$, while modelling of the jet dynamics (using a distance of 5.5 kpc) indicates that $32^\circ \lesssim i \lesssim69^\circ$  (\citealt{wood_2025, burridge_2025}). The absence of ellipsoidal modulations further rules out a high-inclination system \citep{mata_sanchez_2025}.

We report on 47 L-band MeerKAT epochs, taken between 2023 August 27 and 2024 June 2 (MJDs 60183--60463) at $\sim$weekly cadence. Details regarding these observations are found in \citet{hughes_2025_lrlx, hughes_2025_compre}, along with additional radio monitoring using the MeerKAT S-band, e-MERLIN, the Allen Telescope Array (ATA), VLA, ATCA, Atacama Large Millimeter Array (ALMA), and the Submillimeter Array (SMA). A small number of 3$\sigma$ \emph{Swift}/XRT upper limits are reported, derived via count-rate-to-flux conversions assuming a power-law model. Spectral states were classified using X-ray timing and spectral studies \citep{bollemeijer_2023a, bollemeijer_2023b, podgorny_2024}. 

At the start of our MeerKAT campaign, Swift J1727.8–1613 was in the HS, showing a sharp rise in its radio flux density and flat spectral indices indicative of a compact jet. VLBI observations during this period revealed the largest resolved compact jet observed in a BH LMXB to date \citep{wood_2024}. The source remained at $\sim$100 mJy for a few weeks, and then began to fade, showing rapidly evolving spectral indices, and exhibiting a flaring episode around the HS$\rightarrow$SS transition (MJDs $\sim$60207--60249). These features are consistent with the scenario that the compact jet was disrupted, and ejecta were launched -- as confirmed through direct VLBI imaging \citep{wood_2025}. Following the SS transition, the flux density declined, and spectral indices were negative (and highly variable), consistent with the emission originating from ejecta. A modest radio rebrightening was observed during this period, likely due to ejecta-ISM interactions -- although additional ejections remain possible, particularly given the suggestion that the source transitioned several times between the SS and IMS \citep{yu_2023}. Results of the entire SS phase will be presented in Carotenuto et al. (in preparation). In this paper, we show that an extended radio component observed in MeerKAT L-band data was resolved into two distinct point sources in higher resolution S-band data, suggesting that other L-band ejections in our sample may similarly comprise multiple components. After spending $\sim$6 months in the SS, the source returned to the HS, where it remained for $\sim$2 months before returning to quiescence. At this point, the radio emission plateaued, likely due to long-lived ejecta interacting with the ISM.

\subsection*{Swift J1728.9–3613}

\href{https://thunderkat.physics.ox.ac.uk/source/Swift%20J1728.9-3613}{Swift J1728.9–3613} was discovered by \emph{Swift}/BAT in January 2019 \citep{barthelmy_2019}. It is spatially coincident with the supernova remnant G351.9$-$0.9 \citep{whiteoak_1996}, though it is unlikely to be associated with it (\citealt{balakrishnan_2023}). The source's spectral and timing properties indicate a BH accretor \citep{saha_2023}. Although its high column density ($N_H > 10^{22} \text{cm}^{-2}$) complicates distance estimates, \citet{balakrishnan_2023} derived a distance constraint of 7.6 kpc $\leq D \leq$ 9.2 kpc (with a best estimate of 8.4 kpc) by integrating the LOS gas density to match X-ray observations. The only estimates of inclination are via reflection spectroscopy, and give conflicting results \citep{draghis_2023,heiland_2023}. 

During ThunderKAT, we obtained 14 MeerKAT epochs of Swift J1728.9–3613 at $\sim$weekly cadence between 2019 January 31 and May 4 (MJDs 58514--58607), and an additional epoch on 2020 September 24 (MJD 59116). Each epoch had 15 minutes on target, except those on February 12 and March 10 (MJDs 58526 and 58552) which had 45 minutes. J1939–6342 and J1830–3602 were used as the primary and secondary calibrators, respectively. The remnant G351.9–0.9 makes measuring the flux density of Swift J1728.9–3613 more challenging than in isolated fields. In order to reduce the contribution of background emission, we imaged the field with a significant $uv$-plane taper (which down-weights the shortest baselines) using \texttt{wsclean} with a tapered cosine (i.e., Tukey) window, with \texttt{-minuvl=0} and \texttt{-inner-tukey=600}, corresponding to baselines of length $\approx$$140\,\rm{m}$ at $1.28\,\rm{GHz}$. The RMS ($\sigma_\text{rms}$) was measured using a (four-squared-arcminute) annulus centred on the source, and the flux densities of detections (i.e., peak source flux density $>$$3\sigma_\text{rms}$) were extracted using a point source model. Spectral states were classified using the date ranges by \citet{saha_2023}. While no ejecta were resolved in our images, the observed SS emission is likely due to ejecta launched during the HS$\rightarrow$SS transition. 

Only three \emph{Swift}/XRT epochs are quasi-simultaneous with our radio data, all during the HS$\rightarrow$SS transition. The first is a heavily piled-up PC observation and thus unreliable. The remaining two (high-count, WT) observations in the SS yield an $N_H$ that is considerably higher than literature estimates (e.g., \citealt{balakrishnan_2023, draghis_2023}), and while this may simply indicate variable absorption, we opt to exclude these results from our analysis.

\subsection*{Swift J1842.5–1124}

\href{https://thunderkat.physics.ox.ac.uk/source/Swift%20J1842.5-1124}{Swift J1842.5–1124} was first detected in 2008 by \emph{Swift}/BAT \citep{krimm_2008}, and has been classified as a BHC XRB due to its X-ray spectral and timing properties (e.g., \citealt{markwardt_2008, zhao_2016, krimm_2013}). Based on its position on the $L_R$–$L_X$ plane, \citet{zhang_2022} suggested a source distance $\gtrsim$5 kpc, while \citet{abdulghani_2024} estimated $8.12^{+2.2}_{-1.67}$ kpc (statistical) using data for the spectral state transitions. We adopt 8.12 kpc as our nominal estimate, but note that assuming a broader SS$\rightarrow$HS luminosity range (than the 1--4 per cent of Eddington adopted in this study) would affect the estimate (see Section 6.4 in \citealt{abdulghani_2024} for further discussion regarding systematic uncertainties). The inclination angle for this source has not been constrained.

Following reports of a new outburst in May 2020 \citep{shidatsu_2020}, we observed Swift J1842.5–1124 with MeerKAT on a $\sim$weekly basis between 2020 June 1 and June 27 (MJDs 59001--59027), yielding five epochs (\citealt{zhang_2022}). Spectral states were determined using the \emph{Swift}/XRT spectral fits and MAXI HR. No radio emission was detected during the first MeerKAT epoch when the source was in the SS. Prior to the second epoch, \citet{zhang_2022} reported a possible SS$\rightarrow$HS$\rightarrow$SS$\rightarrow$HS sequence, although this remains unconfirmed. After three further epochs in the HS, the source returned to quiescence.

\subsection*{Swift J1858.6–0814}

\href{https://thunderkat.physics.ox.ac.uk/source/Swift%20J1858.6-0814}{Swift J1858.6–0814} was discovered by \emph{Swift}/BAT in October 2018 \citep{krimm_2018}. Type I X-ray bursts have unambiguously identified the accretor as a NS (e.g., \citealt{buisson_2020}). Using the PRE for some of its bursts and adopting likely system parameters (namely, a high inclination and helium atmosphere), \citet{buisson_2020} derived a distance of $12.8^{+0.8}_{-0.6}$ kpc, suggesting a more conservative range of 9–18 kpc to account for systematic uncertainties. The source inclination is proposed to be high ($>$70$^\circ$) based on the duration of its X-ray eclipses \citep{buisson_2021}. 

Weekly MeerKAT monitoring of Swift J1858.6–0814 began on 2018 November 10 (MJD 58432), $\sim$16 days after the outburst was initially reported, and ceased on 2019 January 5 (MJD 58488). A second period of weekly monitoring followed from 2019 February 16 to May 23 (MJDs 58530--58626), with two additional observations on 2019 August 5 and 2020 March 2 (MJDs 58700 and 58910). The data reduction of 25 reported MeerKAT epochs is described in \citet{rhodes_2022}. 

Strong and variable LOS absorption \citep{hare_2020} makes the source's state evolution and intrinsic X-ray luminosity uncertain. In particular, \emph{Swift}/XRT fluxes may underestimate the intrinsic emission by a factor of 2--4 due to absorption \citep{van_den_eijnden_2020}, while additional scattering (probably due to the high inclination) likely further contributes to discrepancies between \emph{Swift}/BAT and XRT results \citep{rhodes_2022}. Therefore, given the unreliability of the XRT fluxes, we exclude this source from our $L_R$--$L_X$ analysis. 

Based on its X-ray properties (although uncertain), the source is assumed to have remained in the HS throughout most of our radio monitoring, except the final epoch that may correspond to the SS (\citealt{buisson_2020b, rhodes_2022}). During this time, the radio flux density was relatively stable with consistently positive spectral indices indicative of a compact jet. However, two optically thick flares were detected during the HS -- one by AMI-LA prior to our MeerKAT campaign and a second with MeerKAT on MJD $\sim$58530. \citet{rhodes_2022} suggested these may be due to transient ejecta launched during a short excursion to the IMS/SS (timescale of $\sim$days) -- which remain unresolved from the core due to the large source distance -- though they may alternatively be a rebrightening of the compact jet. We therefore classify the observation on MJD $\sim$58530 as being in the IMS, although we emphasize that this is uncertain.

\subsection*{XTE J1701–462}

\href{https://thunderkat.physics.ox.ac.uk/source/XTE%20J1701-462}{XTE J1701–462} was first detected with the All-Sky Monitor (ASM) in 2006 \citep{remillard_2006_xte}, and various Type-I X-ray bursts have confirmed a NS accretor \citep{lin_2009b}. The source has been classified as a transitioning Z–atoll source because it exhibited Z-source characteristics during the rise of its 2006/2007 outburst (e.g., \citealt{homan_2007}) and atoll-like behaviour during the decay phase \citep{lin_2009a}. Using the PRE burst, the source distance was estimated to be $8.80 \pm 1.32$ kpc \citep{lin_2009b}. Its inclination is poorly constrained, with some disagreement in the literature. The absence of eclipses or dips suggests that $i\lesssim 75^\circ$, while \citet{lin_2009a} argued that its weak Fe emission line compared to other Z-sources may imply that the inclination is not very low.  

As part of ThunderKAT, we obtained a single MeerKAT non-detection of XTE J1701–462 on 2021 July 26 (MJD 59421). The following year, in September 2022, the source entered a new outburst \citep{iwakiri_2022}, and was observed for 29 epochs with MeerKAT at $\sim$weekly cadence between 2022 September 7 and 2023 March 25 (MJDs 59829--60028). The MeerKAT data reductions are described in \citet{gasealahwe_2024}. 

The \emph{Swift}/XRT spectra were well fit by a simple power-law model, but comparisons with quasi-simultaneous NICER data indicated that they are likely distorted by instrumental effects. The NICER spectra are well described by an absorbed disc blackbody model (\texttt{tbabs*diskbb}), consistent with the results obtained from NuSTAR data \citep{thomas_2024}. These data suggest that the NICER results are more reliable and that all the MeerKAT observations were taken during the SS. To maintain a homogeneous sample in this paper, we nevertheless adopt the \emph{Swift}/XRT fluxes, incorporating an additional 20 per cent systematic uncertainty in quadrature. 

The three optically thin MeerKAT flares detected during our campaign \citep{gasealahwe_2024} are therefore likely due to ejecta launched prior to our observations that are interacting with the ISM.

\subsection*{Vela X-1}

\href{https://thunderkat.physics.ox.ac.uk/source/Vela%20X-1}{Vela X-1} is an eclipsing high-mass XRB (HMXB) discovered in 1967 \citep{chodil_1967}, consisting of the early-type supergiant HD 77581 and a NS. The strong stellar wind of the massive donor star produces a bow shock as it interacts with the ISM, first detected in narrow-band H$\alpha$ imaging \citep{kaper_1997} and later at infrared wavelengths \citep{peri_2015, maiz_2018}. Radio emission from this bow shock was subsequently discovered with MeerKAT, marking the first detection of a stellar-wind-driven radio bow shock around an XRB \citep{van_den_eijden_2022a}. We adopt the source distance of $1.99^{+0.13}_{-0.11}$ kpc, derived by \citet{kretschmar_2021} using Gaia data. The system's inclination remains uncertain despite various estimates, but is assumed to be high ($>$70$^\circ$) due to the presence of eclipses \citep{kretschmar_2021}. 

\citet{van_den_eijden_2022a} focused on the bow shock, describing the MeerKAT data reduction for the three 30-minute epochs taken over 16 days (i.e., multiple binary orbits) during ThunderKAT. While the sensitivity of these individual epochs was insufficient to detect Vela X-1 in its complex field, a combined/stacked image revealed a compact source at its position with a flux density of $96\pm40$ $\upmu$Jy. Although the detection significance is low due to elevated local RMS from the bow-shock and ISM emission, the presence of a clear point source at the XRB position, and the agreement with higher-frequency radio/mm observations, support interpreting this as a detection \citep{van_den_eijnden_2021}.

\emph{Swift}/XRT observed Vela X-1 quasi-simultaneously with the MeerKAT epochs for eight short exposures totalling $\sim$3.1 ks. Since the radio data were combined into a single image, the XRT spectra were likewise combined into one spectrum, which is well described by an absorbed blackbody and Comptonised component: \texttt{tbabs*(bbody+comptt)}. Given that $N_H$ has been reported to vary considerably, we left it free in our fits, and obtained values that are consistent with the literature \citep{kretschmar_2021}. Although Vela X-1 is known to show narrow Fe lines at 6.4–6.96 keV due to X-ray fluorescence in the stellar wind close to the NS, no significant evidence for such features were seen in the XRT spectra, such that the inclusion of additional Gaussian spectral components did not significantly improve the fits.

\section{Sources Not Included}\label{app: not_included}

A few of the XRBs monitored during ThunderKAT were excluded from the sample shared in this paper (and on our website) because they have very few radio observations (often only non-detections) and/or no quasi-simultaneous X-ray observations. These sources are: 3FGL J0427.9–6704, 4U 1728–34, 4U 1730–22, 4U 1755–338, Cygnus X-1, Cygnus X-2, Cygnus X-3, GRO J1655–40, GRS 1758–258, GX 13+1. GX 17+2, GX 301–2, GX 340+0, GX 349+2, HD96670, MAXI J0637–430, MAXI J0655–013, MAXI J0903–531, MAXI J0911–655, MAXI J1535–571, MAXI J1813–095, MAXI J1957+032, PSR J1023+0038, SAX J1712.6–3739, Scorpius X-1, Swift J1357.2–0933, Swift J1658.2–4242, Swift J1713.4–4219, Swift J1910.2–0546, V4641 Sagittarii, XTE J1739–285. While the source KQ Velorum was also observed, recent work has argued against it being an XRB \citep{leto_2022}. Additionally, we do not include the NS XRB GRS 1747–312 in our sample, as it is an eclipsing source located in a crowded field, requiring a more careful analysis which will be presented in future work. The candidate BH MAXI J1848–015 was also observed during ThunderKAT \citep{bahramian_2023}, although only jet ejecta were detected (i.e., no core radio emission). 

\section{Energy Estimate Summary}\label{app: energy_estimates}

\begin{center}
\small
{\captionsetup[table]{labelfont=bf,labelsep=period, font=small, skip=20pt}
\captionof{table}{\label{tab:jet_energy_estimates} For each source, we list an approximate estimate of the kinetic energy of the steady jet during the hard/quiescent-states (HS/QS) monitored during ThunderKAT (and X-KAT for Swift J1727.8–1613, 1A 1744–361, and Aquila X-1), rounded to the nearest half order of magnitude. These were calculated using the methods in Section \ref{sec: energy_estimates} and the nominal distance estimates ($D$) in Table \ref{tab: system_properties} of the main text. To rescale a particular estimate to a different distance ($D_{\text{new}}$), multiply by $(D_{\text{new}}/D)^{24/17}$. Values of $\sim$0 indicate that all the HS/QS measurements for the source (if any) are upper limits.}
}
\vspace{4pt}
    \begin{tabular}{lc}
        \hline 
        Source name (accretor type) & \makecell{Steady jet kinetic\\energy estimate [erg]} \\
        \hline \hline
        1A 1744–361 (NS) & $\sim$$1 \times 10^{42}$  \\
        4U 1543–47 (BH) & $\sim$$5 \times 10^{44}$ \\
        4U 1630–47 (BHC) & $\sim$$1 \times 10^{43}$  \\
        Aquila X-1 (NS) & $\sim$$5 \times 10^{42}$ \\
        Centaurus X-4 (NS) & $\sim$$0$  \\
        EXO 1846–031 (BHC) & $\sim$$1 \times 10^{43}$  \\
        GRS 1739–278 (BHC) & $\sim$$5 \times 10^{42}$  \\
        GX 339–4 (BH)& $\sim$$1 \times 10^{45}$  \\
        H1743–322 (BHC)& $\sim$$5 \times 10^{43}$  \\
        IGR J17091–3624 (BHC)& $\sim$$1 \times 10^{44}$ \\
        MAXI J1348–630 (BHC)& $\sim$$5 \times 10^{43}$  \\
        MAXI J1631–472 (BHC)& $\sim$$5 \times 10^{43}$ \\
        MAXI J1803–298 (BHC)& $\sim$$5 \times 10^{43}$ \\
        MAXI J1807+132 (NS) & $\sim$$0$  \\
        MAXI J1810–222 (NS)& $\sim$$5 \times 10^{43}$ \\
        MAXI J1816–195 (AMXP) & $\sim$0  \\
        MAXI J1820+070 (BH) & $\sim$$1 \times 10^{44}$  \\
        SAX J1808.4–3658 (AMXP) &  $\sim$$5 \times 10^{42}$ \\
        SAX J1810.8–2609 (BHC)& $\sim$$5 \times 10^{42}$   \\
        Swift J1727.8–1613 (BH) & $\sim$$5 \times 10^{44}$  \\
        Swift J1728.9–3613 (BHC)& $\sim$0 \\
        Swift J1842.5–1124 (BHC) & $\sim$$5 \times 10^{42}$  \\
        XTE J1701–462 (NS) & $\sim$0  \\
        \hline 
        \multicolumn{2}{l}{\textbf{Not used for the $L_R{-}L_X$ analysis:}} \\ \hline 
        Circinus X-1 (NS) &  $\sim$0\\
        GRS 1915+105 (BH) &  ? \\
        Swift J1858.6–0814 (NS) & $\sim$$5 \times 10^{43}$  \\
        Vela X-1 (NS HMXB) & $\sim$0 \\
        \hline
    \end{tabular}
\end{center}


\bsp	
\label{lastpage}
\end{document}